\documentclass[11pt]{article}
\pdfoutput=1
\usepackage{color}
\usepackage{mhchem}
\usepackage{xcolor}
\usepackage{cite}
\usepackage{hyperref}
\hypersetup{colorlinks=true,linkcolor=red,anchorcolor=black,citecolor=green}
\usepackage[toc,page]{appendix}
\usepackage{mathtools}
\usepackage{amsfonts}
\usepackage{bbold}
\usepackage{textcomp}
\usepackage[DIV13]{typearea}
\usepackage{amsmath, amsthm, amssymb, mathtools,empheq,latexsym,dsfont}
\usepackage{bbm}
\usepackage{amsmath,amssymb,latexsym,dsfont}
\usepackage{slashed, simplewick}
\usepackage[utf8]{inputenc}
\usepackage{graphicx}
\usepackage{graphicx,placeins}
\usepackage{makeidx}
\usepackage[font=small,labelfont=bf]{caption}
\usepackage{nicefrac}
\usepackage{subfigure}
\usepackage{array, bigdelim,multirow,multicol}
\usepackage{fancybox}
\usepackage{bm}
\usepackage{float}
\usepackage{rotating}
\usepackage{colortbl}
\usepackage{booktabs}
\usepackage{nicefrac}
\usepackage[top=2cm,textwidth=16.6cm,textheight=22.75cm]{geometry}
\usepackage{arydshln}
\usepackage{doi}
\usepackage{tikz}
\usetikzlibrary{mindmap,trees}
\usetikzlibrary{decorations.shapes}
\usepackage{framed}
\usepackage[framemethod=TikZ]{mdframed}
\usepackage{enumitem}
\graphicspath{{immagini/}}

\renewcommand{\theequation}{\thesection.\arabic{equation}}

\definecolor{Gray}{gray}{0.92}

\newcommand{\ignore}[1]{}

\newcommand{\be}{\begin{equation}}
\newcommand{\ee}{\end{equation}}
\newcommand{\bea}{\begin{eqnarray}}
\newcommand{\eea}{\end{eqnarray}}

\newcounter{Thm}[section]
\renewcommand{\theThm}{\arabic{section}.\arabic{Thm}}

\newcounter{nodecount}
\newcommand\tabnode[1]{\addtocounter{nodecount}{1} \tikz \node (\arabic{nodecount}) {#1};}

\tikzstyle{every picture}+=[remember picture,baseline]
\tikzstyle{every node}+=[inner sep=0pt,anchor=base,
text depth=.25ex,outer sep=1.5pt]
\tikzstyle{every path}+=[thick, rounded corners]
\tikzset{
	plabel/.style={inner sep=2pt}
}

\makeatletter
\@addtoreset{equation}{section}
\makeatother

\allowdisplaybreaks
\begin{document}
 \unitlength = 1mm

\setlength{\extrarowheight}{0.2 cm}

\thispagestyle{empty}

\bigskip

\vskip 1cm

\title{
\begin{flushright}
\hfill\mbox{{\small\tt  UCI-TR-2022-29}} \\[5mm]
\begin{minipage}{0.2\linewidth}
\normalsize
\end{minipage}
\end{flushright}
{\Large \bf
Neutrino Mass and Mixing Models with  Eclectic Flavor Symmetry $\Delta (27) \rtimes T'$
\\[2mm]}}
\date{}

\author{
Gui-Jun Ding$^{1}$\footnote{E-mail: {\tt
dinggj@ustc.edu.cn}},  \
Stephen F. King$^{2}$\footnote{E-mail: {\tt king@soton.ac.uk}}, \
Cai-Chang Li$^{3,4,5}$\footnote{E-mail: {\tt
ccli@nwu.edu.cn}},  \
Xiang-Gan Liu$^{6}$\footnote{E-mail: {\tt
xianggal@uci.edu}},  \
Jun-Nan Lu$^{1}$\footnote{E-mail: {\tt
hitman@mail.ustc.edu.cn}}  \
\\*[20pt]
\centerline{
\begin{minipage}{\linewidth}
\begin{center}
$^1${\it \small
Department of Modern Physics, University of Science and Technology of China,\\
Hefei, Anhui 230026, China}\\[2mm]
$^2${\it \small
Physics and Astronomy, University of Southampton, Southampton, SO17 1BJ, U.K.}\\[2mm]
$^3${\it\small School of Physics, Northwest University, Xi'an 710127, China}\\[2mm]
$^4${\it\small Shaanxi Key Laboratory for Theoretical Physics Frontiers, Xi'an 710127, China}\\[2mm]
$^5${\it\small NSFC-SPTP Peng Huanwu Center for Fundamental Theory, Xi'an 710127, China}\\[2mm]
$^6${\it \small Department of Physics and Astronomy, University of California, Irvine, CA 92697-4575, USA}
\end{center}
\end{minipage}}
\\[10mm]}
\maketitle
\thispagestyle{empty}

\begin{abstract}

The K\"ahler potentials of modular symmetry models receive unsuppressed contributions which may be controlled by a flavor symmetry, where the combination of the two symmetry types is referred to as eclectic flavor symmetry. After briefly reviewing the consistency conditions of eclectic flavor symmetry models, including with generalised (g)CP, we perform a comprehensive bottom-up study of eclectic flavor symmetry models based on $\Omega(1)\cong \Delta(27)\rtimes T^\prime$, consisting of the flavor symmetry $\Delta(27)$ in a semi-direct product with the modular symmetry $T^\prime$. The modular transformations of different $\Delta(27)$ multiplets are given by solving the consistency condition. The eight nontrivial singlets of $\Delta(27)$ are related by $T'$ modular symmetry, and they have to be present or absent simultaneously in any $\Omega(1)$ model. The most general forms of the superpotential and K\"ahler potential invariant under $\Omega(1)$ are discussed, and the corresponding fermion mass matrices are presented. Based on the eclectic flavor group $\Omega(1)$, two concrete lepton models which can successfully describe the experimental data of lepton masses and mixing parameters  are constructed. For the two models without gCP, all six mixing parameters vary in small regions. A nearly maximal atmospheric mixing angle $\theta_{23}$ and  Dirac CP phase $\delta_{CP}$ are obtained in the first model. After considering the compatible  gCP symmetry and the assumption of $\Re \tau=0$ in the first model,  the $\mu-\tau$ reflection symmetry is preserved in the charged lepton diagonal basis. As a consequence, the atmospheric mixing angle and  Dirac CP phase are predicted to be maximal, and two Majorana CP phases are predicted to be $\pi$.

\end{abstract}

\newpage


\section{Introduction}

The modular flavor symmetry is a promising approach to address the flavor puzzle of the standard model (SM)~\cite{Feruglio:2017spp}. The superpotential
is strongly constrained by the modular symmetry and all higher-dimensional operators are unambiguously fixed in the limit of unbroken supersymmetry. In particular, the modular symmetry enforces the Yukawa couplings and fermion mass matrices to be modular forms which are holomorphic functions of the complex modulus $\tau$. The modular flavor symmetry has been extensively studied from the bottom-up approach, see the recent review~\cite{Feruglio:2019ybq} and references therein for more details. The modular flavor symmetry allows to construct quite predictive fermion mass models, all the lepton masses and mixing parameters can be described only in terms of six real free parameters in the known minimal model~\cite{Ding:2022nzn}. It is remarkable that the modular invariant models exhibit a universal behavior around the fixed points, and the scaling of fermion masses and mixing parameters is independent of the details of the theory~\cite{Feruglio:2022kea,Feruglio:2023mii}. However, the K\"ahler potential is less constrained by modular symmetry, the most general K\"ahler potential compatible with modular symmetry is of the following form~\cite{Chen:2019ewa,Lu:2019vgm},
\begin{eqnarray}
\label{eq:kahler-MFS}\mathcal{K}=(-i\tau+i\bar{\tau})^{-k_{\psi}}\left(\psi^{\dagger}\psi\right)_{\mathbf{1}}+\sum_{n,\bm{r_1},\bm{r_2}} c^{(n,\bm{r_1},\bm{r_2})} (-i \tau+ i \bar \tau)^{-k_{\psi}+n}
\left(\psi^{\dagger} Y^{(n)\dagger}_{\bm{r_{1}}} Y^{(n)}_{\bm{r_2}} \psi  \right)_{\bm{1}}\,,
\end{eqnarray}
where $\psi$ stands for a generic matter field multiple with modular weight $k_{\psi}$, and $ Y^{(n)}_{\bm{r}}$ is a weight $n\neq0$ modular form in the representation $\bm{r}$ of finite modular group. Usually only the first term of Eq.~\eqref{eq:kahler-MFS} is taken, this is the so called minimal K\"ahler potential. However, we see that the K\"ahler potential generally has
a lot of terms and the couplings $c^{(n,\bm{r_1},\bm{r_2})}$ are not suppressed. Consequently these additional terms can be as important as the first term, and thus the predictive power of modular flavor symmetry would be reduced~\cite{Chen:2019ewa}.

On the other hand, the modular symmetry has been studied in the top-down approach motivated by string theory, and models with modular symmetry have been constructed in heterotic orbifolds~\cite{Baur:2019kwi,Baur:2019iai} and magnetized toroidal compactifications~\cite{Kobayashi:2016ovu,Kobayashi:2018rad,Kobayashi:2018bff,Kariyazono:2019ehj,Ohki:2020bpo,Kikuchi:2020frp,Almumin:2021fbk}.
It is found that the modular symmetry and traditional flavor symmetry appear together in top-down constructions. This leads to the concept of eclectic	flavor group~\cite{Nilles:2020nnc,Nilles:2020kgo}, and it has been developed in a series of papers~\cite{Baur:2019kwi,Baur:2019iai,Nilles:2020nnc,Nilles:2020kgo,Nilles:2020tdp,Nilles:2020gvu}. The eclectic flavor group is a maximal extension of the traditional flavor group by finite modular group, and certain consistency conditions have to be fulfilled in order to consistently combine modular symmetry with traditional flavor symmetry.  The scheme of eclectic flavor group is more predictive than the finite modular group and the traditional flavor group alone, and it combines the advantages of both approaches, in which the superpotential and the  K\"ahler potential would be severely restricted. In particular, the K\"ahler potential would be constrained to be the minimal form plus higher order corrections suppressed by powers of $\langle\Phi\rangle/\Lambda$, where $\Lambda$ denotes the cutoff scale and $\langle\Phi\rangle$ represents the vacuum expectation value (VEV) of flavons breaking the traditional flavor symmetry. It is known that K\"ahler potential is suppressed by powers of $\langle\Phi\rangle/\Lambda$ in the traditional flavor symmetry models~\cite{King:2003xq,King:2004tx,Antusch:2007vw,Chen:2012ha,Chen:2013aya}. Flavon fields are necessary in eclectic flavor models to break the traditional flavor symmetry, otherwise the unbroken flavor group would constrain the fermion mixing matrix to be the unit matrix up to possible row and column permutations. The eclectic flavor group considered here is a nontrivial product of the modular symmetry and traditional flavor symmetry.
The choices of such eclectic symmetries are quite restrictive, with only three choices of
flavor group $Z_3\times Z_3$, $\Delta(27)$ and $\Delta(54)$ catalogued so far allowing for a non-trivial eclectic extension with a CP-like
transformation~\cite{Nilles:2020nnc,Nilles:2020kgo}. This is unlike the quasi-eclectic flavor symmetry which is the direct product of a modular and a traditional flavor symmetry~\cite{Chen:2021prl}, where the modular and traditional flavor groups can be freely chosen, but the K\"ahler potential is also under control due to the presence of traditional flavor symmetry. Moreover, the first string-derived model based on the eclectic flavor group $\Omega(2)\cong[1944, 3448]$ consisting of the traditional flavor group $\Delta(54)$,
the finite modular group $T'$ and a $\mathbb{Z}^R_9$ $R$-symmetry, has been constructed, and it can give a successful fit of flavor in both the quark and lepton sectors~\cite{Baur:2022hma}.

The eclectic flavor group opens a new approach to understand the flavor puzzle of SM. In the present work, we shall study the eclectic flavor group $\Omega(1)\cong[648, 533]$ in a bottom-up way, here we adopt the naming scheme of \texttt{GAP}~\cite{GAP} and the group order of $\Omega(1)$ is 648. The eclectic group $\Omega(1)$ is the semi-direct product of the traditional flavor group $\Delta(27)$ with the finite modular group $\Gamma'_3\cong T'$.  If the generalized CP (gCP) symmetry is included, the eclectic flavor group $\Omega(1)$ would be enlarged to [1296, 2891]~\cite{Nilles:2020nnc}. The finite modular group should be a subgroup of the automorphism group of the traditional flavor group in the scheme of eclectic flavor symmetry. The outer automorphism group of $\Delta(27)$ is $GL(2, 3)$ which really contains the finite modular group $T'$ as a subgroup.   We find that the eight nontrivial singlet representations of $\Delta(27)$ transform into each other under $T'$ modular symmetry. Hence there must be seven other fields in the remaining nontrivial singlet representations of $\Delta(27)$, if a $\Omega(1)$ eclectic flavor model contains a field which is a nontrivial singlet of $\Delta(27)$.
Consequently the $\Omega(1)$ eclectic flavor symmetry is more restrictive than $\Delta(27)$ flavor symmetry alone
and terms allowed by the modular group may be forbidden by the traditional flavor group and vice versa. We shall give the most general form of the superpotential and K\"ahler potential invariant under $\Omega(1)$. The deviations from the canonical kinetic term are found to be suppressed by the VEVs of flavons. Furthermore, we shall construct two concrete lepton models with $\Omega(1)$ eclectic flavor symmetry. The field content, the modular weights and representations of the fields are assigned at will, while they are derived from the underlying geometry in top-down models. Since there are three generations of quarks and leptons, the matter fields can be assigned to triplets/anti-triplets or trivial singlets of $\Delta(27)$, but nontrivial singlets would have to form a reducible octet, as mentioned above, leading to too many families.

The rest of this paper is organized as follows: we review the formalism of eclectic flavor group in section~\ref{sec:EFG}. The eclectic flavor group $\Omega(1)$ is discussed in section~\ref{sec:EFG-Omega1-decomp}.  The modular transformation and flavor symmetry transformation are closely related through the consistency condition in the eclectic flavor group, and the $T'$ transformations of different $\Delta(27)$ multiplets are determined by solving the consistency condition. Moreover, the  gCP transformation matrices are fixed. In section~\ref{sec:mod_general} we present the most general form of the $\Omega(1)$ invariant superpotential and K\"ahler potential  for the fermion masses. Two concrete example models are presented in section~\ref{eq:example-models}, in which the light neutrino masses are generated by the Weinberg operator and type I seesaw mechanism respectively. We summarize our results and give the conclusion in section~\ref{sec:conclusion}. The Appendix~\ref{sec:Delta27_group} and Appendix~\ref{sec:Tp_group} are for the group theory of $\Delta(27)$ and $T'$ respectively, and we present the modular forms of level 3 in Appendix~\ref{sec:Tp_modu_forms}. The superpotential for Majorana fermion mass is derived in an alternative method  by firstly considering $T'$ modular symmetry then $\Delta(27)$ flavor symmetry in Appendix~\ref{app:Majorana-Superpotential-method2}.


\section{\label{sec:EFG}Formalism of eclectic flavor group }

We shall recapitulate the scheme of eclectic flavor group in this section. The modular group $SL(2, \mathbb{Z})$ denoted as $\Gamma$ is the group of two dimensional matrices with integer entries and determinant 1,
\begin{equation}
SL(2, \mathbb{Z})=\left\{\begin{pmatrix}
a  &  b \\
c  &  d
\end{pmatrix}\Bigg|ad-bc=1, a,b,c,d\in\mathbb{Z}\right\}\,.
\end{equation}
The modular group is an infinite discrete group generated by the two elements $S$ and $T$ with
\begin{equation}
S=\begin{pmatrix}
0  & 1\\-1  & 0
\end{pmatrix},~~~~T=\begin{pmatrix}
1 & 1 \\
0  & 1
\end{pmatrix}\,,
\end{equation}
which satisfy the following relations
\begin{equation}
S^4=(ST)^3=\mathbb{1}_2,~~~~S^2T=TS^2\,,
\end{equation}
where $\mathbb{1}_2$ denotes the 2-by-2 unit matrix. Under the action of a modular transformation, the complex modulus $\tau$ and a generic matter field multiplet $\psi$ transform as follow~\cite{Feruglio:2017spp}
\begin{eqnarray}
\tau \stackrel{\gamma}{\longrightarrow}\gamma\tau\equiv\frac{a\tau+b}{c\tau+d},~~~\psi\stackrel{\gamma}{\longrightarrow}(c\tau+d)^{-k_{\psi}}\rho(\gamma)\psi,~~~\gamma=\begin{pmatrix}
a  &  b\\
c  & d
\end{pmatrix}\in SL(2, \mathbb{Z})\,,
\end{eqnarray}
where  $k_{\psi}$ is the modular weight of $\psi$ and $\rho$ is a unitary representation of the finite modular group $\Gamma_N\equiv\Gamma/\pm\Gamma(N)$ or its double covering $\Gamma'_N\equiv\Gamma/\Gamma(N)$, and $\Gamma(N)$ is the so-called principal congruence subgroup of level $N$. For $N\leq5$, the  defining relations of $\Gamma_N$ and $\Gamma'_N$ are given by~\cite{Feruglio:2017spp,Liu:2019khw}
\begin{eqnarray}
\nonumber\Gamma_N&&=\left\{S,T\Big|S^2=(ST)^3=T^N=1\right\}\,,\\
\label{eq:GammaN--prime-defing}\Gamma'_N&&=\left\{S,T\Big|S^4=(ST)^3=T^N=1,~S^2T=TS^2\right\}\,.
\end{eqnarray}
Additional relations are needed to render the group finite for $N\geq 6$~\cite{deAdelhartToorop:2011re}. Notice that the representation matrices $\rho(S)$ and $\rho(T)$ should also satisfy the defining relations of $\Gamma_N$ and $\Gamma'_N$ in Eq.~\eqref{eq:GammaN--prime-defing}. Modular invariance constrains the Yukawa couplings to be modular forms which are holomorphic functions of $\tau$. The level $N$ and integer (even) weight $k_{Y}$ modular forms $Y^{(k_Y)}(\tau)$ can be arranged into multiplets of $\Gamma'_N (\Gamma_N)$~\cite{Feruglio:2017spp,Liu:2019khw}:
\begin{equation}
Y^{(k_Y)}(\tau) \stackrel{\gamma}{\longrightarrow}Y^{(k_Y)}\left(\gamma\tau\right)=\left(c\tau+d\right)^{k_Y}\rho_{Y}(\gamma)\,Y^{(k_Y)}(\tau)\,,
\end{equation}
where $\rho_{Y}(\gamma)$ is a unitary representation of $\Gamma'_N (\Gamma_N)$.

The eclectic flavor group is a nontrivial product of modular group and the traditional flavor group denoted as $G_f$. The modulus $\tau$ as well as modular form $Y^{(k_Y)}(\tau)$ is invariant under the flavor group $G_f$, and the flavor transformation of the matter multiplet $\psi$ is
\begin{eqnarray}
\psi\stackrel{g}{\longrightarrow}\rho(g)\psi,~~~~g\in G_f\,,
\end{eqnarray}
where $\rho(g)$ is a unitary representation of $G_f$. In order to consistently combine a finite modular group with a traditional flavor group, certainly consistency condition has to be fulfilled~\cite{Nilles:2020nnc}. Let us first perform a modular transformation $\gamma\in\Gamma$, subsequently a traditional flavor transformation $g$ and last perform the inverse modular transformation  $\gamma^{-1}$. Since the modulus $\tau$ is invariant under this chain of transformations, the resulting transformation should be another traditional flavor symmetry transformation $g'$~\cite{Nilles:2020nnc}, i.e.    \begin{equation}\label{eq:con_Tra_Mod}
\rho(\gamma) \rho(g) \rho^{-1}(\gamma)=\rho(g^\prime), \qquad g,g^\prime\in G_{f}, \quad \gamma \in \Gamma \,.
\end{equation}
This is the consistency condition between traditional flavor group and finite modular group.
Eq.~\eqref{eq:con_Tra_Mod} implies that the modular transformation $\gamma$ maps the traditional group element $g$ into another element $g^\prime$. Therefore Eq.~\eqref{eq:con_Tra_Mod} defines a homomorphism of the traditional flavor group $G_{f}$.  Furthermore the modular transformation is bijective. Hence modular transformation is an automorphism of the traditional group $G_{f}$. Indeed the possible representation matrices of the finite modular group $\Gamma^\prime_{N}$  form a representation of the automorphism group $\text{Aut}(G_{f})$ of the traditional flavor group $G_{f}$, i.e.
\begin{equation}\label{eq:con_Tra_Mod-2}
\rho(\gamma) \rho(g) \rho^{-1}(\gamma)=\rho(u_{\gamma}(g)) \,, \qquad \forall g\in G_{f}\,,
\end{equation}
where $\rho(\gamma)$ represents the automorphism $u_{\gamma}:G_{f}\rightarrow G_{f}$. If $u_{\gamma}$ is the trivial identity automorphism with $u_{\gamma}=1$ for any modular transformation $\gamma$ so that $u_{\gamma}(g)=g$, the modular symmetry transformation would commute with the flavor symmetry transformation and consequently the eclectic flavor group is the direct product $G_f\times \Gamma_N$ or $G_f\times \Gamma'_N$~\cite{Nilles:2020nnc}. Hence one can freely choose both $\Gamma_N$, $\Gamma'_N$ and $G_f$, the resulting flavor model would be more complex than the traditional flavor symmetry models and modular models. If $u_{\gamma}$ is an inner automorphism of $G_f$, the modular symmetry transformation $\rho(\gamma)$ would coincide with certain flavor symmetry transformation besides the automorphy factor $(c\tau+d)^{-k_{\psi}}$. Hence $u_{\gamma}$ is required to be outer automorphism of $G_f$ in the paradigm of eclectic flavor group~\cite{Nilles:2020nnc}. As a consequence, the mathematical structure of the group comprising  the traditional flavor symmetry $G_{f}$ and the finite modular group $\Gamma^\prime_{N}$ ($\Gamma_N$) is in general a semi-direct product $G_{f}\rtimes \Gamma^{\prime}_{N}$ ($G_{f}\rtimes \Gamma_{N}$). Since the finite modular groups $\Gamma_N$ and $\Gamma^\prime_{N}$ can be generated by the two generators $S$ and $T$, it is sufficient to only discuss the two outer automorphisms $u_{S}$ and $u_{T}$ and the consistency condition in Eq.~\eqref{eq:con_Tra_Mod-2} reduces to
\begin{equation}\label{eq:ST_Cons}
\rho(S)\,\rho(g)\,\rho^{-1}(S)= \rho(u_{S}(g)), \qquad \rho(T)\,\rho(g)\,\rho^{-1}(T)~=~ \rho(u_{T}(g))\;.
\end{equation}
Since $\rho(S)$ and $\rho(T)$ are representation matrices of the finite modular group, the outer automorphisms $u_{S}$ and $u_{T}$ should also satisfy the multiplication rules Eq.~\eqref{eq:GammaN--prime-defing} of the finite modular group $\Gamma^\prime_N$ ($\Gamma_N$):
\begin{equation}\label{eq:uS_uT_rules}
\left(u_{S}\right)^{N_s} =\left(u_{T}\right)^N =\left(u_{S} \circ u_{T}\right)^3=1,  \qquad \left(u_{S}\right)^2  \circ u_{T} = u_{T} \circ \left(u_{S}\right)^2\;,
\end{equation}
with $N_s=4$ for $\Gamma'_N$ and $N_s=2$ for $\Gamma_N$, additional relations are necessary for level $N\geq6$. Therefore the finite modular group $\Gamma^\prime_N$ ($\Gamma_N$) must be a subgroup of the full automorphism group of the traditional flavor group $G_{f}$.

\subsection{\label{subsec:EFG-GCP}Eclectic flavor group and  gCP}

The modular symmetry group $\Gamma$ can be consistently combined with the gCP symmetry. Then a new generator $K_{*}$ corresponding to gCP transformation could be introduced and the modular group  $\Gamma\cong SL(2,\mathbb{Z})$ is enhanced to  $\Gamma^*\cong GL(2,\mathbb{Z})$~\cite{Novichkov:2019sqv}:
\begin{equation}
\Gamma^{*}= \Big\{ \tau \stackrel{S}{\longrightarrow} -1/\tau, ~~ \tau \stackrel{T}{\longrightarrow} \tau + 1, ~~  \tau \stackrel{K_*}{\longrightarrow} -\tau^{*} \Big\}\,,
\end{equation}
where the action of the CP transformation $K_{*}$ on the modulus $\tau$ is represented by the matrix
\begin{equation}
K_{*}=\left(\begin{array}{cc}
		1  &  0  \\
		0  &  -1
	\end{array}\right)\,.
\end{equation}
As a consequence, the action of $\Gamma^{*}$ on the complex modulus $\tau$
is given by
\begin{equation}
	\left(\begin{array} {cc}a & b \\ c &d \end{array}\right)
	\in \Gamma^*:\quad
	\begin{cases}
		\tau \rightarrow \cfrac{a\tau + b}{c\tau + d} &\quad \text{for} \quad ad - bc = 1\,, \\
		\tau \rightarrow \cfrac{a\tau^{*} + b}{c\tau^{*} + d}
		&\quad \text{for} \quad ad - bc = -1\,.
	\end{cases}
\end{equation}
On the other hand, the CP transformation acts on the matter field and the modular form multiplets as
\begin{equation}\label{eq:K*_to_matter}
	\psi(x)\stackrel{K_{*}}{\longrightarrow} \rho(K_*)[\psi^{\dagger}(t, -{\bm x})]^T, \qquad Y(\tau)\stackrel{K_{*}}{\longrightarrow} Y(-\tau^{*})= \rho(K_*)Y^{*}(\tau)\,,
\end{equation}
where the unitary matrix $\rho(K_*)$ is the  gCP transformation matrix of a generic matter field $\psi$. Note that the obvious action of CP on the possible spinor indices has been suppressed in Eq.~\eqref{eq:K*_to_matter}. As the CP transformation $K_{*}$ is order 2, then one can find
\begin{equation}\label{eq:rho_K*_order2}
\rho(K_*)\,\rho^*(K_*) = \mathbb{1}\quad \text{or equivalently} \quad \rho^*(K_*)= \rho^{-1}(K_*)\,.
\end{equation}
The consistency between the modular symmetry and gCP symmetry requires the following consistency condition has to be satisfied~\cite{Novichkov:2019sqv,Ding:2021iqp}
\begin{equation}\label{eq:consistency_cond}
\rho(K_*)\rho^{*}(\gamma)	\rho^{-1}(K_*)=\rho(\mathfrak{u}(\gamma)), \qquad \gamma\in\Gamma\,,
\end{equation}
where $\mathfrak{u}(\gamma)$ denotes an outer automorphism of the modular group with
\begin{equation}
	\gamma=\left(\begin{array}{cc}
		a  &  b \\
		c  &  d
	\end{array}\right)\rightarrow \mathfrak{u}(\gamma)=K_{*}\gamma K^{-1}_{*}=\left(\begin{array}{cc}
		a  &  -b \\
		-c  &  d
	\end{array}\right)\,.
\end{equation}
It is easy to check that $\mathfrak{u}(S)=S^{-1}$ and $\mathfrak{u}(T)=T^{-1}$. It is sufficient to impose the consistency condition in Eq.~\eqref{eq:consistency_cond} on the generators $S$ and $T$ of the modular group $\Gamma$, then one can obtain~\cite{Novichkov:2019sqv,Ding:2021iqp,Qu:2021jdy}
\begin{equation}\label{eq:cons_Gn_CP}
\rho(K_*)\rho^{*}(S)\rho^{-1}(K_*)=\rho^{-1}(S),\qquad 	\rho(K_*)\rho^{*}(T)\rho^{-1}(K_*)=\rho^{-1}(T)\,.
\end{equation}
If one works in the basis where both representation matrices  $\rho(S)$ and $\rho(T)$ are unitary and symmetric, $\rho(K_*)$ would reduce to the canonical CP transformation,
\begin{equation}
\rho(K_*)=\mathbb{1}\,,
\end{equation}
where an arbitrary overall phase is omitted.

Furthermore,  it is known that the following consistency condition has to be satisfied when combining the traditional flavor $G_{f}$ with the gCP symmetry~\cite{Grimus:1995zi,Feruglio:2012cw,Ding:2013hpa,Holthausen:2012dk,Chen:2014tpa}:
\begin{equation}\label{eq:con_uK}
\rho(K_*)\rho^*(g)\rho^{-1} (K_*)= \rho(u_{{K}_*}(g))\,, \qquad \forall g\in G_{f}\,,
\end{equation}
where $u_{{K}_*}$ is an automorphism of the traditional flavor group $G_{f}$. Furthermore, the CP transformation $\rho(K_*)$ with order 2 must be subject to the consistency condition in Eq.~\eqref{eq:cons_Gn_CP}. Consequently, the automorphisms $u_{{K}_*}$, $u_{S}$ and $u_{T}$ of the traditional flavor group $G_{f}$ should satisfy the following multiplication rules
\begin{eqnarray}
\nonumber &&\left(u_{S}\right)^{N_{s}} =\left(u_{T}\right)^N =\left(u_{S} \circ u_{T}\right)^3=1, \qquad \left(u_{S}\right)^2  \circ u_{T} = u_{T} \circ \left(u_{S}\right)^2\,, \\
		\label{eq:auto_Ga*_rules}&&(u_{{K}_*})^2=1,\qquad u_{{K}_*}\circ u_{S}\circ u_{{K}_*}=u_{{S}}^{-1},\qquad u_{{K}_*}\circ u_{T}\circ u_{{K}_*}=u_{{T}}^{-1}\,,
\end{eqnarray}
for $N\leq5$ and additional multiplication rules should be imposed for $N>6$.

From the general formalism sketched above, one could find all possible corresponding eclectic flavor group and gCP for a given traditional flavor group $G_{f}$. It is necessary to find out all outer automorphisms of the traditional flavor group $G_{f}$ and determine whether there exist particular outer automorphisms $u_{S}$, $u_{T}$ and $u_{K_{*}}$ satisfying the multiplication rules in Eq.~\eqref{eq:auto_Ga*_rules}. For a field multiplet transforming in the representation $\rho(g)$ of the traditional flavor group $G_f$, one can not freely assign the modular transformation of the fields as in modular symmetry alone. Nevertheless, the modular transformation matrices $\rho(S)$, $\rho(T)$ and the gCP transformation $\rho(K_{*})$ should be determined by solving the consistency conditions in Eqs.~\eqref{eq:ST_Cons}, \eqref{eq:cons_Gn_CP} and \eqref{eq:con_uK}.
In general, the representation matrices $\rho(S)$, $\rho(T)$ span a reducible representation of the finite modular group $\Gamma^\prime_N$ ($\Gamma_N$). In the following, we shall study the eclectic flavor group $\Omega(1)$ which is semi-direct product of the traditional flavor group $\Delta(27)$ with the finite modular group $T^\prime$.

\section{\label{sec:EFG-Omega1-decomp} The eclectic flavor group $\Omega(1)\cong\Delta(27)\rtimes T^\prime$ }

In the present work, we shall consider the eclectic flavor group $\Omega(1)$ which is the extension of the traditional flavor symmetry $\Delta(27)$ by  finite modular group $T^\prime$. The mathematical structure of the group $\Omega(1)$ arising from the combination of the flavor symmetry $\Delta(27)$ and the $T'$ modular symmetry is a semi-direct product of the form $\Omega(1)\cong\Delta(27)\rtimes T^\prime$. The group theories of the traditional flavor group $\Delta(27)$ and the finite modular group $T^\prime$ are given in Appendix~\ref{sec:Delta27_group} and Appendix~\ref{sec:Tp_group}, respectively. The finite modular group compatible with $\Delta(27)$  must be generated by the outer automorphisms of $\Delta(27)$.  We find that the outer automorphism group of $\Delta(27)$ is  $GL(2,3)$\footnote{$GL(2,3)$ is the group of all $2\times2$ invertible matrices whose elements are from the three-member ring $\mathbb{Z}_3$.}:
\begin{equation}
	\mathrm{Out}\left(\Delta(27)\right) ~\cong~ \text{Aut}\left(\Delta(27)\right)/\text{Inn}\left(\Delta(27)\right) ~\cong~ GL(2,3)\,,
\end{equation}
where $\text{Aut}\left(\Delta(27)\right)$ and $\text{Inn}\left(\Delta(27)\right)$ represent the automorphism group and inner automorphism group of $\Delta(27)$, respectively. Their group ID in \texttt{GAP} system are $\text{Aut}\left(\Delta(27)\right) \cong [432, 734]$ and $\text{Inn}\left(\Delta(27)\right)\cong Z_{3}\times Z_3 \cong [9,2]$. The group $GL(2,3)$ contains two finite modular groups $\Gamma_2\cong S_3$ and $\Gamma^\prime_3\cong T^\prime\cong SL(2,3)$ which can be generated by the outer automorphisms. This implies that the traditional flavor group $\Delta(27)$ can be extended to eclectic flavor group in two ways: $\Delta(27)\rtimes S_{3}$ or $\Delta(27)\rtimes T^\prime$~\cite{Nilles:2020nnc}.  In the present work, we are concerned with the traditional flavor group $\Delta(27)$ and its eclectic extension by $\Gamma^\prime_3\cong T^\prime$, and the scenarios with/without gCP will be studied.

\subsection{\label{sec:Tp_extention}Finite modular group $\Gamma^\prime_{3}\cong T^{\prime}$ and gCP }

The two outer automorphisms $u_{S}$ and $u_{T}$ can be taken to be~\cite{Nilles:2020nnc}
\begin{equation}\label{eq:TpfromDelta27}
	u_{S}(A) = B^2 {A},\qquad  u_{S}(B)  = B^2A^2 ,\qquad
	   u_{T}(A) =BA, \qquad   u_{T}(B) =B\;,
\end{equation}
where $A$ and $B$ are the generators of $\Delta(27)$, see the Appendix~\ref{sec:Delta27_group} for the group theory of $\Delta(27)$.
One can check that the outer automorphisms $u_{S}$ and $u_{T}$ in Eq.~\eqref{eq:TpfromDelta27} satisfy the multiplication rules in Eq.~\eqref{eq:uS_uT_rules} with $N_s=4$ and $N=3$. As a consequence, the $\Delta(27)$ flavor symmetry could be extended by the finite modular group is $\Gamma^\prime_{3}\cong T^{\prime}$ in the paradigm of eclectic flavor group. If gCP symmetry is considered further, the corresponding automorphism $u_{K_*}$ should be included. The three automorphisms $u_{S}$, $u_{T}$ and $u_{K_*}$ should satisfy the following multiplication rules
\begin{eqnarray}
	\nonumber &&\left(u_{S}\right)^{4} = \left(u_{T}\right)^3 =\left(u_{S} \circ u_{T}\right)^3=1, \qquad \left(u_{S}\right)^2  \circ u_{T} = u_{T} \circ \left(u_{S}\right)^2, \\
\label{eq:auto_Tp*_rules}	&&(u_{{K}_*})^2=1,\qquad u_{{K}_*}\circ u_{S}\circ u_{{K}_*}=u_{{S}}^{-1},\qquad u_{{K}_*}\circ u_{T}\circ u_{{K}_*}=u_{{T}}^{-1}\,,
\end{eqnarray}
where the outer automorphism $u_{{K}_*}$ acts on the generators $A$ and $B$ as follows
\begin{equation}\label{eq:CPfromDelta27}
u_{{K}_*}(A) = A^2 B,\qquad
u_{{K}_*}(B) =  A^2B A\,.
\end{equation}
Then the resulting eclectic flavor group is [1296, 2891], it has 1296 group elements. Moreover, the gCP transformation matrix $\rho_{\bm{r}}(K_*)$ must satisfy Eq.~\eqref{eq:rho_K*_order2} and the consistency condition in Eq.~\eqref{eq:cons_Gn_CP}. In our basis given in Eq.~\eqref{eq:Tp_irre},  the group generators $S$ and $T$ are represented by symmetric matrices for all irreducible representations of $T^\prime$. As a consequence, the expressions of $\rho_{\bm{r}}(K_*)$ are identity matrices for all irreducible representations up to an overall phase, i.e.
\begin{equation}\label{eq:rho_K_Tp}
 \rho_{\bm{r}}(K_*)=\mathbb{1}_{\bm{r}}, \qquad \bm{r}=\bm{1}, \,\bm{1}', \,\bm{1}'',\,\bm{2}, \,\bm{2}', \,\bm{2}'', \, \bm{3}\,.
\end{equation}
The modular forms of level $N=3$ up to weight 10 are given in Appendix~\ref{sec:Tp_modu_forms}. From the discussion of Appendix~\ref{sec:Tp_modu_forms}, we find that the CP transformation acts on  the modular form multiplets $Y^{(k)}_{\bm{r}}(\tau)$ as follow~\cite{Novichkov:2019sqv,Ding:2021iqp,Qu:2021jdy}
\begin{equation}\label{eq:Ytau_K_Tra}	Y^{(k_{Y})}_{\bm{r}}(\tau)~\stackrel{K_*}{\longrightarrow}~Y^{(k_{Y})}_{\bm{r}}(-\bar{\tau})=\left(Y^{(k_{Y})}_{\bm{r}}(\tau)\right)^*\,,
\end{equation}
where $Y^{(k_{Y})}_{\bm{r}}(\tau)$ represents the weight $k_{Y}$ and level 3 modular multiplets in the irreducible representation $\bm{r}$ of the finite modular group $T^\prime$.

\subsection{$T^\prime$ modular transformations of $\Delta(27)$ multiplets }

Automorphisms are mappings from a group into itself, and they can be divided into inner automorphisms and outer automorphisms. It is known that all inner automorphism maps each conjugacy class into itself, while certain conjugacy classes are mapped into different ones by the outer automorphisms. In order to determine the explicit form of the modular transformations $\rho(S)$, $\rho(T)$ and the gCP transformation $\rho(K_*)$ corresponding to the outer automorphisms $u_{S}$, $u_{T}$ and $u_{K_*}$, one needs to know the actions of the three automorphisms $u_{S}$, $u_{T}$ and $u_{K_*}$ on the conjugacy classes and representations of $\Delta(27)$. It is necessary to consider how these three automorphisms act on the character table of $\Delta(27)$, the results are summarized in table~\ref{tab:character_Delta27}.

Firstly, we discuss how the three automorphisms $u_{S}$, $u_{T}$ and $u_{K_*}$ act on the eleven conjugacy classes of $\Delta(27)$. From Eqs.~\eqref{eq:TpfromDelta27} and \eqref{eq:D27CC}, we see that the outer automorphism $u_S:(A,\,B)\rightarrow (B^2A,\,B^2A^2)$ acts on the all conjugacy classes elements of $\Delta(27)$ as
\begin{equation}\label{eq:uS_CC}
u_S:~ 3C^{(3)}_3\rightarrow 3C^{(4)}_3\rightarrow 3C^{(6)}_3\rightarrow 3C^{(2)}_3\rightarrow 3C^{(3)}_3\,,
\quad 3C^{(8)}_3\rightarrow 3C^{(1)}_3\rightarrow 3C^{(7)}_3\rightarrow 3C^{(5)}_3\rightarrow 3C^{(8)}_3\,,
\end{equation}
and the elements of the remaining three conjugacy classes map to the elements of the same conjugacy classes under the action of the  automorphism $u_S$. Now let us discuss the action of the outer automorphism $u_T$ on the eleven conjugacy classes of $\Delta(27)$. From Eqs.~\eqref{eq:TpfromDelta27} and \eqref{eq:D27CC} we find that the results of the outer automorphism $u_T$ acting on the conjugacy classes of $\Delta(27)$ are given by
\begin{equation}
u_T:~ 3C^{(3)}_3\rightarrow 3C^{(5)}_3\rightarrow 3C^{(4)}_3\rightarrow 3C^{(3)}_3, \qquad 3C^{(1)}_3\rightarrow 3C^{(2)}_3\rightarrow 3C^{(6)}_3\rightarrow 3C^{(1)}_3\,.
\end{equation}
Furthermore, the outer automorphism $u_T$ acting on the other five conjugacy classes of $\Delta(27)$  will make them unchanged. If gCP symmetry is imposed, we need to consider the actions of the outer automorphism $u_{K^*}$ on the eleven conjugacy classes.
We find that the outer automorphism $u_{K_*}$ acts on the conjugacy classes as
\begin{equation}
 u_{K_*}:~1C^{(1)}_3\leftrightarrow 1C^{(2)}_3, \quad 3C^{(1)}_3\leftrightarrow 3C^{(5)}_3,\quad 3C^{(2)}_3 \leftrightarrow 3C^{(3)}_3, \quad 3C^{(4)}_3\leftrightarrow 3C^{(6)}_3,
\end{equation}
where the symbol ``$\leftrightarrow$'' denotes that the outer automorphism $u_{K_*}$ interchanges the two conjugacy classes on the two sides of it. The remaining three conjugacy classes of $\Delta(27)$ are invariant under $u_{K_*}$.

Secondly, let us show how the two automorphisms $u_{S}$ and $u_{T}$ act on the eleven irreducible representations of $\Delta(27)$. The consistency condition Eq.~\eqref{eq:con_Tra_Mod-2} may be understood as the action of an outer automorphism $u_{\gamma}$ on an irreducible representation $\rho$ of traditional flavor symmetry as follow
\begin{equation}\label{eq:rep_trans}
u:\rho\rightarrow\rho^\prime,~~~ \text{with}~~~\rho^\prime=\rho\circ u_{\gamma}\,.
\end{equation}
Then we have
\begin{equation}
\rho^\prime(u_{\gamma}^{-1}(g))=\rho(g)\,.
\end{equation}
This implies that automorphisms are symmetries of the character table. The action of the automorphism on the group elements and irreducible representations is to exchange the rows and columns of the character table. Eq.~\eqref{eq:rep_trans} indicates that a modular transformation connects the irreducible representation $\rho$ of traditional flavor symmetry to another irreducible representation $\rho^\prime=\rho\circ u_{\gamma}$. If $\rho^\prime\neq \rho$, then $\rho$ does not contain all representations that are connected via the outer automorphism $u_{\gamma}$, there is no modular transformation that fulfils the consistency condition \eqref{eq:con_Tra_Mod-2}. To implement this modular transformation, the representation $\rho$ therefore has to be enlarged by the missing representations, i.e. we should extend the irreducible representation $\rho$ to a reducible representation  which contains the irreducible representations $\rho$ and $\rho^\prime$ at the same time. This is the reason why we have to define modular transformation on the reducible representation of $\Delta(27)$ sometimes, as shown in the following. Now let us give the actions of outer automorphisms $u_{S}$ and $u_T$ on irreducible representations of $\Delta(27)$ and show how to choose the representations in which modular transformation fulfils the consistency conditions Eq.~\eqref{eq:ST_Cons}.  The outer automorphism $u_S$ acts on the eight nontrivial singlet irreducible representations as
\begin{equation}
\label{eq:uS-D27-singlets}u_S:~ \bm{1_{0,1}}\rightarrow \bm{1_{2,2}}\rightarrow\bm{1_{0,2}}\rightarrow\bm{1_{1,1}}\rightarrow\bm{1_{0,1}}\,, \qquad \bm{1_{1,0}}\rightarrow \bm{1_{1,2}}\rightarrow\bm{1_{2,0}}\rightarrow\bm{1_{2,1}}\rightarrow\bm{1_{1,0}} \,.
\end{equation}
In a similar way, we find that the outer automorphism $u_T$ acts on the eight nontrivial singlet representations as
\begin{equation}
\label{eq:uT-D27-singlets}\hskip-0.1in u_T:~ \bm{1_{0,1}}\rightarrow\bm{1_{1,1}}\rightarrow \bm{1_{2,1}}\rightarrow\bm{1_{0,1}}, ~~ \bm{1_{0,2}}\rightarrow\bm{1_{2,2}}\rightarrow \bm{1_{1,2}}\rightarrow\bm{1_{0,2}}, ~~ \bm{1_{1,0}}\rightarrow\bm{1_{1,0}},\quad  \bm{1_{2,0}}\rightarrow\bm{1_{2,0}}\,.
\end{equation}
The action of the automorphisms $u_{S}$ and $u_{T}$ on the character table of $\Delta(27)$ is shown in table~\ref{tab:character_Delta27}. The automorphisms $u_S$ and $u_{T}$ act on the character table by interchanging its the columns and rows, such that the table keeps invariant, as an outer automorphism should do.

In short, the eight nontrivial one-dimensional representations $\bm{1_{(r,s)}}$ of $\Delta(27)$ with $\bm{r,s}\neq0$ are related by the $T'$ modular symmetry, in order to fulfil the consistency condition Eq.~\eqref{eq:ST_Cons} with  $u_S$ and $u_T$ given by Eq.~\eqref{eq:TpfromDelta27}, the eight nontrivial one-dimensional representations must be extended to a eight-dimensional  reducible representation
\begin{equation}\label{eq:D27_8DR_def} \bm{8}=\bm{1_{0,1}}\oplus\bm{1_{0,2}}\oplus\bm{1_{1,0}}\oplus\bm{1_{1,1}}\oplus\bm{1_{1,2}}\oplus\bm{1_{2,0}}\oplus\bm{1_{2,1}}\oplus\bm{1_{2,2}}.
\end{equation}
Without loss of generality, the eight-dimensional reducible representation matrices  $\rho_{\bm{8}}(A)$ and $\rho_{\bm{8}}(B)$ can take the following form
\begin{eqnarray}
\nonumber \rho_{\bm{8}}(A)&=&\text{diag}(\rho_{\bm{1_{0,1}}}(A),\rho_{\bm{1_{0,2}}}(A),\rho_{\bm{1_{1,0}}}(A),\rho_{\bm{1_{1,1}}}(A),\rho_{\bm{1_{1,2}}}(A),\rho_{\bm{1_{2,0}}}(A),\rho_{\bm{1_{2,1}}}(A),\rho_{\bm{1_{2,2}}}(A))\\
\nonumber &=& \text{diag}(1,1,\omega,\omega,\omega,\omega^2,\omega^2,\omega^2)\,,\\
\nonumber \rho_{\bm{8}}(B)&=&\text{diag}(\rho_{\bm{1_{0,1}}}(B),\rho_{\bm{1_{0,2}}}(B),\rho_{\bm{1_{1,0}}}(B),\rho_{\bm{1_{1,1}}}(B),\rho_{\bm{1_{1,2}}}(B),\rho_{\bm{1_{2,0}}}(B),\rho_{\bm{1_{2,1}}}(B),\rho_{\bm{1_{2,2}}}(B))\\
\label{eq:8D_AB}	&=& \text{diag}(\omega,\omega^2,1,\omega,\omega^2,1,\omega,\omega^2)\,.
\end{eqnarray}
Analogously the outer automorphisms $u_{S}$ and $u_{T}$ map the trivial singlet representation $\bm{1_{0,0}}$ to itself. Furthermore the two three-dimensional representations $\bm{3}$ and $\bm{\bar{3}}$ of $\Delta(27)$ are invariant under the actions of $u_{S}$ and $u_{T}$. Hence the three irreducible representations $\bm{1_{0,0}}$, $\bm{3}$ and $\bm{\bar{3}}$ need not be enlarged to other reducible representations to implement the outer automorphisms $u_{S}$ and $u_{T}$.
In order to obtain the solutions of the consistency conditions in  Eq.~\eqref{eq:ST_Cons} for the outer automorphisms in Eq.~\eqref{eq:TpfromDelta27}, we should find the solutions of the following equalities
\begin{eqnarray}
\nonumber &&\rho_{\bm{r}}(S)\,\rho_{\bm{r}}(A)\,\rho^{-1}_{\bm{r}}(S) ~=~ \rho_{\bm{r}}(B^2A),\qquad \rho_{\bm{r}}(S)\,\rho_{\bm{r}}(B)\,\rho^{-1}_{\bm{r}}(S)~=~ \rho_{\bm{r}}(B^2\,A^2) \,,\\
\label{eq:cons_D27_Tp}&&\rho_{\bm{r}}(T)\,\rho_{\bm{r}}(A)\,\rho^{-1}_{\bm{r}}(T)~=~ \rho_{\bm{r}}(BA), \qquad \rho_{\bm{r}}(T)\,\rho_{\bm{r}}(B)\,\rho^{-1}_{\bm{r}}(T)~=~ \rho_{\bm{r}}(B)\,,
\end{eqnarray}
where $\bm{r}$ can take the three irreducible representations  $\bm{1_{0,0}}$, $\bm{3}$ and $\bm{\bar{3}}$,  and the reducible eight-dimensional representation $\bm{8}$ in Eq.~\eqref{eq:D27_8DR_def} of $\Delta(27)$. As elements $S$ and $T$ are the generators of the finite modular group $\Gamma^\prime_{3}$, then the solutions of $\rho_{\bm{r}}(S)$ and $\rho_{\bm{r}}(T)$ must satisfy the multiplication rules in Eq.~\eqref{eq:Tp_Mult_rules}, i.e.
\begin{equation}\label{eq:Tp_mul_rules_rhoST}
\rho^4_{\bm{r}}(S)=\rho^3_{\bm{r}}(T)=\left[\rho_{\bm{r}}(S)\rho_{\bm{r}}(T)\right]^3=\mathbb{1}_{\bm{r}},\qquad \rho^2_{\bm{r}}(S)\rho_{\bm{r}}(T)=\rho_{\bm{r}}(T)\rho^2_{\bm{r}}(S)\,.
\end{equation}
For the trivial singlet $\bm{r}=\bm{1_{0,0}}$, it is easy to see $\rho_{\bm{1_{0,0}}}(S)=1$ and $\rho_{\bm{1_{0,0}}}(T)=1\,, \omega\,\text{or}\,\omega^2$. As a consequence, a field invariant under $\Delta(27)$ should be $T'$ singlet $\bm{1}$, $\bm{1^\prime}$ or $\bm{1^{\prime\prime}}$. For the reducible octet representation $\bm{r}=\bm{8}$ of $\Delta(27)$, we find that three solutions for $\rho_{\bm{r}}(S)$ and $\rho_{\bm{r}}(T)$ which fulfill both the consistency conditions in Eq.~\eqref{eq:cons_D27_Tp} and the multiplication rules in Eq.~\eqref{eq:Tp_mul_rules_rhoST}. In the basis of Eq.~\eqref{eq:8D_AB}, the three solutions of $\rho_{\bm{r}}(S)$ and $\rho_{\bm{r}}(T)$ take the following form:
\begin{equation}
\rho_{\bm{8_k}}(S)=\left(
\begin{array}{cccccccc}
	0 & 0 & 0 & 0 & 0 & 0 & 0 & 1 \\
	0 & 0 & 0 & 1 & 0 & 0 & 0 & 0 \\
	0 & 0 & 0 & 0 & 1 & 0 & 0 & 0 \\
	1 & 0 & 0 & 0 & 0 & 0 & 0 & 0 \\
	0 & 0 & 0 & 0 & 0 & 1 & 0 & 0 \\
	0 & 0 & 0 & 0 & 0 & 0 & 1 & 0 \\
	0 & 0 & 1 & 0 & 0 & 0 & 0 & 0 \\
	0 & 1 & 0 & 0 & 0 & 0 & 0 & 0 \\
\end{array}
\right)\,, ~~\rho_{\bm{8_k}}(T)=\omega^{k}\left(
\begin{array}{cccccccc}
	0 & 0 & 0 & 1 & 0 & 0 & 0 & 0 \\
	0 & 0 & 0 & 0 & 0 & 0 & 0 & 1 \\
	0 & 0 & 1 & 0 & 0 & 0 & 0 & 0 \\
	0 & 0 & 0 & 0 & 0 & 0 & 1 & 0 \\
	0 & 1 & 0 & 0 & 0 & 0 & 0 & 0 \\
	0 & 0 & 0 & 0 & 0 & 1 & 0 & 0 \\
	1 & 0 & 0 & 0 & 0 & 0 & 0 & 0 \\
	0 & 0 & 0 & 0 & 1 & 0 & 0 & 0 \\
\end{array}
\right)\,,
\end{equation}
with $k=0,1,2$. For the three solutions, the eight-dimensional representation matrices $\rho_{\bm{8_{k}}}(S)$ and $\rho_{\bm{8_{k}}}(T)$  generate the finite modular group $\Gamma^\prime_3 \cong T^\prime$ and the corresponding eight-dimensional representations decompose into different irreducible representations of $T^\prime$ as follows:
\begin{equation}
\bm{8_k}=\bm{1^k} \oplus\bm{2^{[k+1]}}\oplus\bm{2^{[k+2]}}\oplus\bm{3}\,,
\end{equation}
in notation of Appendix~\ref{sec:Tp_group}. The representation matrices of the reducible octet $\bm{8_k}$ can be make block-diagonal by performing the similarity transformations $\Omega_{\bm{8_k}}\rho_{\bm{8_k}}(\gamma)\Omega^\dagger_{\bm{8_k}}$ with the unitary matrices $\Omega_{\bm{8_k}}$ given by

\begin{equation}
	 \Omega_{\bm{8_k}}=\frac{U^{k}_{8}}{2\sqrt{6}}\left(
	\begin{array}{cccccccc}
		\sqrt{3} & \sqrt{3} & \sqrt{3} & \sqrt{3} & \sqrt{3} & \sqrt{3} & \sqrt{3} & \sqrt{3} \\
		2 \omega  & -2 \omega  & 0 & 2 \omega ^2 & -2 & 0 & 2 & -2 \omega ^2 \\
		-\sqrt{2} & \sqrt{2} & -i \sqrt{6} & -\sqrt{2} & \sqrt{2} & i \sqrt{6} & -\sqrt{2} & \sqrt{2} \\
		-\sqrt{2}\,\omega  & \sqrt{2}\, \omega  & i \sqrt{6}\, \omega  & -\sqrt{2}\, \omega  & \sqrt{2}\, \omega  & -i \sqrt{6} \,\omega  & -\sqrt{2} \,\omega  & \sqrt{2}\, \omega  \\
		-2 & 2 & 0 & -2 \omega ^2 & 2 \omega  & 0 & -2 \omega  & 2 \omega ^2 \\
		-\omega ^2 & -\omega ^2 & 3 \omega ^2 & -\omega ^2 & -\omega ^2 & 3 \omega ^2 & -\omega ^2 & -\omega ^2 \\
		2 \omega  & 2 \omega  & 0 & 2 & 2 \omega ^2 & 0 & 2 \omega ^2 & 2 \\
		2 & 2 & 0 & 2 \omega  & 2 \omega ^2 & 0 & 2 \omega ^2 & 2 \omega  \\
	\end{array}
	\right)\,,
\end{equation}
with
\begin{equation}
	U_{8}=\left(\begin{array}{cc} \mathbb{1}_{5} & \mathbb{0}  \\
	 \mathbb{0}  & P
	\end{array}\right)\,,
\end{equation}
where the matrix $P$ is given in Eq.~\eqref{eq:per_mat}. Using these similarity transformation, we can obtain the decomposition of the octet
$\Phi_{\bm{8_{k}}}=(\phi_{1},\phi_{2},\phi_{3},\phi_{4},\phi_{5},\phi_{6},\phi_{7},\phi_{8})^T$  into one singlet, two doublets and one triplet of the finite modular symmetry $T^\prime$
\begin{equation}
 \left\{\begin{array}{ll}
\bm{1^k}  : & \frac{1}{2\sqrt{2}}\left(\phi_{1}+\phi_{2}+\phi_{3}+\phi_{4}+\phi_{5}+\phi_{6}+\phi_{7}+\phi_{8}\right)\,,\\
\bm{2^{[k+1]}}  : & \frac{1}{2\sqrt{3}}\left(\begin{array}{c}\sqrt{2} \left[\omega  (\phi_{1}-\phi_{2})+\omega^2(  \phi_{4}- \phi_{8})-\phi_{5}+\phi_{7}\right]\\
-\phi_{1}+\phi_{2}-i \sqrt{3} \phi_{3}-\phi_{4}+\phi_{5}+i \sqrt{3} \phi_{6}-\phi_{7}+\phi_{8}\end{array}\right) \,,\\
\bm{2^{[k+2]}}  : & \frac{1}{2\sqrt{3}}\left(\begin{array}{c}\omega  \left(-\phi_{1}+\phi_{2}+i \sqrt{3} \phi_{3}-\phi_{4}+\phi_{5}-i \sqrt{3} \phi_{6}-\phi_{7}+\phi_{8}\right)\\
\sqrt{2} \left[-\phi_{1}+\phi_{2}-\omega^2 (\phi_{4}-\phi_{8})+\omega(\phi_{5}-\phi_{7}) \right]\end{array}\right) \,,\\
\bm{3}  : & \frac{P^{k}}{2\sqrt{6}}\left(\begin{array}{c} -\omega ^2(\phi_{1}+\phi_{2}-3 \phi_{3}+\phi_{4}+\phi_{5}-3 \phi_{6}+\phi_{7}+\phi_{8})\\
2 \left[\omega  (\phi_{1}+\phi_{2})+\phi_{4}+\phi_{8}+\omega^2 (\phi_{5}+\phi_{7})\right]\\
2 \left[\phi_{1}+\phi_{2}+\omega  (\phi_{4}+\phi_{8})+\omega^2 (\phi_{5}+\phi_{7})\right]\end{array}\right)\,.
\end{array}
\right.
\end{equation}
Furthermore, for the three-dimensional representation $\bm{3}$ of $\Delta(27)$ in Eq.~\eqref{eq:Delta27_irre}, the modular transformation $\rho_{\bm{r}}(S)$ and $\rho_{\bm{r}}(T)$ which must fulfil the consistency conditions in Eq.~\eqref{eq:cons_D27_Tp} and the multiplication rules in Eq.~\eqref{eq:Tp_mul_rules_rhoST} are determined to be
\begin{equation}\label{eq:3DTpfromDelta27}
\rho_{\bm{3_k}}({S}) = \frac{i}{\sqrt{3}}\left(
\begin{array}{ccc}
	\omega ^2 & \omega  & \omega  \\
	\omega  & \omega ^2 & \omega  \\
	\omega ^2 & \omega ^2 & 1 \\
\end{array}
\right) \;,\qquad
\rho_{\bm{3_k}}({T}) = \omega^{k}\left(
\begin{array}{ccc}
	\omega  & 0 & 0 \\
	0 & \omega  & 0 \\
	0 & 0 & 1 \\
\end{array}
\right) \,,
\end{equation}
for $k = 0,1,2$. The three dimensional representation $\rho_{\bm{3_k}}({S})$ and $\rho_{\bm{3_k}}({T})$ generate the finite modular group $\Gamma^\prime_3 \cong T^\prime$, and it can be reduced to the direct sum of a one-dimensional and a two-dimensional irreducible representations of $T'$,
\begin{equation}
\label{eq:delta27-triplet-decom-Tp}\bm{3_k}=\bm{1^{[k+1]}} \oplus\bm{2^{[k+2]}}\,.
\end{equation}
The reducible triplets $\Phi_{\bm{3_k}}=(\phi_{1},\phi_{2},\phi_{3})$ transforming as $\rho_{\bm{3_k}}$ under $T'$ will decompose to one singlet and one doublet of $T^\prime$,
 which can be obtained by performing the following similarity transformation
\begin{equation}
\Omega_{\bm{3_k}}=\frac{1}{\sqrt{2}}\left(
\begin{array}{ccc}
 1 & -1 & 0 \\
 0 & 0 & \sqrt{2}\omega \\
 1 & 1 & 0 \\
\end{array}
\right)\,.
\end{equation}
 Then the corresponding irreducible singlet and doublet are given by
\begin{equation}
 \bm{1^{[k+1]}}:\frac{1}{\sqrt{2}}(\phi_{1}-\phi_{2}), \qquad \text{and} \qquad \bm{2^{[k+2]}}:\frac{1}{\sqrt{2}}\left(\begin{array}{c}\sqrt{2}\omega  \phi_{3}\\
\phi_{1}+\phi_{2}\end{array}\right)\,.
\end{equation}
For the $\bm{\bar{3}}$ representation, the compatible modular transformations $\rho_{\bm{r}}(S)$ and $\rho_{\bm{r}}(T)$ are denoted as $\rho_{\bm{\bar{3}_k}}({S})$ and $\rho_{\bm{\bar{3}_k}}({T})$ which are the complex conjugate of $\rho_{\bm{3_k}}({S})$ and $\rho_{\bm{3_k}}({T})$ in Eq.~\eqref{eq:3DTpfromDelta27}, respectively. The three reducible three-dimensional representations of $T^\prime$ are labelled as $\bm{\bar{3}_k}$ and we find the following decompositions
\begin{equation}
\bm{\bar{3}_k}=\bm{1^{[2-k]}} \oplus\bm{2^{[1-k]}}\,.
\end{equation}
The corresponding irreducible singlet and doublet for the three reducible triplets $\bm{\bar{3}_{k}}$ are all given by the linear combinations
\begin{equation}
\bm{1^{[2-k]}}:\frac{1}{\sqrt{2}}(\phi_{1}-\phi_{2}), \qquad \text{and} \qquad  \bm{2^{[1-k]}}:\frac{1}{\sqrt{2}}\left(\begin{array}{c}\phi_{1}+\phi_{2}\\
			-\sqrt{2}\omega^2  \phi_{3}\end{array}\right)\,.
\end{equation}
Finally, we shall consider the action of  the automorphism $u_{K_*}$ on the eleven representations of $\Delta(27)$. Note that the consistency condition Eq.~\eqref{eq:con_uK} may be read as a similarity transformation between the representations $\rho$ and $\rho\circ u_{K_*}$. Similar to the two automorphisms $u_{S}$ and $u_{T}$, we define the action of the outer automorphism $u_{K_*}$ on representation as $u_{K_*}:\rho\rightarrow \rho\circ u_{K_*}$. Following the discussion above, we find that the automorphism $u_{K_*}$ in Eq.~\eqref{eq:CPfromDelta27} acts on the irreducible representations of $\Delta(27)$ as
\begin{equation}\label{eq:reps_tra_uK*}
	u_{K_*}:~ \bm{1_{0,1}}\leftrightarrow \bm{1_{1,1}}, \qquad \bm{1_{0,2}}\leftrightarrow\bm{1_{2,2}}, \qquad \bm{1_{1,0}}\leftrightarrow\bm{1_{2,0}}, \qquad \bm{3}\leftrightarrow \bm{\bar{3}}\,,
\end{equation}
and the other irreducible representations are invariant under the action of the outer automorphism $u_{K_*}$.  The action of the automorphism $u_{K_*}$ on the conjugacy classes and the representations of $\Delta(27)$ is shown in the table~\ref{tab:character_Delta27}, and the character table of $\Delta(27)$ is invariant under $u_{K_*}$. In order to obtain the solution of the gCP transformation $\rho({K_*})$, one should  solve the following consistency conditions
\begin{eqnarray}
\nonumber && \rho_{\bm{r}}(K_*)\,\rho^*_{\bm{r}}(S)\,\rho^{-1}_{\bm{r}}(K_*)  =  \rho^{-1}_{\bm{r}}(S),\quad  \rho_{\bm{r}}(K_*)\,\rho^*_{\bm{r}}(T)\,\rho^{-1}_{\bm{r}}(K_*) = \rho^{-1}_{\bm{r}}(T),\quad
\rho^*_{\bm{r}}(K_*)  = \rho^{-1}_{\bm{r}}(K_*)\,,\\
\label{eq:cons_D27_uK*} && \rho_{\bm{r}}(K_*)\rho^*_{\bm{r}}(A)\rho^{-1}_{\bm{r}}(K_*)= \rho_{\bm{r}}(A^2B) \,,\quad
\rho_{\bm{r}}(K_*)\rho^*_{\bm{r}}(B)\rho^{-1}_{\bm{r}}(K_*)= \rho_{\bm{r}}(A^2BA) \,.
\end{eqnarray}
The automorphism $u_{K_*}$ in combination with the automorphism $u_{S}$ and $u_{T}$ implies that the solutions of $\rho_{\bm{r}}({S})$, $\rho_{\bm{r}}({T})$ and $\rho_{\bm{r}}({K_*})$ in the consistency conditions Eq.~\eqref{eq:cons_D27_Tp} and Eq.~\eqref{eq:cons_D27_uK*} can be obtained in the representations $\bm{1_{0,0}}$, $\bm{3}$ , $\bm{\bar{3}}$ and $\bm{8}$ of $\Delta(27)$. The possible gCP transformation acting on the fields $\Phi_{\bm{8_{k}}}$ which is the reducible octet representation $\bm{8}$ of $\Delta(27)$ is found to be of the following form
\begin{equation}\label{eq:exp_rohK_8}
\rho_{\bm{8_k}}(K_*)=\left(\begin{array}{cccc} 0 & \mathbb{0}& \mathbb{0}  & 1 \\
0 & P^\prime& \mathbb{0} & 0 \\
0 & \mathbb{0} &P^\prime & 0 \\
1 & \mathbb{0}& \mathbb{0}  & 0
\end{array}\right), \qquad \text{with} \qquad
P^\prime=\left(
\begin{array}{ccc}
0 & 0 & 1  \\
0 & 1 & 0  \\
1 & 0  & 0\\
\end{array}
\right)\,,
\end{equation}
up to an overall phase.  For the representations $\bm{3_k}$, the expressions of $\rho_{\bm{3_k}}({K_*})$  are given by
\begin{equation}\label{eq:exp_rohK_3}
	\rho_{\bm{3_k}}(K_*)=\left(
	\begin{array}{ccc}
		0 & 1 & 0 \\
		1 & 0 & 0 \\
		0 & 0 & \omega \\
	\end{array}
	\right)\,.
\end{equation}
For the $\bm{\bar{3}_{k}}$ representations, the CP transformation $\rho_{\bm{\bar{3}_{k}}}(K_*)$ takes the complex conjugate of $\rho_{\bm{3_k}}(K_*)$ in Eq.~\eqref{eq:exp_rohK_3}. We see that the gCP transformation matrix $\rho_{\bm{8_k}}(K_*)$ and $\rho_{\bm{3_k}}(K_*)$ are not diagonal, and the different nontrivial singlets of $\Delta(27)$ are related by gCP symmetry.

\section{\label{sec:mod_general}Effective action invariant under $\Omega(1)$ eclectic flavor symmetry  }

As shown in Eqs.~(\ref{eq:uS-D27-singlets}, \ref{eq:uT-D27-singlets}), the eight nontrivial singlets of $\Delta(27)$ are related by the $T'$ modular symmetry. As a consequence, all the eight nontrivial singlets carrying the same SM quantum numbers must be present or absent in a $\Omega(1)$ eclectic model. In other words, if a field is assigned to a nontrivial singlet representation of $\Delta(27)$, then the $\Omega(1)$ eclectic model has to contain other seven fields in the remaining nontrivial singlets of $\Delta(27)$ while transforming in the same way under SM gauge group. Consequently the three generations of quarks and leptons fields can only be assigned to triplets $\bm{3}$, $\bm{\bar{3}}$ or trivial singlet $\bm{1_{0, 0}}$ of $\Delta(27)$, and the Higgs fields can only be in $\bm{1_{0, 0}}$ if no more Higgs fields are introduced. Nevertheless, the flavons  can be trivial singlet, triplet and reducible octet $\bm{8}$ of the traditional flavor symmetry $\Delta(27)$. The left-handed (LH) quark/lepton doublet, the right-handed (RH) quark/lepton fields and the flavon field are denoted as $\psi$, $\psi^c$ and $\Phi$, respectively. For illustration, we assume that both $\psi$ and $\psi^{c}$ transform as $\bm{3}$ under $\Delta(27)$ while the Higgs $H_u$ and $H_d$ are invariant under $\Delta(27)$. As shown in Eqs.~(\ref{eq:3DTpfromDelta27},\ref{eq:delta27-triplet-decom-Tp}), the $T'$ transformation of both $\psi$ and $\psi^{c}$ would be the reducible triplet $\bm{3_{0}}$, $\bm{3_{1}}$ or $\bm{3_{2}}$, in the following $\bm{3_{0}}$ is chosen as an example. The results for other assignments of matter fields can be fixed analogously.

In the following, we shall preform a general analysis for the superpotential $\mathcal{W}$ and the K\"ahler potential $\mathcal{K}$ invariant under the eclectic flavor group $\Omega(1)\cong\Delta(27)\rtimes T^\prime$, and the framework of $\mathcal{N}=1$ global supersymmetry is adopted. We assume that the level 3 modular form multiplets at weight $k_Y$ comprise all possible irreducible multiplets of $T^\prime$, i.e.
\begin{eqnarray}
\nonumber &&\hskip-0.3in Y^{(k_Y)}_{\bm{1}}(\tau)=Y_1,\quad
Y^{(k_Y)}_{\bm{1'}}(\tau)=Y_2,\quad
Y^{(k_Y)}_{\bm{1''}}(\tau)=Y_3\,,\\
\label{eq:Yn_def}&&\hskip-0.3in Y^{(k_Y)}_{\bm{2}}(\tau)=\begin{pmatrix}
Y_4 \\
Y_5
\end{pmatrix},\quad
Y^{(k_Y)}_{\bm{2'}}(\tau)=\begin{pmatrix}
Y_6 \\Y_7
\end{pmatrix},\quad
Y^{(k_Y)}_{\bm{2''}}(\tau)=\begin{pmatrix}
Y_8 \\
Y_9
\end{pmatrix},\quad
Y^{(k_Y)}_{\bm{3}}(\tau)=\begin{pmatrix}
Y_{10} \\
Y_{11} \\
Y_{12}
\end{pmatrix}\,.
\end{eqnarray}
If certain modular multiplets are absent, one can set the corresponding modular forms $Y_i$ to be zero. If there are several linearly independent modular form multiplets at a given weight, their contributions to the effective action are of similar form and they can be straightforwardly read out from the general results in the following subsections. We shall firstly give the general form of the superpotential which is a holomorphic function of the matter fields and flavons and the couplings are generally expressed in terms of  modular forms. Then we shall perform a detail analysis of the K\"ahler potential.

\subsection{Superpotential}

Given the assignment $\psi, \psi^{c}\sim\left(\bm{3}, \bm{3_{0}}\right)$\footnote{In the notation $(\bm{r},\bm{r^\prime})$, the two representations $\bm{r}$ and $\bm{r^\prime}$ refer to the transformations under $\Delta(27)$ and $T^\prime$, respectively. For any $\Delta(27)$ representation $\bm{r}$, the consistency condition of Eq.~\eqref{eq:cons_D27_Tp} allows to fix the modular transformation $\bm{r^{\prime}}$, as discussed in section~\ref{sec:EFG-Omega1-decomp}.}, from the Kronecker products of $\Delta(27)$ in Eq.~\eqref{eq:KP_Delta27}, we find that the flavon $\Phi\equiv(\phi_1, \phi_2, \phi_3)^{T}$ transforming as $\bm{3}$ under $\Delta(27)$ must be introduced to obtain $\Omega(1)$ invariant Yukawa superpotential, and it should form a $T'$ reducible triplet $\bm{3_{k}}$. The modular transformation of $\Phi$ is taken to be $\bm{3_{0}}$ for example. Here the flavon $\Phi$ with modular weight $k_{\Phi}$ can be understood as a single flavon or combination of several flavons. Then the superpotential for the charged lepton/quark field $\psi$ can be written as
\begin{equation}
\mathcal{W}_{D}=\frac{1}{\Lambda}\sum_{\bm{r}, s} c_{\bm{r},s}\left(Y^{(k_Y)}_{\bm{r}}\Phi \psi^{c}\psi\right)_{(\bm{1_{0,0}},\bm{1}), s}H_{u/d}\,,
\end{equation}
where $k_Y=k_{\psi}+k_{\psi^{c}}+k_{\Phi}$, the subscripts $\bm{1_{0,0}}$ and $\bm{1}$ refer to trivial singlet of $\Delta(27)$ and $T^\prime$ respectively, and one has to sum over all modular multiplets at weight $k_Y$ and all independent singlet contractions $s$. Invariance under the action of $\Delta(27)$ requires $\mathcal{W}_{D}$ should be of the following form
\begin{eqnarray}
	\nonumber \mathcal{W}_{D}&=&\frac{1}{\Lambda}\sum_{\bm{r}} \Big(Y^{(k_Y)}_{\bm{r}}\big[c_{\bm{r},1} (\psi^c_1 \psi_{1} \phi_{1}+\psi^c_2 \psi_{2} \phi_{2}+\psi^c_3 \psi_{3} \phi_{3})\\
	\nonumber &&+c_{\bm{r},2} (\psi^c_1 \psi_{2} \phi_{3}+\psi^c_1 \psi_{3} \phi_{2}+\psi^c_2 \psi_{3} \phi_{1}+\psi^c_2 \psi_{1} \phi_{3}+\psi^c_3 \psi_{1} \phi_{2}+\psi^c_3 \psi_{2} \phi_{1})\\
	&&+c_{\bm{r},3}(\psi^c_1 \psi_{2} \phi_{3}-\psi^c_1 \psi_{3} \phi_{2}+\psi^c_2 \psi_{3} \phi_{1}-\psi^c_2 \psi_{1} \phi_{3}+\psi^c_3 \psi_{1} \phi_{2}-\psi^c_3\psi_{2}\phi_{1})\big]\Big)_{\bm{1}}H_{u,d}\,.
\end{eqnarray}
We denote the above three $\Delta(27)$ invariant contractions as $\mathcal{I}_1$, $\mathcal{I}_2$ and $\mathcal{I}_3$ with
\begin{eqnarray}
\nonumber\mathcal{I}_1&=&\psi^c_1 \psi_{1} \phi_{1}+\psi^c_2 \psi_{2} \phi_{2}+\psi^c_3 \psi_{3} \phi_{3}=\left(\left(\psi^{c}\psi\right)_{\bm{\bar{3}_{S,1}}}\Phi\right)_{\bm{1_{0,0}}}\,,\\
\nonumber\mathcal{I}_2&=&\psi^c_1 \psi_{2} \phi_{3}+\psi^c_1 \psi_{3} \phi_{2}+\psi^c_2 \psi_{3} \phi_{1}+\psi^c_2 \psi_{1} \phi_{3}+\psi^c_3 \psi_{1} \phi_{2}+\psi^c_3 \psi_{2} \phi_{1}=\left(\left(\psi^{c}\psi\right)_{\bm{\bar{3}_{S,2}}}\Phi\right)_{\bm{1_{0,0}}}\,,\\
\mathcal{I}_3&=&\psi^c_1 \psi_{2} \phi_{3}-\psi^c_1 \psi_{3} \phi_{2}+\psi^c_2 \psi_{3} \phi_{1}-\psi^c_2 \psi_{1} \phi_{3}+\psi^c_3 \psi_{1} \phi_{2}-\psi^c_3 \psi_{2} \phi_{1}=\left(\left(\psi^{c}\psi\right)_{\bm{\bar{3}_{A}}}\Phi\right)_{\bm{1_{0,0}}}\,.
\end{eqnarray}
Since all the three fields $\psi\equiv\left(\psi_1, \psi_2, \psi_3\right)^T$, $\psi^{c}=\left(\psi^c_1, \psi^c_2, \psi^c_3\right)^{T}$, $\Phi\equiv(\phi_1, \phi_2, \phi_3)^{T}$ are $T'$ reducible triplet $\bm{3_{0}}$, we find that the modular transformations of $\mathcal{I}_{1,2,3}$ under $S$ and $T$ are given by
\begin{eqnarray}
\nonumber&&\mathcal{I}_1\stackrel{S}{\longrightarrow}-\frac{i}{\sqrt{3}}\mathcal{I}_1-\frac{i}{\sqrt{3}}\omega \mathcal{I}_2,~~~~\mathcal{I}_1\stackrel{T}{\longrightarrow}\mathcal{I}_1\,,\\
\nonumber&&\mathcal{I}_2\stackrel{S}{\longrightarrow}-\frac{2i}{\sqrt{3}}\omega^2\mathcal{I}_1+\frac{i}{\sqrt{3}}\mathcal{I}_2,
~~~~\mathcal{I}_2\stackrel{T}{\longrightarrow}\omega^2\mathcal{I}_2\,, \\
&&\mathcal{I}_3\stackrel{S}{\longrightarrow}\mathcal{I}_3,~~~~\mathcal{I}_3\stackrel{T}{\longrightarrow}\omega^2\mathcal{I}_3\,.
\end{eqnarray}
Hence the combination $\mathcal{I}_3$ transforms as the singlet $\bm{1^{\prime\prime}}$ under $T'$, $\mathcal{I}_1$ and $\mathcal{I}_2$ span a two-dimensional representation space and they can be arranged into $T'$ doublet $\bm{2^{\prime}}$, i.e.
\begin{equation}
\begin{pmatrix}
\mathcal{I}_2\\
-\sqrt{2}\,\omega^2\mathcal{I}_1
\end{pmatrix}\sim\bm{2^{\prime}},~~~~~\mathcal{I}_3\sim\bm{1^{\prime\prime}}\,.
\end{equation}
Therefore only the modular form multiplets $Y^{(k_Y)}_{\bm{1'}}=Y_2$ and $Y^{(k_Y)}_{\bm{2''}}=\left(Y_8, Y_9\right)^T$ can contract with $\mathcal{I}_{1,2,3}$ to form modular invariant combinations under the finite modular group $T'$. Thus the eclectic symmetry group $\Omega(1)$ severely constrains the superpotential to be
\begin{eqnarray}
\nonumber \mathcal{W}_{D}&=&\frac{1}{\Lambda}\Big[i\omega\alpha_1 Y_{2} (\psi^c_1\psi_{2} \phi_{3}-\psi^c_1\psi_{3} \phi_{2}+\psi^c_2\psi_{3} \phi_{1}-\psi^c_2\psi_{1} \phi_{3}+\psi^c_3\psi_{1} \phi_{2}-\psi^c_3 \psi_{2} \phi_{1})\\
\nonumber &&+\sqrt{2}\, \alpha_{2} Y_{8} (\psi^c_1 \psi_{1} \phi_{1}+\psi^c_2 \psi_{2} \phi_{2}+\psi^c_3 \psi_{3} \phi_{3})\\
&&+\omega \alpha_{2} Y_{9} (\psi^c_1\psi_{2} \phi_{3}+\psi^c_1\psi_{3} \phi_{2}+\psi^c_2\psi_{3} \phi_{1}+\psi^c_2\psi_{1} \phi_{3}+\psi^c_3\psi_{1} \phi_{2}+\psi^c_3 \psi_{2} \phi_{1})\Big]H_{u,d}\,,
\end{eqnarray}
where $c_{\bm{2^{\prime\prime}}, 1}=-\sqrt{2}\,\omega^2c_{\bm{2^{\prime\prime}}, 2}=-\sqrt{2}\,\alpha_2$ and $c_{\bm{1^{\prime}}, 3}=i\omega\alpha_1$. With the above convention for coupling constants, both $\alpha_{1}$ and $\alpha_{2}$ would be enforced to be real by the gCP transformation $K_*$, see Eq.~\eqref{eq:couplings_phases}. After the electroweak symmetry breaking, one can read off the mass matrix of the fermion $\psi$ as
\begin{equation}\label{eq:Ch_mass1}
M_{\psi}= \frac{v_{u,d}}{\Lambda}\left[i\omega\alpha_1 Y_{2}\left(
\begin{array}{ccc}
	0 &  \phi_{3}  &  -\phi_{2}  \\
	 -\phi_{3}  & 0 &  \phi_{1}  \\
	 \phi_{2}  &  -\phi_{1}  & 0 \\
\end{array}
\right)+\alpha_{2}\left(
\begin{array}{ccc}
	\sqrt{2} \phi_{1} Y_{8} & \omega \phi_{3} Y_{9} & \omega\phi_{2} Y_{9} \\
	\omega \phi_{3} Y_{9} & \sqrt{2} \phi_{2} Y_{8} & \omega \phi_{1} Y_{9} \\
	\omega \phi_{2} Y_{9} & \omega \phi_{1} Y_{9} & \sqrt{2} \phi_{3} Y_{8} \\
\end{array}
\right)\right]\,,
\end{equation}
where the mass matrix $M_{\psi}$ is given in the convention $\psi^{c}M_{\psi}\psi$.
Note that the first term proportional to $\alpha_1$ arises from the coupling with $Y^{(k_Y)}_{\bm{1'}}$ and $k_Y$ should be even, and the second term proportional to $\alpha_2$ arises from the coupling with $Y^{(k_Y)}_{\bm{2''}}$ for odd $k_Y$.

When the gCP transformation $ K_*$ is included in the theory, the transformation properties of modular multiplets, matter fields and flavons can be obtained from Eqs.~\eqref{eq:Ytau_K_Tra} and \eqref{eq:exp_rohK_3}, i.e.
\begin{eqnarray}
\nonumber  K_*:&& Y^{(k_Y)}_{\bm{1^{\prime}}}\rightarrow \left(Y^{(k_Y)}_{\bm{1^{\prime}}}\right)^\dagger,\quad Y^{(k_Y)}_{\bm{2^{\prime\prime}}}\rightarrow \left(Y^{(k_Y)}_{\bm{2^{\prime\prime}}}\right)^{\dagger T},\\
 &&  \left(\begin{array}{c} \psi_{1} \\ \psi_{2} \\ \psi_{3}\end{array}\right)\rightarrow \left(\begin{array}{c} \psi^{\dagger}_{2} \\ \psi^{\dagger}_{1} \\ \omega \psi^{\dagger}_{3}\end{array}\right),\quad
 \left(\begin{array}{c} \psi^{c}_1 \\ \psi^c_2 \\ \psi^c_3 \end{array}\right)\rightarrow \left(\begin{array}{c} \psi^{c\dagger}_2 \\ \psi^{c\dagger}_{1}  \\ \omega \psi^{c\dagger}_3 \end{array}\right),\quad
 \left(\begin{array}{c} \phi_{1} \\ \phi_{2} \\ \phi_{3}\end{array}\right)\rightarrow \left(\begin{array}{c} \phi^{\dagger}_{2} \\ \phi^{\dagger}_{1} \\ \omega \phi^{\dagger}_{3}\end{array}\right)\,.
\end{eqnarray}
Consequently under the action of $K_{*}$ the superpotential $\mathcal{W}_{D}$ transforms into
\begin{eqnarray}
\nonumber \mathcal{W}_{D}&\stackrel{K_{*}}{\longrightarrow}&\frac{1}{\Lambda}\Big\{i\omega^2\alpha_1   Y^{\dagger}_{2} \left[\psi^{c\dagger}_2\psi^*_{1} \phi^*_{3}-\psi^{c\dagger}_2\psi^*_{3} \phi^{\dagger}_{1}+\psi^{c\dagger}_1\psi^{\dagger}_{3} \phi^{\dagger}_{2}-\psi^{c\dagger}_1\psi^{\dagger}_{2} \phi^{\dagger}_{3}+\psi^{c\dagger}_3\psi^{\dagger}_{2} \phi^{\dagger}_{1}-\psi^{c\dagger}_3\psi^{\dagger}_{1} \phi^{\dagger}_{2}\right]\\
\nonumber &&+\sqrt{2}\, \alpha_{2} Y^{\dagger}_{8} \left[\psi^{c\dagger}_2\psi^{\dagger}_{2} \phi^{\dagger}_{2}+\psi^{c\dagger}_1\psi^{\dagger}_{1} \phi^{\dagger}_{1}+\psi^{c\dagger}_3\psi^{\dagger}_{3} \phi^{\dagger}_{3}\right]+\omega^2 \alpha_{2} Y^{\dagger}_{9} \left[\psi^{c\dagger}_2\psi^{\dagger}_{1} \phi^{\dagger}_{3}+\psi^{c\dagger}_2\psi^{\dagger}_{3} \phi^{\dagger}_{1}\right.\\
&&\left.+\psi^{c\dagger}_1\psi^{\dagger}_{3} \phi^{\dagger}_{2}+\psi^{c\dagger}_1\psi^{\dagger}_{2} \phi^{\dagger}_{3}+\psi^{c\dagger}_3 \psi^{\dagger}_{2} \phi^{\dagger}_{1}+\psi^{c\dagger}_3 \psi^{\dagger}_{1} \phi^{\dagger}_{2}\right] \Big\} H^{\dagger}_{u,d}\,,
\end{eqnarray}
which must be equal to $\mathcal{W}^{\dagger}_{D}$ due to gCP invariance. As a result, the couplings $\alpha_1$ and $\alpha_2$ should fulfill
\begin{equation}
\alpha_1=\alpha^*_1, \qquad \alpha_{2}=\alpha^*_{2}\,.
\end{equation}
Consequently both $\alpha_1$ and $\alpha_2$ are real, i.e.
\begin{equation}\label{eq:couplings_phases}
\text{arg}(\alpha_1)=0~(\text{mod}~\pi),~~~\text{arg}(\alpha_2)=0~(\text{mod}~\pi)\,.
\end{equation}
In the same fashion, we can write out the $\Omega(1)$ invariant superpotential and subsequently read off the mass matrix for other assignments of matter fields and flavons. For instance, when the flavon fields are nontrivial singlets of $\Delta(27)$ with $\psi\sim \left(\bm{3},~\bm{3_0}\right)$, $\psi^c\sim \left(\bm{\bar{3}},~\bm{\bar{3}_0}\right)$ and $\Phi\sim \left(\bm{8},~\bm{8_0} \right)$ under $\Omega(1)$, the fermion mass matrix is determined to be
{\footnotesize
\begin{eqnarray}
\nonumber M^\prime_{\psi}&=&  \frac{v_{u,d}}{\Lambda}\left[i\alpha_{1}Y_{1}\left(
\begin{array}{ccc}
  \omega\left(\phi_{3}+\omega\phi_{6}\right) & \phi_{2}+\phi_{5}+\phi_{8}  &  \phi_{1}+\omega^2 \phi_{4}+\omega\phi_{7} \\
  \phi_{1}+\phi_{4}+\phi_{7} &  \omega^2\phi_{3}+\omega\phi_{6} &  \phi_{2}+\omega\phi_{5}+\omega^2 \phi_{8} \\
  \phi_{2}+\omega^2 \phi_{5}+\omega \phi_{8} & \phi_{1}+\omega\phi_{4}+\omega^2\phi_{7}  & \phi_{3}+\phi_{6} \\
\end{array}
\right)\right. \\
\nonumber &&\hskip-0.6in\left. +\omega^2\alpha_{2}\left(
\begin{array}{ccc}
 -3 Y_{10} \left(\phi_{6}+\omega^2 \phi_{3}\right) & \omega Y_{10} \phi_{258}-2Y_{11}\phi^\prime_{852}-2 Y_{12}\phi^\prime_{258}  & Y_{10} \phi^\prime_{417}-2 \omega^2 Y_{11} \phi_{147}-2 Y_{12} \phi^\prime_{147} \\
  \omega Y_{10} \phi_{147}-2  Y_{11}\phi^\prime_{471}-2 Y_{12}\phi^\prime_{174}  & -3 Y_{10} \left(\phi_{3}+\omega^2 \phi_{6}\right) & Y_{10} \phi^\prime_{825}-2 \omega^2 Y_{11} \phi_{258}-2 Y_{12} \phi^\prime_{285} \\
 Y_{10} \phi^\prime_{528}- 2Y_{11} \phi^\prime_{582}-2 Y_{12} \phi_{258}  & Y_{10}\phi^\prime_{714}-2 Y_{11} \phi^\prime_{741}-2 Y_{12} \phi_{147} & -3 \omega   Y_{10} (\phi_{3}+\phi_{6}) \\
\end{array}
\right)\right]\,,\\
\label{eq:Me_p1}
\end{eqnarray}}
for even $k_{Y}=k_{\psi}+k_{\psi^c}+k_{\Phi}$  and
\begin{eqnarray}
\nonumber \hskip-0.6in M^{\prime\prime}_{\psi}&=& \frac{v_{u,d}}{\Lambda}\left[i\alpha_{1}\left(
\begin{array}{ccc}
  \sqrt{3}\,i\omega Y_{7} \left(- \phi_{3}+\omega \phi_{6}\right) & - Y_{7} \phi_{258}+ \sqrt{2}\, Y_{6} \phi^\prime_{582}  &  -\sqrt{2}\, Y_{6} \phi^\prime_{471}+ Y_{7} \phi^\prime_{174} \\
 Y_{7} \phi_{147}- \sqrt{2}\, Y_{6} \phi^\prime_{741} & \sqrt{3}\,i\omega Y_{7} \left(-\omega \phi_{3}+\phi_{6}\right) & \sqrt{2}\, Y_{6} \phi^\prime_{852}- Y_{7} \phi^\prime_{258} \\
  \sqrt{2}\,\omega^2 Y_{6} \phi_{258}- Y_{7} \phi^\prime_{285}  &- \sqrt{2}\, \omega^2 Y_{6} \phi_{147}+ Y_{7} \phi^\prime_{147}  &  \sqrt{3}\,i Y_{7} (-\phi_{3}+\phi_{6}) \\
\end{array}
\right) \right.\\
\label{eq:Me_p2}&&\hskip-0.25in\left.+i\omega\alpha_{2}\left(
\begin{array}{ccc}
 \sqrt{3}\,i Y_{8} \left(   \phi_{3}-\omega \phi_{6}\right) & - \omega^2  Y_{8} \phi_{258}-\sqrt{2}\,Y_{9} \phi^\prime_{285}  & \sqrt{2}\,Y_{9} \phi_{147}+ Y_{8} \phi^\prime_{741} \\
 \omega^2 Y_{8} \phi_{147}+\sqrt{2}\, Y_{9} \phi^\prime_{147} &  \sqrt{3}\,i Y_{8} \left(\omega \phi_{3}- \phi_{6}\right) &  - Y_{8} \phi^\prime_{582}-\sqrt{2}\, Y_{9} \phi_{258} \\
- Y_{8} \phi^\prime_{852}-\sqrt{2}\,Y_{9} \phi^\prime_{258}   & Y_{8} \phi^\prime_{471}+\sqrt{2}\,Y_{9} \phi^\prime_{174} &  \sqrt{3}\,i\omega^2 Y_{8} (\phi_{3}-\phi_{6}) \\
\end{array}
\right)\right]\,,
\end{eqnarray}
for odd $k_{Y}=k_{\psi}+k_{\psi^c}+k_{\Phi}$. Here the functions $\phi_{ijk}$ and $\phi^\prime_{ijk}$ are defined as
\begin{equation}
\phi_{ijk}=\phi_i+ \phi_j+\phi_k\,, \qquad \phi^\prime_{ijk}=\phi_i+ \omega\phi_j+\omega^2\phi_k\,.
\end{equation}
Furthermore, we summarize the fermion mass matrices for all independent assignments of quarks/leptons and flavons in table~\ref{tab:CL_mms}. The mass matrix would become its transpose, if the transformation properties of $\psi$ and $\psi^c$ are exchanged.

\begin{table}[t!]
\centering
\renewcommand{\arraystretch}{1.1}
\begin{tabular}{|c|c|c|c|c|c|c|c|c|c|c|c|}  \hline\hline
$\mathcal{W}$ & $\left(\Delta(27),T^\prime\right)$   & $Y^{(k_Y)}_{\bm{r}}$ & $M_{\psi}$   \\  \hline
\multirow{3}{*}{$\mathcal{W}_{D1}$} & \multirow{2}{*}{$\psi\sim(\bm{3},\bm{3_{0}})$, $\Phi\sim(\bm{\bar{3}},\bm{\bar{3}_{0}})$}  & \multirow{3}{*}{$Y^{(k_Y)}_{\bm{1^{k}}}$} & \multirow{3}{*}{$\frac{v_{u,d}Y_{k}}{\Lambda}\left(
\begin{array}{ccc}
	\alpha_{1} \phi_{1} & \alpha_{1} \phi_{2} & \alpha_{1} \phi_{3} \\
	\alpha_{2} \phi_{1} & \alpha_{2} \phi_{2} & \alpha_{2} \phi_{3} \\
	\alpha_{3} \phi_{1} & \alpha_{3} \phi_{2} & \alpha_{3} \phi_{3} \\
\end{array}
\right)$}  \\
& \multirow{2}{*}{$(\psi^c_1,\psi^c_2,\psi^c_3)\sim(\bm{1_{(0,0)}},\bm{1^{[3-k]}})$} & &  \\
&&& \\  \hline

\multirow{3}{*}{$\mathcal{W}_{D2}$} & $\psi\sim(\bm{3},\bm{3_{0}})$,   & $Y^{(k_Y)}_{\bm{1}}$ & \multirow{3}{*}{$\frac{v_{u,d}}{\Lambda}P^{k}\left(
	\begin{array}{ccc}
		\alpha_{1}Y_{1} \phi_{1} & \alpha_{1}Y_{1} \phi_{2} & \alpha_{1}Y_{1} \phi_{3} \\
		\alpha_{2}Y_{3} \phi_{1} & \alpha_{2}Y_{3} \phi_{2} & \alpha_{2}Y_{3} \phi_{3} \\
		\alpha_{3}Y_{2} \phi_{1} & \alpha_{3}Y_{2} \phi_{2} & \alpha_{3}Y_{2} \phi_{3} \\
	\end{array}
	\right)$}   \\
& $\psi^c_{m}\sim(\bm{1_{(0,0)}},\bm{1^{m-1}})$, & $Y^{(k_Y)}_{\bm{1^{\prime}}}$ &  \\
& $\Phi\sim(\bm{\bar{3}},\bm{\bar{3}_{k}})$ & $Y^{(k_Y)}_{\bm{1^{\prime\prime}}}$&  \\  \hline

$\mathcal{W}_{D3}$ & $\psi,\,\psi^c\sim(\bm{3},\bm{3_{0}})$, $\Phi\sim(\bm{3},\bm{3_{0}})$   & $Y^{(k_Y)}_{\bm{1^{\prime}}}$, $Y^{(k_Y)}_{\bm{2^{\prime\prime}}}$  & $M_{\psi}$ in Eq.~\eqref{eq:Ch_mass1} \\  \hline

$\mathcal{W}_{D4}$ & $\psi,\,\psi^c\sim(\bm{3},\bm{3_{0}})$, $\Phi\sim(\bm{3},\bm{3_{1}})$   & $Y^{(k_Y)}_{\bm{1}}$, $Y^{(k_Y)}_{\bm{2^{\prime}}}$ & $M_{\psi}\left(Y_2\to Y_1,Y_{8}\to Y_{6},Y_{9}\to Y_{7}\right)$  \\  \hline

$\mathcal{W}_{D5}$ & $\psi,\,\psi^c\sim(\bm{3},\bm{3_{0}})$, $\Phi\sim(\bm{3},\bm{3_{2}})$   & $Y^{(k_Y)}_{\bm{1^{\prime\prime}}}$, $Y^{(k_Y)}_{\bm{2}}$ & $M_{\psi}\left(Y_2\to Y_3,Y_{8}\to Y_{4},Y_{9}\to Y_{5}\right)$  \\  \hline

$\mathcal{W}_{D6}$ & $\psi,\,\psi^c\sim(\bm{\bar{3}},\bm{\bar{3}_{0}})$, $\Phi\sim(\bm{\bar{3}},\bm{\bar{3}_{0}})$   & $Y^{(k_Y)}_{\bm{1^{\prime\prime}}}$, $Y^{(k_Y)}_{\bm{2^{\prime}}}$ & $M_{\psi}\left(Y_2\to \omega Y_{3},Y_{8}\to  Y_{7},Y_{9}\to -\omega Y_{6}\right)$  \\  \hline

$\mathcal{W}_{D7}$ & $\psi,\,\psi^c\sim(\bm{\bar{3}},\bm{\bar{3}_{0}})$, $\Phi\sim(\bm{\bar{3}},\bm{\bar{3}_{1}})$   & $Y^{(k_Y)}_{\bm{1}}$, $Y^{(k_Y)}_{\bm{2^{\prime\prime}}}$ & $M_{\psi}\left(Y_2\to  \omega Y_{1},Y_{8}\to  Y_{9},Y_{9}\to -\omega Y_{8}\right)$  \\  \hline

$\mathcal{W}_{D8}$ & $\psi,\,\psi^c\sim(\bm{\bar{3}},\bm{\bar{3}_{0}})$, $\Phi\sim(\bm{\bar{3}},\bm{\bar{3}_{2}})$   & $Y^{(k_Y)}_{\bm{1^{\prime}}}$, $Y^{(k_Y)}_{\bm{2}}$ & $M_{\psi}\left(Y_2\to  \omega Y_{2},Y_{8}\to  Y_{5},Y_{9}\to -\omega Y_{4}\right)$  \\  \hline

\multirow{2}{*}{$\mathcal{W}_{D9}$} & $\psi\sim(\bm{3},\bm{3_{0}})$, $\psi^c\sim(\bm{\bar{3}},\bm{\bar{3}_{0}})$,    & $Y^{(k_Y)}_{\bm{1}}$, $Y^{(k_Y)}_{\bm{3}}$  & $M^\prime_{\psi}$ in Eq.~\eqref{eq:Me_p1} for even $k_Y$ \\ \cline{3-4}
 &  $\Phi\sim(\bm{8},\bm{8_{0}})$  & $Y^{(k_Y)}_{\bm{2^{\prime}}}$, $Y^{(k_Y)}_{\bm{2^{\prime\prime}}}$ &  $M^{\prime\prime}_{\psi}$ in Eq.~\eqref{eq:Me_p2} for odd $k_Y$  \\ \hline

\multirow{2}{*}{$\mathcal{W}_{D10}$} & $\psi\sim(\bm{3},\bm{3_{0}})$, $\psi^c\sim(\bm{\bar{3}},\bm{\bar{3}_{0}})$,   & $Y^{(k_Y)}_{\bm{1^{\prime\prime}}}$, $Y^{(k_Y)}_{\bm{3}}$  & $M^\prime_{\psi}\left(Y_1\to Y_3,Y^{(k_Y)}_{\bm{3}}\to PY^{(k_Y)}_{\bm{3}}\right)$   \\ \cline{3-4}
 &   $\Phi\sim(\bm{8},\bm{8_{1}})$  & $Y^{(k_Y)}_{\bm{2}}$, $Y^{(k_Y)}_{\bm{2^{\prime}}}$ &  $M^{\prime\prime}_{\psi}\left(Y^{(k_Y)}_{\bm{2^{\prime}}}\to Y^{(k_Y)}_{\bm{2}},Y^{(k_Y)}_{\bm{2^{\prime\prime}}}\to Y^{(k_Y)}_{\bm{2^{\prime}}}\right)$  \\ \hline

\multirow{2}{*}{$\mathcal{W}_{D11}$} & $\psi\sim(\bm{3},\bm{3_{0}})$, $\psi^c\sim(\bm{\bar{3}},\bm{\bar{3}_{0}})$,    & $Y^{(k_Y)}_{\bm{1^{\prime}}}$, $Y^{(k_Y)}_{\bm{3}}$  & $M^\prime_{\psi}\left(Y_1\to Y_2,Y^{(k_Y)}_{\bm{3}}\to  P^2Y^{(k_Y)}_{\bm{3}}\right)$  \\ \cline{3-4}
 & $\Phi\sim(\bm{8},\bm{8_{2}})$   & $Y^{(k_Y)}_{\bm{2}}$, $Y^{(k_Y)}_{\bm{2^{\prime\prime}}}$ &  $M^{\prime\prime}_{\psi}\left(Y^{(k_Y)}_{\bm{2^{\prime}}}\to \omega Y^{(k_Y)}_{\bm{2^{\prime\prime}}},Y^{(k_Y)}_{\bm{2^{\prime\prime}}}\to \omega^2 Y^{(k_Y)}_{\bm{2}}\right)$  \\ \hline\hline

\end{tabular}
\caption{\label{tab:CL_mms} The mass matrices of the charged leptons/quarks $\psi$ for all independent assignments of $\psi$, $\psi^{c}$ and $\Phi$  under the eclectic symmetry group $\Omega(1)\equiv\Delta (27) \rtimes T'$, where $P$ is permutation matrix given in Eq.~\eqref{eq:per_mat}. In the third column, we list the modular form multiplets which give nonzero contributions.  In our convention, all couplings in $M_{\psi}$ are fixed to be real by the gCP transformation $K_*$.  }
\end{table}

\subsubsection{\label{subsubsec:superpot-Majorana}The superpotential of Majorana mass terms }

The Majorana mass term for the fermion field $\psi^{c}$ (e.g., the right-handed neutrinos) can be generally written as
\begin{equation}
\label{eq:W-Maj-expr1}\mathcal{W}_{M}=\sum_{\bm{r}, s} d_{\bm{r},s}\left(Y^{(k_Y)}_{\bm{r}}\psi^c\psi^c\Phi\right)_{(\bm{1_{0,0}},\bm{1}),s}\,,
\end{equation}
where $k_Y=2k_{\psi^{c}}+k_{\Phi}$, and $\Phi$ can be interpreted as one/several flavons or as combination of Higgs fields with flavons\footnote{For example, $\Phi$ is the combination of two Higgs fields and flavon in the Weinberg operator, $\psi^c$ would be the LH lepton doublet.}. For illustration, we consider the case that both $\psi^{c}$ and $\Phi$ transform as $(\bm{3},\bm{3_0})$ under the eclectic symmetry group $\Omega(1)$.

From the tensor product $\bm{3}\otimes \bm{3}=\bm{\bar{3}_{S,1}}\oplus\bm{\bar{3}_{S,2}}\oplus\bm{\bar{3}_{A}}$ of $\Delta(27)$ in Appendix~\ref{sec:Delta27_group}, we know the contraction of $\psi^{c}\psi^c$ gives rise to two  symmetric triplets:
\begin{eqnarray}
\left(\psi^{c}\psi^{c}\right)_{\bm{\bar{3}_{S,1}}}=\begin{pmatrix}
(\psi^{c}_1)^2 \\
(\psi^{c}_2)^2 \\
(\psi^{c}_3)^2
\end{pmatrix},~~~~\left(\psi^{c}\psi^{c}\right)_{\bm{\bar{3}_{S,2}}}=\begin{pmatrix}
\psi^{c}_2\psi^{c}_3 \\
\psi^{c}_1\psi^{c}_3 \\
\psi^{c}_1\psi^{c}_2
\end{pmatrix}\,.
\end{eqnarray}
Consequently there are only two $\Delta(27)$ invariant contractions for the Majorana mass term $\psi^c\psi^c\Phi$,
\begin{eqnarray}
\nonumber\left(\left(\psi^{c}\psi^{c}\right)_{\bm{\bar{3}_{S,1}}}\Phi\right)_{\bm{1_{0,0}}} &=&(\psi_{1}^{c})^2\phi_{1}+(\psi_{2}^{c})^2 \phi_{2}+(\psi_{3}^{c})^2 \phi_{3}\,,\\
\left(\left(\psi^{c}\psi^{c}\right)_{\bm{\bar{3}_{S,2}}}\Phi\right)_{\bm{1_{0,0}}} \label{eq:Majoran-mass-D27-con}&=&\psi^c_{1}\psi^c_{2}\phi_{3}+\psi^c_{1}\psi^c_{3}\phi_{2}+\psi^c_{2}\psi^c_{3}\phi_{1}\,.
\end{eqnarray}
The modulus and modular forms are invariant under $\Delta(27)$ flavor symmetry, thus $\Delta(27)$ invariance constrains the superpotential $\mathcal{W}_M$ to be
\begin{equation}
	\mathcal{W}_{M}=\sum_{\bm{r}} \Big(Y^{(k_Y)}_{\bm{r}}\left[d_{\bm{r},1} \left((\psi_{1}^{c})^2\phi_{1}+(\psi_{2}^{c})^2 \phi_{2}+(\psi_{3}^{c})^2 \phi_{3}\right)+ d_{\bm{r},2} (\psi^c_{1}\psi^c_{2}\phi_{3}+\psi^c_{1}\psi^c_{3}\phi_{2}+\psi^c_{2}\psi^c_{3}\phi_{1})\right]\Big)_{\bm{1}}\,.
\end{equation}
The two $\Delta(27)$ singlet contractions in Eq.~\eqref{eq:Majoran-mass-D27-con} can be arranged into a $T'$ doublet $\bm{2^\prime}$:

\begin{equation}
\begin{pmatrix}
\psi^c_{1}\psi^c_{2}\phi_{3}+\psi^c_{1}\psi^c_{3}\phi_{2}+\psi^c_{2}\psi^c_{3}\phi_{1} \\
-\omega^2\left((\psi_{1}^{c})^2\phi_{1}+(\psi_{2}^{c})^2 \phi_{2}+(\psi_{3}^{c})^2 \phi_{3}\right)/\sqrt{2}
\end{pmatrix}\sim\bm{2^\prime}\,.
\end{equation}
Consequently modular invariance of $\mathcal{W}_{M}$ requires $d_{\bm{2^{\prime\prime}},2}=-\sqrt{2}\,\omega d_{\bm{2^{\prime\prime}},1}$ and only the modular multiplet $Y^{(k_Y)}_{\bm{2''}}(\tau)=\left(Y_8, Y_9\right)^{T}$ is relevant. Thus the eclectic flavor symmetry $\Omega(1)$ fixes the superpotential $\mathcal{W}_{M}$ to be
\begin{equation}
\label{eq:W-Maj-fin}\mathcal{ W}_{M} =\alpha\left\{\sqrt{2}  Y_{8} \left[(\psi_{1}^{c})^2\phi_{1}+(\psi_{2}^{c})^2 \phi_{2}+(\psi_{3}^{c})^2 \phi_{3}\right]+2\omega Y_{9} (\psi^c_{1} \psi^c_{2} \phi_{3}+\psi^c_{1} \psi^c_{3} \phi_{2}+\psi^c_{2} \psi^c_{3} \phi_{1})\right\}\,,
\end{equation}
where $\alpha=\omega^2 d_{\bm{2^{\prime\prime}},2}/2$ is a general complex number. Thus we can read out the Majorana mass matrix of $\psi^c$ as follow
\begin{equation}\label{eq:M_psic}
M_{\psi^{c}}=\alpha\left(
\begin{array}{ccc}
 \sqrt{2}\,Y_{8} \phi_{1} & \omega Y_{9} \phi_{3} & \omega Y_{9} \phi_{2} \\
\omega Y_{9} \phi_{3} &  \sqrt{2}\, Y_{8} \phi_{2} & \omega Y_{9} \phi_{1} \\
\omega Y_{9} \phi_{2} & \omega Y_{9} \phi_{1} &  \sqrt{2}\, Y_{8} \phi_{3} \\
\end{array}
\right)\,.
\end{equation}
If the gCP transformation $K_*$ is considered, the superpotential $\mathcal{ W}_{M}$ transforms as
\begin{small}
\begin{eqnarray}
\hskip-0.1in \mathcal{W}_{M}\stackrel{K_{*}}{\longrightarrow} \alpha\left\{\sqrt{2}\, Y^{\dagger}_{8} \left[(\psi^{c\dagger}_{2})^2 \phi^{\dagger}_{2}+(\psi^{c\dagger}_{1})^2 \phi^{\dagger}_{1}+(\psi^{c\dagger}_{3})^2 \phi^{\dagger}_{3}\right]+2\omega^2 Y^{\dagger}_{9} (\psi^{c\dagger}_{2} \psi^{c\dagger}_{1} \phi^{\dagger}_{3}+\psi^{c\dagger}_{2} \psi^{c\dagger}_{3} \phi^{\dagger}_{1}+\psi^{c\dagger}_{1} \psi^{c\dagger}_{3} \phi^{\dagger}_{2})\right\}\,,
\end{eqnarray}
\end{small}
which must equal to $\mathcal{W}^{\dagger}_{M}$. Then we have $\alpha^*=\alpha$ and consequently
\begin{equation}
\text{arg}(\alpha)=0~(\text{mod}~\pi)\,.
\end{equation}
For other assignments of $\psi^c$ and $\Phi$, the corresponding predictions for the mass matrix of $\psi^c$ are listed in table~\ref{tab:WO_mms}.

\begin{table}[t!]
\centering
\renewcommand{\arraystretch}{1.1}
\begin{tabular}{|c|c|c|c|c|c|c|c|c|c|c|c|}  \hline\hline
$\mathcal{W}$ & $\left(\Delta(27),T^\prime\right)$  & $Y^{(k_Y)}_{\bm{r}}$ & $M_{\psi^{c}}$  \\  \hline

$\mathcal{W}_{M1}$ & $\psi^c\sim(\bm{3},\bm{3_{0}})$, $\Phi\sim(\bm{3},\bm{3_{0}})$  & $Y^{(k_Y)}_{\bm{2^{\prime\prime}}}$ & $M_{\psi^{c}}$ in Eq.~\eqref{eq:M_psic} \\  \hline

$\mathcal{W}_{M2}$  & $\psi^c\sim(\bm{3},\bm{3_{0}})$, $\Phi\sim(\bm{3},\bm{3_{1}})$  & $Y^{(k_Y)}_{\bm{2^{\prime}}}$ & $M_{\psi^{c}}\left(Y_{8}\to Y_{6},Y_{9}\to Y_{7}\right)$ \\  \hline

$\mathcal{W}_{M3}$ & $\psi^c\sim(\bm{3},\bm{3_{0}})$, $\Phi\sim(\bm{3},\bm{3_{2}})$  & $Y^{(k_Y)}_{\bm{2}}$ & $M_{\psi^{c}}\left(Y_{8}\to Y_{4},Y_{9}\to Y_{5}\right)$  \\  \hline

$\mathcal{W}_{M4}$ & $\psi^c\sim(\bm{\bar{3}},\bm{\bar{3}_{0}})$, $\Phi\sim(\bm{\bar{3}},\bm{\bar{3}_{0}})$  & $Y^{(k_Y)}_{\bm{2^{\prime}}}$ & $M_{\psi^{c}}\left(Y_{8}\to Y_{7},Y_{9}\to -\omega Y_{6}\right)$ \\  \hline

$\mathcal{W}_{M5}$ & $\psi^c\sim(\bm{\bar{3}},\bm{\bar{3}_{0}})$, $\Phi\sim(\bm{\bar{3}},\bm{\bar{3}_{1}})$  & $Y^{(k_Y)}_{\bm{2^{\prime\prime}}}$ & $M_{\psi^c}\left(Y_{8}\to  Y_{9},Y_{9}\to -\omega Y_{8}\right)$  \\  \hline

$\mathcal{W}_{M6}$ & $\psi^c\sim(\bm{\bar{3}},\bm{\bar{3}_{0}})$, $\Phi\sim(\bm{\bar{3}},\bm{\bar{3}_{2}})$  & $Y^{(k_Y)}_{\bm{2}}$ & $M_{\psi^c}\left(Y_{8}\to  Y_{5},Y_{9}\to -\omega Y_{4}\right)$\\  \hline\hline
				
\end{tabular}
\caption{\label{tab:WO_mms} The mass matrices of the Majorana fermion $\psi^c$ for possible independent assignments of $\psi^{c}$ and $\Phi$ under the eclectic symmetry group $\Omega(1)\equiv\Delta (27) \rtimes T'$. The singlet assignments of $\psi^{c}$ and $\Phi$ are omitted, as $\Omega(1)$ would play less rule. In the third column, we list the modular form multiplets giving nonzero contributions. In our convention, all couplings in $M_{\psi^c}$ are fixed to be real by the gCP transformation $K_*$. }
\end{table}

\subsection{\label{sec:Kahlerpotential} K\"ahler potential}

The K\"ahler potential is a hermitian function of the modulus $\tau$, the chiral superfields of quarks/leptons $\psi$, $\psi^c$ and the flavon superfield $\Phi$ as well as their complex conjugates $\tau^{\dagger}$, $\psi^{\dagger}$, $\psi^{c\dagger}$ and $\Phi^{\dagger}$. As shown in Eq.~\eqref{eq:kahler-MFS}, many terms on the same footing as the minimal K\"ahler are compatible with modular symmetry~\cite{Chen:2019ewa}. They can affect the fermion masses and mixing parameters significantly, see Refs.~\cite{Chen:2019ewa,Lu:2019vgm} for explicit examples. In the paradigm of eclectic flavor symmetry, the K\"ahler potential has to be invariant under both modular symmetry and the traditional flavor symmetry transformations. In particular, the traditional flavor symmetry can impose severe constraints on the K\"ahler potential so that it can be expanded in power of $\langle\Phi\rangle/\Lambda$, where $\Phi$ denotes the VEVs of the flavon fields. As explained in previous section, the matter fields $\psi$ and $\psi^c$ can only be singlet $\bm{1_{0,0}}$ or triplets $\bm{3}$ and $\bm{\bar{3}}$ of $\Delta(27)$. As an example, we assume that both $\psi$ and $\psi^c$ transform as $(\bm{3},\bm{3_{0}})$ under $\Omega(1)$ eclectic flavor group in this section. Since the K\"ahler potential of $\psi$ has a very similar form to that of $\psi^{c}$, consequently we shall focus on $\psi$ in the following.

At leading order, the K\"ahler potential for the matter fields $\psi$ can be written as\footnote{The modular transformation of $-i\tau + i \bar{\tau}$ is
\begin{equation*}
-i\tau + i \bar{\tau} \stackrel{\gamma}{\longrightarrow}|c\,\tau+d|^{-2} (-i\tau + i\bar{\tau} )\;.
\end{equation*}}
\begin{eqnarray}
\label{eq:kahler-eclectic}   \mathcal{K}_{\rm LO} &=& \sum_{n,\bm{r_1},\bm{r_2},s} (-i \tau+ i \bar \tau)^{-k_{\psi}+n}
	\left( Y^{(n)\dagger}_{\bm{r_1}} Y^{(n)}_{\bm{r_2}}  \psi^{\dagger} \psi\right)_{(\bm{1_{0,0}},\bm{1}),s}\,,
\end{eqnarray}
where the couplings of each independent term are dropped, and we should sum over the weights $n\in\mathbbm{N}$ and representation $\bm{r}$ of all linearly independent modular multiplets $Y^{(n)}_{\bm{r}}(\tau)$ and all $\Omega(1)$ singlet contractions labelled by the index $s$. The modular form with weight zero is a constant and it is taken to be $Y^{(0)}_{\bm{r}}=1$ without loss of generality. The minimal K\"ahler potential corresponds to the term of $n=0$. Because the modular forms are invariant under $\Delta(27)$ flavor symmetry, $\psi^{\dagger}\psi$ has to be a $\Delta(27)$ trivial singlet, i.e.
\begin{equation}
 \left(\psi^{\dagger}\psi\right)_{\bm{1_{0,0}}}=\psi^{\dagger}_1\psi_1+\psi^{\dagger}_2\psi_2+\psi^{\dagger}_3\psi_3=\left(\psi^{\dagger}\psi\right)_{(\bm{1_{0,0}},\bm{1})}\,,
\end{equation}
which is also an invariant singlet $\bm{1}$ of the $T'$ modular symmetry. Hence the invariance of $\mathcal{K}$ under $\Delta(27)$ requires the leading order K\"ahler potential in Eq.~\eqref{eq:kahler-eclectic} to be of the following form
\begin{eqnarray}
\mathcal{K}_{\rm LO} &=& \sum_{n,\bm{r_1},\bm{r_2}}  (-i \tau+ i \bar \tau)^{-k_{\psi}+n}
	\left( Y^{(n)\dagger}_{\bm{r_1}} Y^{(n)}_{\bm{r_2}} \right)_{(\bm{1_{0,0}},\bm{1})} \left(\psi^{\dagger} \psi\right)_{(\bm{1_{0,0}},\bm{1})}\,.
\end{eqnarray}
Note that the non-vanishing contraction $\left( Y^{(n)\dagger}_{\bm{r_1}} Y^{(n)}_{\bm{r_2}} \right)_{(\bm{1_{0,0}},\bm{1})}$ requires that $\bm{r_1}$ and $\bm{r_2}$ should be the same representation $\bm{r_1}=\bm{r_2}=\bm{r}$ of $T'$. Moreover, from the Kronecker products and Clebsch-Gordon (CG) coefficients of $T^\prime$ in Appendix~\ref{sec:Tp_group}, we see that there is only a unique singlet contraction  $\left( Y^{(n)\dagger}_{\bm{r}} Y^{(n)}_{\bm{r}} \right)_{(\bm{1_{0,0}},\bm{1})}$ for any irreducible representation of $T^\prime$, i.e.
\begin{equation}
\left( Y^{(n)\dagger}_{\bm{r}} Y^{(n)}_{\bm{r}} \right)_{(\bm{1_{0,0}},\bm{1})} =Y^{(n)\dagger}_{\bm{r},1} Y^{(n)}_{\bm{r},1}+Y^{(n)\dagger}_{\bm{r},2} Y^{(n)}_{\bm{r},2}+\ldots\,.
\end{equation}
Then the most general leading order K\"ahler potential invariant under $\Omega(1)$ for $\psi$ takes the form
\begin{eqnarray}
\nonumber\mathcal{K}_{\rm LO} &=& \sum_{n,\bm{r}} (-i \tau+ i \bar \tau)^{-k_{\psi}+n}
	\left( Y^{(n)\dagger}_{\bm{r}} Y^{(n)}_{\bm{r}} \right)_{(\bm{1_{0,0}},\bm{1})} \left(\psi^{\dagger} \psi\right)_{(\bm{1_{0,0}},\bm{1})}\\
\label{eq:LO-Kaehler-final}&=&\sum_{n,\bm{r}}  (-i \tau+ i \bar \tau)^{-k_{\psi}+n}
	\left( Y^{(n)\dagger}_{\bm{r}} Y^{(n)}_{\bm{r}} \right)_{(\bm{1_{0,0}},\bm{1})} \left(\psi^{\dagger}_1\psi_1+\psi^{\dagger}_2\psi_2+\psi^{\dagger}_3\psi_3\right)\,.
\end{eqnarray}
Hence the K\"ahler metric is diagonal and it is proportional to a unit matrix at leading order. The overall factor of $\mathcal{K}_{\rm LO}$ can be absorbed by redefinition of field, thus leading order K\"ahler potential invariant under $\Omega(1)$ doesn't introduce any corrections to fermion masses and mixing due to the traditional flavor symmetry $\Delta(27)$. If the three generations of matter fields $\psi$ transform as $(\bm{1_{0,0}}, \bm{1}^{k})$ under the eclectic flavor group $\Delta(27)\rtimes T'$, they are generally distinguished by the different charges under auxiliary abelian symmetry, so that the K\"ahler potential would be still diagonal.

Then we proceed to consider the next-to-leading-order (NLO) corrections to the K\"ahler potential of $\psi$. It involve only one flavon $\Phi$ and its most general form is given by
\begin{eqnarray}
\label{eq:DeltaK-1}  \mathcal{K}_{\rm NLO} &=& \sum_{m, n,\bm{r_1},\bm{r_2}, s} (-i \tau+ i \bar \tau)^{-k_{\psi}+m}
	\left( Y^{(m)\dagger}_{\bm{r_1}} Y^{(n)}_{\bm{r_2}}  \psi^{\dagger} \psi\Phi\right)_{(\bm{1_{0,0}},\bm{1}),s}+\text{h.c.}\,,
\end{eqnarray}
where the modular weight $k_{\Phi}$ of the flavon $\Phi$ should fulfill
\begin{equation}
k_{\Phi}=n-m\,.
\end{equation}
If some auxiliary symmetry is imposed, obviously the flavon $\Phi$ should be chargeless under the auxiliary group. From the $\Delta(27)$ tensor product $\bm{\bar{3}}\otimes \bm{3}=\sum^{2}_{r,s=0} \bm{1_{r,s}}$ in Eq.~\eqref{eq:KP_Delta27}, we know the contractions of $\psi^{\dagger}\psi$ are
\begin{eqnarray}
\nonumber && \left(\psi^{\dagger}\psi\right)_{(\bm{1_{0,0}},\bm{1})}=\psi^{\dagger}_1\psi_1 +\psi^{\dagger}_2\psi_2+\psi^{\dagger}_3\psi_3\,, \\
\label{eq:LLb-contract}	&& \left(\psi^{\dagger}\psi\right)_{(\bm{8}, \bm{8_0})}=\left(\begin{array}{c}
\psi^{\dagger}_1\psi_2 +\psi^{\dagger}_2\psi _3+\psi^{\dagger} _3\psi_1 \\
\psi^{\dagger}_1\psi_3+\psi^{\dagger}_2\psi_1+\psi^{\dagger}_3\psi_2 \\
\psi^{\dagger}_1 \psi_1+ \omega^2\psi^{\dagger}_2\psi_2 + \omega\psi^{\dagger}_3\psi_3 \\
\omega^2\psi^{\dagger}_1\psi_2+\omega\psi^{\dagger}_2\psi_3+\psi^{\dagger}_3\psi_1\\
\omega\psi^{\dagger}_1\psi_3+\psi^{\dagger}_2\psi_1+\omega^2\psi^{\dagger}_3\psi_2 \\
\psi^{\dagger}_1 \psi_1+  \omega\psi^{\dagger}_2\psi_2 + \omega^2\psi^{\dagger}_3 \psi_3 \\
\omega\psi^{\dagger}_1\psi_2+\omega^2 \psi^{\dagger}_2\psi_3+\psi^{\dagger}_3\psi_1\\
\omega^2 \psi^{\dagger}_1\psi_3+\psi^{\dagger}_2\psi_1+\omega\psi^{\dagger}_3\psi_2
\end{array}\right)\,.
\end{eqnarray}
Invariance under $\Delta(27)$ require $\psi^{\dagger}\psi\Phi$ should be a trivial singlet of $\Delta(27)$, so that the flavon $\Phi_{i}$ must transform as $\bm{1_{0,0}}$ or $\bm{8}$ under $\Delta(27)$. In the case of $\Phi\sim(\bm{1_{0,0}},\bm{1^k})$, we find
\begin{eqnarray}
\nonumber\left(\psi^{\dagger}\psi\Phi\right)_{(\bm{1_{0,0}},\bm{1^k})}&=&\left(\psi^{\dagger}\psi\right)_{(\bm{1_{0,0}},\bm{1})}\Phi=\left(\psi^{\dagger}_1\psi_1+\psi^{\dagger}_2\psi_2+\psi^{\dagger}_3\psi_3\right)\Phi\,,\\
\left( Y^{(m)\dagger}_{\bm{r_1}} Y^{(n)}_{\bm{r_2}}  \psi^{\dagger} \psi\Phi\right)_{(\bm{1_{0,0}},\bm{1})}&=&\left( Y^{(m)\dagger}_{\bm{r_1}} Y^{(n)}_{\bm{r_2}} \right)_{(\bm{1_{0,0}},\bm{1^{[3-k]}})}\left( \psi^{\dagger} \psi\right)_{(\bm{1_{0,0}},\bm{1})}\Phi\,,
\end{eqnarray}
which is proportional to $\psi^{\dagger}_1\psi_1+\psi^{\dagger}_2\psi_2+\psi^{\dagger}_3\psi_3$. As a result, the corresponding contribution can be absorbed into the LO K\"ahler potential in Eq.~\eqref{eq:LO-Kaehler-final}, and it yields no corrections to fermion mass matrices extracted from the superpotential. If the flavon transform as $\Phi\sim(\bm{8},\bm{8_k})$, the  NLO K\"ahler potential $\mathcal{K}_{\rm NLO}$ invariant under the eclectic flavor group $\Omega(1)$ would be of the following form
\begin{eqnarray}
\nonumber \mathcal{K}_{\rm NLO} &=& \sum_{m, n,\bm{r_1},\bm{r_2}}(-i \tau+ i \bar \tau)^{-k_{\psi}+m}
	\left( Y^{(m)\dagger}_{\bm{r_1}} Y^{(n)}_{\bm{r_2}} \right)_{(\bm{1_{0,0}},\bm{1^{[3-k]}})} \left(\left(\psi^{\dagger} \psi\right)_{(\bm{8}, \bm{8_0})}\Phi\right)_{(\bm{1_{0,0}},\bm{1^{[k]}})}\\
\nonumber+&&\hskip-0.3in\sum_{m, n,\bm{r_1},\bm{r_2}}(-i \tau+ i \bar \tau)^{-k_{\psi}+m}
\left(\left( Y^{(m)\dagger}_{\bm{r_1}} Y^{(n)}_{\bm{r_2}} \right)_{(\bm{1_{0,0}},\bm{2^{[2-k]}})} \left(\left(\psi^{\dagger} \psi\right)_{(\bm{8}, \bm{8_0})}\Phi\right)_{(\bm{1_{0,0}},\bm{2^{[1+k]}})}\right)_{(\bm{1_{0,0}},\bm{1})}\\
\nonumber+&&\hskip-0.3in\sum_{m, n,\bm{r_1},\bm{r_2}}(-i \tau+ i \bar \tau)^{-k_{\psi}+m}
\left(\left( Y^{(m)\dagger}_{\bm{r_1}} Y^{(n)}_{\bm{r_2}} \right)_{(\bm{1_{0,0}},\bm{2^{[1-k]}})} \left(\left(\psi^{\dagger} \psi\right)_{(\bm{8}, \bm{8_0})}\Phi\right)_{(\bm{1_{0,0}},\bm{2^{[2+k]}})}\right)_{(\bm{1_{0,0}},\bm{1})} \\
\nonumber +&&\hskip-0.3in\sum_{m, n,\bm{r_1},\bm{r_2}}(-i \tau+ i \bar \tau)^{-k_{\psi}+m}
\left(\left( Y^{(m)\dagger}_{\bm{r_1}} Y^{(n)}_{\bm{r_2}} \right)_{(\bm{1_{0,0}},\bm{3})} \left(\left(\psi^{\dagger} \psi\right)_{(\bm{8}, \bm{8_0})}\Phi\right)_{(\bm{1_{0,0}},\bm{3})}\right)_{(\bm{1_{0,0}},\bm{1})}+\text{h.c.}\,. \\
\end{eqnarray}
The general formulas for the $\Delta(27)$ trivial singlet contractions of two octets $\chi\sim(\bm{8},\bm{8_k})$ and $\varphi\sim(\bm{8},\bm{8_{k^\prime}})$ are given by:
\begin{eqnarray}
\nonumber \left(\chi\varphi\right)_{(\bm{1_{0,0}},\bm{1^{[k+k^\prime]}})}&=&\chi_{1}\varphi_{2}+\chi_{2}\varphi_{1} +\chi_{3}\varphi_{6}+\chi_{4}\varphi_{8}+\chi_{5}\varphi_{7}+\chi_{6}\varphi_{3} +\chi_{7}\varphi_{5}+\chi_{8}\varphi_{4} \,, \\
\nonumber \left(\chi\varphi\right)_{(\bm{1_{0,0}},\bm{2^{[k+k^\prime+1]}})}&=&\left(\begin{array}{c}
\sqrt{2}\, \left(\omega\chi_{1}\varphi_{2}-\omega\chi_{2}\varphi_{1}+\omega^2\chi_{4}\varphi_{8} -\chi_{5}\varphi_{7} +\chi_{7}\varphi_{5} -\omega^2\chi_{8}\varphi_{4}\right)\\
-\chi_{1}\varphi_{2} +\chi_{2}\varphi_{1} -i \sqrt{3}\, \chi_{3} \varphi_{6}-\chi_{4}\varphi_{8}+\chi_{5}\varphi_{7}  +i \sqrt{3}\, \chi_{6}\varphi_{3}-\chi_{7}\varphi_{5} +\chi_{8}\varphi_{4}
\end{array}\right)\,, \\
\nonumber \left(\chi\varphi\right)_{(\bm{1_{0,0}},\bm{2^{[k+k^\prime+2]}})}&=&\left(\begin{array}{c}
\chi_{1}\varphi_{2} -\chi_{2}\varphi_{1} -i \sqrt{3}\,\chi_{3}\varphi_{6}+\chi_{4}\varphi_{8} -\chi_{5}\varphi_{7} +i \sqrt{3}\, \chi_{6}\varphi_{3} +\chi_{7}\varphi_{5} -\chi_{8}\varphi_{4} \\
\sqrt{2}\left(\omega^2\chi_{1}\varphi_{2} -\omega^2\chi_{2}\varphi_{1} +\omega\chi_{4}\varphi_{8} -\chi_{5}\varphi_{7} +\chi_{7}\varphi_{5} -\omega\chi_{8}\varphi_{4} \right)
\end{array}\right)\,, \\
\nonumber \left(\chi\varphi\right)_{(\bm{1_{0,0}},\bm{3})}&=&P^{k+k^\prime}\left(\begin{array}{c}
\chi_{1}\varphi_{2} +\chi_{2}\varphi_{1} -3 \chi_{3}\varphi_{6} +\chi_{4}\varphi_{8} +\chi_{5}\varphi_{7}-3 \chi_{6}\varphi_{3} +\chi_{7}\varphi_{5} +\chi_{8}\varphi_{4} \\
-2(\omega^2\chi_{1}\varphi_{2} + \omega^2\chi_{2}\varphi_{1} +\omega\chi_{4}\varphi_{8} + \chi_{5}\varphi_{7} + \chi_{7}\varphi_{5}+ \omega\chi_{8}\varphi_{4})\\
-2 (\omega\chi_{1}\varphi_{2} +\omega\chi_{2}\varphi_{1}+\omega^2\chi_{4}\varphi_{8} +\chi_{5}\varphi_{7}  +\chi_{7}\varphi_{5}+\omega^2\chi_{8}\varphi_{4}  )
\end{array}\right)\,, \\
\label{eq:R8-R8-cont}
\end{eqnarray}
with the permutation matrix $P$ given in Eq.~\eqref{eq:per_mat}.
Using the contraction rules in Eq.~\eqref{eq:R8-R8-cont}, we can write out the explicit expression of $\mathcal{K}_{\rm NLO}$, yet it is too lengthy to show here. We find that $\mathcal{K}_{\rm NLO}$ will give rise to off-diagonal elements of the K\"ahler metric and they are suppressed by $\langle\Phi\rangle/\Lambda$ in comparison with $\mathcal{K}_{\rm LO}$.

If the $\Omega(1)$ eclectic model does not contain a flavon $\Phi$ transforming as $(\bm{8},\bm{8_k})$ under $\Delta(27)\rtimes T^\prime$, the off-diagonal contributions to the K\"ahler metric of the matter field $\psi$ could arise from the next-to-next-to-leading order (NNLO) terms which can be written as\footnote{Depending on the transformation properties of the matter fields and flavons under $\Omega(1)$, $\mathcal{K}_{\rm NNLO}$ can possibly admit the terms of form $\sum_{m, n,\bm{r_1},\bm{r_2}, s}\; (-i \tau+ i \bar \tau)^{-k_{\psi}-k_{\Theta}+m}\left( Y^{(m)\dagger}_{\bm{r_1}} Y^{(n)}_{\bm{r_2}}  \psi^{\dagger} \psi\Theta\Phi\right)_{(\bm{1_{0,0}},\bm{1}),s}+\text{h.c.}$, which can be analyzed in a similar way. }:
\begin{eqnarray}
\label{eq:DeltaK-2}  \mathcal{K}_{\rm NNLO} &=& \sum_{m, n,\bm{r_1},\bm{r_2}, s} (-i \tau+ i \bar \tau)^{-k_{\psi}-k_{\Theta}+m}
	\left( Y^{(m)\dagger}_{\bm{r_1}} Y^{(n)}_{\bm{r_2}}  \psi^{\dagger}\psi\Theta^{\dagger}\Phi\right)_{(\bm{1_{0,0}},\bm{1}),s}+\text{h.c.}\,.
\end{eqnarray}
We see that two generic flavons $\Phi$ and $\Theta$ are involved and they could be identical with $\Phi=\Theta$. The total modular weight of each term should be vanishing and consequently the modular weight should fulfill the following relation
\begin{equation}
m-n=k_{\Theta}-k_{\Phi}\,.
\end{equation}
The traditional flavor symmetry $\Delta(27)$ requires the combination $\psi^{\dagger} \psi\Theta^{\dagger}\Phi$ be a $\Delta(27)$ singlet $\bm{1_{0,0}}$, thus we have the contractions
\begin{eqnarray}
\label{eq:psi-psiDag-theta-phi}\left(\psi^{\dagger} \psi\Theta^{\dagger}\Phi\right)_{(\bm{1_{0,0}},\bm{r})}=\left(\psi^{\dagger}\psi\right)_{(\bm{1_{0,0}},\bm{1})}\left(\Theta^{\dagger}\Phi\right)_{(\bm{1_{0,0}},\bm{r})}
~~~\text{or}~~~\left[\left(\psi^{\dagger}\psi\right)_{(\bm{8},\bm{8_0})}\left(\Theta^{\dagger}\Phi\right)_{(\bm{8},\bm{8_k})}
\right]_{(\bm{1_{0,0}},\bm{r})}\,,
\end{eqnarray}
which contracts with the modular form $Y^{(m)\dagger}_{\bm{r_1}} Y^{(n)}_{\bm{r_2}}$ to form $T'$ invariants, here $\bm{r}$ are certain representations of $T'$ modular group. As can be seen from the expressions of $\left(\psi^{\dagger}\psi\right)_{(\bm{1_{0,0}},\bm{1})}$ and $\left(\psi^{\dagger}\psi\right)_{(\bm{8},\bm{8_0})}$ in Eq.~\eqref{eq:LLb-contract}, the first contraction $\left(\psi^{\dagger}\psi\right)_{(\bm{1_{0,0}},\bm{1})}\left(\Theta^{\dagger}\Phi\right)_{(\bm{1_{0,0}},\bm{r})}$ leads to a diagonal K\"ahler metric and its contribution can be absorbed in the overall normalization while the second contraction $\left[\left(\psi^{\dagger}\psi\right)_{(\bm{8},\bm{8_0})}\left(\Theta^{\dagger}\Phi\right)_{(\bm{8},\bm{8_k})}
\right]_{(\bm{1_{0,0}},\bm{r})}$ gives rise to a off-diagonal K\"ahler metric. Moreover, we see that $\Theta^{\dagger}\Phi$ should be the invariant singlet $\bm{1_{0,0}}$ or the reducible octet $\bm{8}$ of the $\Delta(27)$ flavor symmetry. The flavons $\Theta$ and $\Phi$ can transform as $\bm{1_{0,0}}$, $\bm{3}$, $\bm{\bar{3}}$ or $\bm{8}$ under $\Delta(27)$, and they can contribute to the K\"ahler potential $\mathcal{K}_{\rm NNLO}$ only for the following assignments.
\begin{itemize}
\item[i)] $\Theta\sim\bm{1_{0,0}}$, $\Phi\sim\bm{1_{0,0}}$ \\
In this case, only the first contraction in Eq.~\eqref{eq:psi-psiDag-theta-phi} is allowed, and the resulting K\"ahler potential is of the same form as the minimal K\"ahler potential.

\item[ii)] $\Theta, \Phi\sim\bm{3}$ or $\Theta, \Phi\sim\bm{\bar{3}}$  \\
In terms of the components of the two $\Delta(27)$ triplets $\Theta=\left(\vartheta_1, \vartheta_2, \vartheta_3\right)^T$, $\Phi=\left(\phi_1, \phi_2, \phi_3\right)^{T}$, we have
\begin{eqnarray}
\nonumber && \left(\Theta^{\dagger}\Phi\right)_{(\bm{1_{0,0}}, \bm{1^k})}=\vartheta^{\dagger}_1\phi_1 +\vartheta^{\dagger}_2\phi_2+\vartheta^{\dagger}_3\phi_3\,, \\
\nonumber&& \left(\Theta^{\dagger}\Phi\right)_{(\bm{8}, \bm{8_k})}=\left(\begin{array}{c}
\vartheta^{\dagger}_1\phi_2 +\vartheta^{\dagger}_2\psi _3+\vartheta^{\dagger} _3\phi_1 \\
\vartheta^{\dagger}_1\phi_3+\vartheta^{\dagger}_2\phi_1+\vartheta^{\dagger}_3\phi_2 \\
\vartheta^{\dagger}_1 \phi_1+ \omega^2\vartheta^{\dagger}_2\phi_2 + \omega\vartheta^{\dagger}_3\phi_3 \\
\omega^2\vartheta^{\dagger}_1\phi_2+\omega\vartheta^{\dagger}_2\phi_3+\vartheta^{\dagger}_3\phi_1\\
\omega\vartheta^{\dagger}_1\phi_3+\vartheta^{\dagger}_2\phi_1+\omega^2\vartheta^{\dagger}_3\phi_2 \\
\vartheta^{\dagger}_1 \phi_1+  \omega\vartheta^{\dagger}_2\phi_2 + \omega^2\vartheta^{\dagger}_3 \phi_3 \\
\omega\vartheta^{\dagger}_1\phi_2+\omega^2 \vartheta^{\dagger}_2\phi_3+\vartheta^{\dagger}_3\phi_1\\
\omega^2 \vartheta^{\dagger}_1\phi_3+\vartheta^{\dagger}_2\phi_1+\omega\vartheta^{\dagger}_3\phi_2
\end{array}\right)~~~\text{for}~~~\Theta, \Phi\sim\bm{3}\,,\\
\label{eq:Theta-dag-Phi-33}&&\left(\Theta^{\dagger}\Phi\right)_{(\bm{8},\bm{8_k})}=\left(\begin{array}{c}
		\vartheta^{\dagger}_{1} \phi_{3}+ \vartheta^{\dagger}_{2} \phi_{1}+\vartheta^{\dagger}_{3} \phi_{2} \\
		\vartheta^{\dagger}_{1} \phi_{2}+\vartheta^{\dagger}_{2} \phi_{3}+\vartheta^{\dagger}_{3} \phi_{1} \\
		\vartheta^{\dagger}_{1} \phi_{1} + \omega^2 \vartheta^{\dagger}_{2} \phi_{2} + \omega \vartheta^{\dagger}_{3} \phi_{3}\\
		\vartheta^{\dagger}_{1} \phi_{3}+\omega^2 \vartheta^{\dagger}_{2} \phi_{1}+\omega \vartheta^{\dagger}_{3} \phi_{2}\\
		\vartheta^{\dagger}_{1} \phi_{2}+\omega^2\vartheta^{\dagger}_{2} \phi_{3}+\omega \vartheta^{\dagger}_{3} \phi_{1} \\
		\vartheta^{\dagger}_{1} \phi_{1} +  \omega \vartheta^{\dagger}_{2} \phi_{2} + \omega^2 \vartheta^{\dagger}_{3} \phi_{3}\\
		\vartheta^{\dagger}_{1} \phi_{3}+\omega \vartheta^{\dagger}_{2} \phi_{1}+\omega^2 \vartheta^{\dagger}_{3} \phi_{2}\\
		\vartheta^{\dagger}_{1} \phi_{2}+\omega \vartheta^{\dagger}_{2} \phi_{3}+\omega^2\vartheta^{\dagger}_{3} \phi_{1}
	\end{array}\right) ~~~\text{for}~~~\Theta, \Phi\sim\bm{\bar{3}}\,.
\end{eqnarray}
Then both singlet and octet contractions in Eq.~\eqref{eq:psi-psiDag-theta-phi} are allowed, and the explicit expressions can be obtained from Eqs.~(\ref{eq:LLb-contract},\ref{eq:R8-R8-cont},\ref{eq:Theta-dag-Phi-33}). Obviously these contributions depend on the flavon VEVs and are suppressed by $\langle\Phi\rangle^2/\Lambda^2$ with respect to the leading order term $\mathcal{K}_{\rm LO}$.

\item[iii)] $\Theta\sim\bm{1_{0,0}}, \Phi\sim\bm{8}$ or $\Theta\sim\bm{8}, \Phi\sim\bm{1_{0,0}}$ \\

Only the octet contractions in Eq.~\eqref{eq:psi-psiDag-theta-phi} is allowed in this case, and it leads to off-diagonal K\"ahler metric of the matter fields.

\item[iv)] $\Theta\sim\bm{8}$ and $\Phi\sim\bm{8}$  \\

For this assignment, eight independent singlet contractions $\left(\Theta^{\dagger}\Phi\right)_{(\bm{1_{0,0}}, \bm{1^k})}$ and seven independent octet contractions $\left(\Theta^{\dagger}\Phi\right)_{(\bm{8},\bm{8_k})}$ can be built. They generally induce corrections to the canonical kinetic terms of matter fields.

\end{itemize}

In short, generally the K\"ahler potential $\mathcal{K}_{\rm NNLO}$ can yield deviations from canonical kinetic terms of quark/lepton fields after the flavons develop VEVs, unless all the flavons $\Theta$  and $\Phi$ are invariant singlet of $\Delta(27)$. However, the induced corrections to the quark/lepton mixing parameters are suppressed by $\langle\Phi\rangle^2/\Lambda^2$ which are usually small enough to be negligible. In the following example models, we do not introduce a flavon transforming as $(\bm{8}, \bm{8}_k)$ under $\Delta(27)\rtimes T'$. Hence the corrections to the minimal K\"ahler potential arise from $\mathcal{K}_{\rm NNLO}$.

\section{\label{eq:example-models} $\Omega(1)\cong \Delta(27)\rtimes T'$ eclectic models of lepton mass and mixing }

\begin{table}[t!]
	\centering
	\begin{tabular}{|c|c|c|c|c|c|c|c|c|}
		\hline \hline
		$Z^{g}_{3}$ & Res. CP &  $\langle\phi_{\bm{3}}\rangle/v_{\phi}$  &    $\langle\phi_{\bm{\bar{3}}}\rangle/v_{\phi}$ &  $\langle\phi_{\bm{8}}\rangle/v_{\phi}$ \\ \hline

		$Z^{A}_{3}$ & ---  & $(1,1,1)^T$  & $(1,1,1)^T$ &  $v_{8_1}=(1,x,0,0,0,0,0,0)^T$  \\ \hline
		\multirow{3}{*}{$Z^{B}_{3}$} & $A$ & $(1,0,0)^T$ & $(1,0,0)^T$  &  $v_{8_2}=(0,0,\omega^2,0,0,-\omega x,0,0)^T$  \\ \cline{2-5}
		& $AB$ & $(\omega^2,0,0)^T$ & $(\omega,0,0)^T$  &  $v_{8_2}$  \\ \cline{2-5}
		& $BAB^2$ & $(\omega,0,0)^T$ & $(\omega^2,0,0)^T$  &  $v_{8_2}$  \\ \hline
		
		\multirow{3}{*}{$Z^{AB}_{3}$}  &  $BAB^2A^2$ & $(1,\omega^2,1)^T$  & $(1,\omega,1)^T$  &  $v_{8_3}=(0,0,0,0,1-ix,0,1+ix,0)^T$  \\ \cline{2-5}
		&  $ABA^2B^2$ & $\omega(1,\omega^2,1)^T$  & $\omega^2(1,\omega,1)^T$  &  $v_{8_3}$  \\ \cline{2-5}
		&  $AB$ & $\omega^2(1,\omega^2,1)^T$  & $\omega(1,\omega,1)^T$  &  $v_{8_3}$  \\ \hline
		
		\multirow{3}{*}{$Z^{BA}_{3}$} &  $BAB^2A^2$ & $\omega(1,1,\omega^2)^T$  & $\omega^2(1,1,\omega)^T$ &  $v_{8_3}$ \\ \cline{2-5}
		&  $ABA^2B^2$ & $\omega^2(1,1,\omega^2)^T$  & $\omega(1,1,\omega)^T$ &  $v_{8_3}$ \\ \cline{2-5}
		&  $BA$ & $(1,1,\omega^2)^T$  & $(1,1,\omega)^T$ &  $v_{8_3}$ \\ \hline
		
		$Z^{ABA}_{3}$ & --- & $(1,\omega^2,\omega^2)^T$  & $(1,\omega,\omega)^T$ &  $v_{8_4}=(0,0,0,1,0,0,0,x)^T$ \\ \hline
		$Z^{A^2B}_{3}$ & --- & $(1,1,\omega)^T$ & $(1,1,\omega^2)^T$ &  $v_{8_4}$ \\ \hline
		
		$Z^{BA^2}_{3}$ & --- & $(1,\omega,1)^T$  & $(1,\omega^2,1)^T$ &  $v_{8_4}$ \\ \hline
		$Z^{BAB^2}_{3}$ & --- & $(1,\omega,\omega^2)^T$ & $(1,\omega^2,\omega)^T$ &  $v_{8_1}$ \\ \hline
		
		$Z^{B^2AB}_{3}$ & --- & $(1,\omega^2,\omega)^T$  & $(1,\omega,\omega^2)^T$  &  $v_{8_1}$\\ \hline
		
		\multirow{3}{*}{$Z^{A^2BA}_{3}$} & $A^2$  & $(0,1,0)^T$ & $(0,1,0)^T$ &  $v^*_{8_2}$ \\ \cline{2-5}
		& $B^2A^2B$  & $\omega(0,1,0)^T$ & $\omega^2(0,1,0)^T$ &  $v^*_{8_2}$ \\ \cline{2-5}
		& $BA^2B^2$  & $\omega^2(0,1,0)^T$ & $\omega(0,1,0)^T$ &  $v^*_{8_2}$ \\ \hline
		
		\multirow{3}{*}{$Z^{ABA^2}_{3}$} & $BAB^2A^2$  & $(0,0,1)^T$  &  $(0,0,1)^T$ &  $v_{8_5}=(0,0,1,0,0, x,0,0)^T$ \\ \cline{2-5}
		& $ABA^2B^2$  & $\omega(0,0,1)^T$  &  $\omega^2(0,0,1)^T$ &  $v_{8_5}$ \\ \cline{2-5}
		& $AB^2A^2$  & $\omega^2(0,0,1)^T$  &  $\omega(0,0,1)^T$ &  $v_{8_5}$ \\ \hline
		
		\multirow{3}{*}{$Z^{A^2BA^2}_{3}$} & $BAB^2A^2$  & $(\omega^2,1,1)^T$ & $(\omega,1,1)^T$ &  $v_{8_3}$ \\ \cline{2-5}
		& $ABA^2B^2$  & $\omega(\omega^2,1,1)^T$ & $\omega^2(\omega,1,1)^T$ &  $v_{8_3}$ \\ \cline{2-5}
		& $A^2BA^2$  & $\omega^2(\omega^2,1,1)^T$ & $\omega(\omega,1,1)^T$ &  $v_{8_3}$ \\ \hline
		
		$Z^{BAB^{2}A^2}_{3}$  & ---  & --- & --- &  general 8-D vector \\ \hline\hline

\end{tabular}
\caption{\label{tab:VEVs_subs}
The most general vacuum alignment invariants under the $Z_3$ subgroups of $\Delta(27)$ and residual CP symmetry, where $\phi_{\bm{3}}$, $\phi_{\bm{\bar{3}}}$ and $\phi_{\bm{8}}$ denote the flavons which transform as $\bm{3}$, $\bm{\bar{3}}$ and reducible $\bm{8}$ respectively under $\Delta(27)$, and $x$ is a real free parameter. The residual CP symmetry in the second column refers to $\rho(g)\rho(K_{*})$ with $g\in\Delta(27)$, and we only present the element $g$ for simplicity. The overall VEV $v_{\phi}$ is constrained to be real by the residual CP symmetry. }
\end{table}

In this section, we shall construct two concrete lepton models based on the eclectic flavor group $\Omega(1)\cong \Delta(27)\rtimes T'$ in the framework of rigid $\mathcal{N}=1$ global supersymmetry. Then modular invariance requires that the modular weight of each term in the Lagrangian should be vanishing. In the two example models, the neutrinos are assumed to be Majorana particles and their masses are generated through the Weinberg operator and the type I seesaw mechanism, respectively. The model construction is guided by the principle of minimality and simplicity, and consequently we shall introduce as few flavon fields as possible besides the complex modulus $\tau$. The VEVs of flavons spontaneously break the $\Delta(27)$ flavor symmetry, and the breaking of $T'$ modular symmetry can arise from the VEVs of both flavons and modulus $\tau$. Without loss of generality, the VEV of $\tau$ is limited in the fundamental domain $\mathcal{D}$ of $SL(2,\mathbb{Z})$:
\begin{equation}
	\label{eq:fundomain_Gamma}
 \mathcal{D}=\left\{\tau ~\big|~ |\Re{\tau}|\leq1/2\,,~\Im{\tau}>0\,,~ |\tau|\geq1 \right\}\,.
\end{equation}
Other values of $\tau$ are related by modular transformations to those in $\mathcal{D}$. In both example models, the Higgs doublets $H_{u,d}$ are invariant under both $\Delta(27)$ and $T^\prime$ with vanishing modular weights. The flavon fields transform as trivial singlet $\bm{1_{0,0}}$ or triplet $\bm{3}$, $\bm{\bar{3}}$ under $\Delta(27)$, and we don't introduce any flavon in nontrivial singlet representation of $\Delta(27)$ otherwise there would be a flavon octant, as explained in section~\ref{sec:EFG-Omega1-decomp}. Extensive studies of the discrete flavor symmetry approach to the flavor problem have revealed that the flavon fields have to develop a set of particularly VEVs~\cite{Feruglio:2017spp}. Hence we will consider the vacuum configuration of flavon which preserve certain subgroup of $\Delta(27)$ and the most general invariant VEVs of all $Z_{3}$ subgroups of $\Delta(27)$ are given in table~\ref{tab:VEVs_subs}. It is known that rather elaborate scalar potentials of flavons should be designed to get the correct vacuum alignment in traditional discrete flavor symmetry approach. On the other hand, it is still an open question to determine the VEV of $\tau$ from first principles in modular flavor symmetry, this is the so-called modulus stabilisation problem~\cite{Ishiguro:2020tmo,Novichkov:2022wvg}. The presence of both flavons and complex modulus in eclectic flavor symmetry exacerbates the vacuum selection problem mentioned above. Hence we will not attempt to build the most general supersymmetric and eclectic symmetry invariant scalar potential for the modulus $\tau$ and flavons. Both $\tau$ and flavons will be treated as spurions, and the VEV of $\tau$ is freely varied in the fundamental domain to adjust the agreement with the data.

In previous section, we have performed a general analysis for the $\Omega(1)$ invariant superpotential and K\"ahler potential. Given the representation assignments of the lepton fields and flavons in the two example models, we find that the nontrivial corrections to the minimal K\"ahler potential arise at the NNLO. The corresponding contributions to the lepton masses and mixing parameters are suppressed by $\langle\Phi\rangle^2/\Lambda^2$ such that they are negligible. We shall present the two example models based on the eclectic flavor symmetry $\Omega(1)\cong \Delta(27)\rtimes T'$ in the following.

\subsection{\label{sec:_model_WO}  Example model 1 }

\begin{table}[t!]
	\renewcommand{\tabcolsep}{1.05mm}
	\centering
	\begin{tabular}{|c|c|c|c|c||c|c|c|c|c|}
		\hline \hline
		Fields &   $L$    &  $E^{c}$    &  $H_{u}$ & $H_{d}$ &  $\phi$  &  $\varphi$  &  $\chi$  & $\xi$  & $Y^{(k_Y)}_{\bm{r}}$ \\ \hline
		
		$\text{SU}(2)_L\times \text{U}(1)_{Y}$ & $(\bm{2},-\frac{1}{2})$ & $(\bm{1},1)$ & $(\bm{2},\frac{1}{2})$ & $(\bm{2},-\frac{1}{2})$ &  $(\bm{1},0)$ &  $(\bm{1},0)$ &  $(\bm{1},0)$ &  $(\bm{1},0)$ &  $(\bm{1},0)$ \\ \hline
		
		$\Delta(27)$  &   $\bm{3}$  &  $\bm{3}$  &  $\bm{1_{0,0}}$   & $\bm{1_{0,0}}$   &  $\bm{3}$   & $\bm{3}$  & $\bm{3}$ &  $\bm{1_{0,0}}$  & $\bm{1_{0,0}}$   \\ \hline

		$\Gamma^\prime_3\cong T^\prime$ & $\bm{3_0}$ & $\bm{3_0}$ & $\bm{1}$  & $\bm{1}$ & $\bm{3_1}$ & $\bm{3_0}$ & $\bm{3_1}$ & $\bm{1}$  & $\bm{r}$ \\ \hline
		
		modular weight & $0$ & $0$ & $0$ & $0$ & $5$ & $5$ & $7$ & $-1$ & $k_Y$ \\  \hline
		$Z_2$ & $1$ & $-1$ & $1$ & $1$ & $-1$ & $1$ & $1$  & $1$   & $1$  \\ \hline
		$Z_3$ & $\omega$ & $\omega^2$ & $1$ & $1$ & $1$ & $\omega$ &  $\omega$ & $1$  & $1$   \\ \hline \hline
		
\end{tabular}
\caption{\label{tab:model_fields_WO}The matter fields, flavon fields of the example model 1 and their transformation properties under $\Delta(27)$ and $T^\prime$, where $Y^{(k_Y)}_{\bm{r}}$ refers to the level 3 and weight $k_Y$ modular multiplet in the irreducible representation $\bm{r}$ of $T^\prime$. }
\end{table}

The light neutrino masses are generated by the Weinberg operator in this model. All the fields and their classifications under the traditional flavor group $\Delta(27)$ and the finite modular group $T^\prime$ are summarized in table~\ref{tab:model_fields_WO}. The LH leptons $L$, RH charged leptons $E^c$, the flavon $\phi$ in the charged lepton and the flavons $\varphi$ and $\chi$ in the neutrino sector are all assumed to be triplet $\bm{3}$ of $\Delta(27)$. Then these fields must transform as reducible three-dimensional representations $\bm{3_0}$, $\bm{3_1}$ or $\bm{3_2}$, and the corresponding representation matrices of generators $S$ and $T$ are given in Eq.~\eqref{eq:3DTpfromDelta27}. Furthermore, one $\Delta(27)$ trivial singlet flavon $\xi$ is introduced in the model. In order to eliminate the unwanted operators, we introduce the auxiliary symmetry $Z_{2}\times Z_{3}$ in the model. Then we can read out the superpotential for the charged lepton and neutrino masses, which is invariant under the symmetry of the model,
\begin{eqnarray}
\nonumber \mathcal{W}&=& \frac{\alpha}{\Lambda}  \left(E^c L\phi Y^{(5)}_{\bm{2^{\prime}}}\right)_{(\bm{1_{0,0}},\bm{1})} H_d +\frac{\beta}{\Lambda^2}  \left(E^c L\xi\phi Y^{(4)}_{\bm{1}}\right)_{(\bm{1_{0,0}},\bm{1})} H_d \\
\label{eq:SPP_WO}	&&+\frac{g_1}{2\Lambda^2}\left(LL\varphi Y^{(5)}_{\bm{2^{\prime\prime}}}\right)_{(\bm{1_{0,0}},\bm{1})}H_{u}H_{u}
+\frac{g_2}{2\Lambda^2}\left(LL\chi Y^{(7)}_{\bm{2^{\prime}}}\right)_{(\bm{1_{0,0}},\bm{1})}H_{u}H_{u}\,.
\end{eqnarray}
The first and second rows of Eq.~\eqref{eq:SPP_WO} are the mass terms of charged leptons and neutrinos, respectively. The charged lepton superpotential is the case of $\mathcal{W}_{D4}$ in table~\ref{tab:CL_mms},  and the two neutrino mass terms correspond to $\mathcal{W}_{M1}$  and $\mathcal{W}_{M2}$ in table~\ref{tab:WO_mms}, respectively.  Notice that the second term contains two flavons and it is suppressed by $\langle\xi\rangle/\Lambda$ with respect to the first term. We assume that the traditional flavor symmetry $\Delta(27)$ is broken down to $Z_3^{A^2 BA^2}$ in the charged lepton sector by the VEV of $\phi$, and the VEVs of the flavons $\varphi$ and $\chi$ entering in neutrino sector are invariant under $Z_3^{ABA^2}$ and $Z_3^{B}$, respectively. As a consequence, the residual traditional flavor symmetry $Z_3^{A^2 BA^2}$ is preserved in the charged lepton sector after symmetry breaking, while the $\Delta(27)$ flavor symmetry is completely broken in the neutrino sector\footnote{The residual symmetry of the Majorana neutrino mass matrix can only be $Z_2$ or $K_4$ for  three nonvanishing light neutrino masses~\cite{Chen:2014wxa,Chen:2015nha}.  Since the $\Delta(27)$ flavor group doesn't have $Z_2$ and $K_{4}$ subgroups, no residual flavor symmetry should be preserved in the neutrino sector.}.  As shown in table~\ref{tab:VEVs_subs}, the assumed residual symmetry allows us to fix  VEVs of flavons $\xi$, $\phi$, $\varphi$ and $\chi$ as follows,
\begin{equation}\label{eq:M_WO_VEVs}
\langle\xi\rangle=v_{\xi}, \qquad \langle\phi\rangle=(\omega^2,1,1)^Tv_{\phi}, \qquad \langle\varphi\rangle=(0,0,1)^Tv_{\varphi}, \qquad \langle\chi\rangle=(1,0,0)^Tv_{\chi}\,,
\end{equation}
where $v_{\xi}$, $v_{\phi}$, $v_{\varphi}$ and $v_{\chi}$ are undetermined and generally complex. After the electroweak and flavor symmetry breaking in the way of Eq.~\eqref{eq:M_WO_VEVs}, we obtain the charged lepton mass matrix and the light neutrino mass matrix as follows
\begin{eqnarray}
	\nonumber &&\hskip-0.2in m_{l}=\frac{\alpha v_{\phi}v_d}{\Lambda} \left(
	\begin{array}{ccc}
		\sqrt{2}\omega^2  Y^{(5)}_{\bm{2^{\prime}},1} & \omega  Y^{(5)}_{\bm{2^{\prime}},2} &   \omega Y^{(5)}_{\bm{2^{\prime}},2} \\
		\omega   Y^{(5)}_{\bm{2^{\prime}},2} & \sqrt{2}  Y^{(5)}_{\bm{2^{\prime}},1} &   Y^{(5)}_{\bm{2^{\prime}},2} \\
		\omega Y^{(5)}_{\bm{2^{\prime}},2} &   Y^{(5)}_{\bm{2^{\prime}},2} & \sqrt{2}  Y^{(5)}_{\bm{2^{\prime}},1}\\
	\end{array}
	\right)+\frac{i \beta Y^{(4)}_{\bm{1}}v_{\xi} v_{\phi}v_d}{\Lambda^2}\left(
	\begin{array}{ccc}
		0 & \omega  & -\omega  \\
		-\omega   & 0 & 1  \\
		\omega   & -1   & 0 \\
	\end{array}
	\right) ,\\
	\label{eq:lep_mass_WO}&&\hskip-0.2in m_{\nu}=\frac{g_1v_{\varphi}v^2_u}{\Lambda^2}\left(
	\begin{array}{ccc}
		0 & \omega  Y^{(5)}_{\bm{2^{\prime\prime}},2} & 0 \\
		\omega  Y^{(5)}_{\bm{2^{\prime\prime}},2} & 0 & 0  \\
		0 & 0  &  \sqrt{2} Y^{(5)}_{\bm{2^{\prime\prime}},1} \\
	\end{array}
	\right)+\frac{g_2v_{\chi}v^2_u}{\Lambda^2}\left(
	\begin{array}{ccc}
		\sqrt{2} Y^{(7)}_{\bm{2^{\prime}},1}  & 0 & 0 \\
		0 & 0 & \omega  Y^{(7)}_{\bm{2^{\prime}},2}  \\
		0 & \omega  Y^{(7)}_{\bm{2^{\prime}},2}  & 0 \\
	\end{array}
	\right)\,,
\end{eqnarray}
where we use $Y^{(k_Y)}_{\bm{r},i}$ to represent the $i$th component of modular multiple $Y^{(k_Y)}_{\bm{r}}$.
Obviously we have $\rho^\dagger_{\bm{3}}(A^2 BA^2)m^\dagger_{l}m_{l}\rho_{\bm{3}}(A^2 BA^2)=m^\dagger_{l}m_{l}$ which implies that the charged lepton mass matrix is invariant under the residual flavor symmetry $Z_3^{A^2 BA^2}$.  Therefore the unitary transformation $U_{l}$ diagonalizing the  hermitian combination $m^\dagger_{l}m_{l}$ via $U_{l}^{\dagger} m^\dagger_{l}m_{l}U_{l}=\text{diag}(m^2_e,m^2_{\mu},m^2_{\tau})$ is of the form:
\begin{equation}\label{eq:Ue1}
U_{l}=\frac{1}{\sqrt{3}}\left(
\begin{array}{ccc}
 1 & \omega^2 & \omega \\
  \omega^2 & \omega^2 & \omega^2 \\
1 & \omega & \omega^2 \\
\end{array}
\right)\,.
\end{equation}
The charged lepton masses are given by
\begin{eqnarray}
\nonumber && m_{e}=\left|\sqrt{2} Y^{(5)}_{\bm{2^{\prime}},1}- Y^{(5)}_{\bm{2^{\prime}},2}- \frac{\sqrt{6}\beta v_{\xi} Y^{(4)}_{\bm{1}}  }{\alpha\Lambda} \right|\frac{\alpha v_{\phi}v_d}{\Lambda} , \\
\nonumber && m_{\mu}=\left|\sqrt{2} Y^{(5)}_{\bm{2^{\prime}},1}- Y^{(5)}_{\bm{2^{\prime}},2}+\frac{\sqrt{6}\beta v_{\xi} Y^{(4)}_{\bm{1}}  }{\alpha\Lambda} \right|\frac{ \alpha v_{\phi}v_d}{\Lambda}, \\
&& m_{\tau}=\left|\sqrt{2}Y^{(5)}_{\bm{2^{\prime}},1}+2 Y^{(5)}_{\bm{2^{\prime}},2}\right|\frac{\alpha v_{\phi}v_d}{\Lambda}\,.
\end{eqnarray}
Since $L$ and $E^{c}$ are assigned to be triplet of $\Delta(27)$, both couplings $\alpha$ and $\beta$ are relevant to the three charged lepton masses. Consequently some fine-tuning in the values of $\tau$ and $\beta v_{\xi}/(\alpha\Lambda)$ is necessary to accommodate charged lepton mass hierarchies. We transfer to the charged lepton diagonal basis by redefining the LH lepton fields $L\rightarrow U_{l}L$, then the light neutrino mass matrix is given by
\begin{eqnarray}
\nonumber\hskip-0.3in m^\prime_{\nu }&=&U^T_{l}m_{\nu }U_{l}\\
\nonumber\hskip-0.3in &=&\frac{g_1v_{\varphi}v^2_u}{\Lambda^2}\left(
	\begin{array}{ccc}
		\sqrt{2}  Y^{(5)}_{\bm{2^{\prime\prime}},1}+2  Y^{(5)}_{\bm{2^{\prime\prime}},2} &  \omega \left(\sqrt{2} Y^{(5)}_{\bm{2^{\prime\prime}},1}-Y^{(5)}_{\bm{2^{\prime\prime}},2}\right) & \omega^2  \left(\sqrt{2} Y^{(5)}_{\bm{2^{\prime\prime}},1}-Y^{(5)}_{\bm{2^{\prime\prime}},2}\right) \\
		\omega \left(\sqrt{2} Y^{(5)}_{\bm{2^{\prime\prime}},1}-Y^{(5)}_{\bm{2^{\prime\prime}},2}\right) & \omega^2  \left(\sqrt{2} Y^{(5)}_{\bm{2^{\prime\prime}},1}+2 Y^{(5)}_{\bm{2^{\prime\prime}},2}\right) & \sqrt{2} Y^{(5)}_{\bm{2^{\prime\prime}},1}-Y^{(5)}_{\bm{2^{\prime\prime}},2} \\
		\omega^2  \left(\sqrt{2} Y^{(5)}_{\bm{2^{\prime\prime}},1}-Y^{(5)}_{\bm{2^{\prime\prime}},2}\right) &\sqrt{2} Y^{(5)}_{\bm{2^{\prime\prime}},1} -Y^{(5)}_{\bm{2^{\prime\prime}},2} & \omega \left(\sqrt{2} Y^{(5)}_{\bm{2^{\prime\prime}},1}+2 Y^{(5)}_{\bm{2^{\prime\prime}},2}\right) \\
	\end{array}
	\right) \\
	\label{eq:nu_mass_WO2}\hskip-0.3in&&+\frac{g_2v_{\chi}v^2_u}{\Lambda^2}\left(
	\begin{array}{ccc}
		\sqrt{2} Y^{(7)}_{\bm{2^{\prime}},1}+2  Y^{(7)}_{\bm{2^{\prime}},2} &  \omega^2 \left(\sqrt{2} Y^{(7)}_{\bm{2^{\prime}},1}-Y^{(7)}_{\bm{2^{\prime}},2}\right) & \omega  \left(\sqrt{2} Y^{(7)}_{\bm{2^{\prime}},1}-Y^{(7)}_{\bm{2^{\prime}},2}\right) \\
		\omega^2 \left(\sqrt{2} Y^{(7)}_{\bm{2^{\prime}},1}-Y^{(7)}_{\bm{2^{\prime}},2}\right) & \omega \left(\sqrt{2} Y^{(7)}_{\bm{2^{\prime}},1}+2 Y^{(7)}_{\bm{2^{\prime}},2}\right) & \sqrt{2} Y^{(7)}_{\bm{2^{\prime}},1}-Y^{(7)}_{\bm{2^{\prime}},2} \\
		\omega  \left(\sqrt{2} Y^{(7)}_{\bm{2^{\prime}},1}-Y^{(7)}_{\bm{2^{\prime}},2}\right) & \sqrt{2} Y^{(7)}_{\bm{2^{\prime}},1}-Y^{(7)}_{\bm{2^{\prime}},2} & \omega^2 \left(\sqrt{2} Y^{(7)}_{\bm{2^{\prime}},1}+2 Y^{(7)}_{\bm{2^{\prime}},2}\right) \\
	\end{array}
	\right)\,.
\end{eqnarray}
Let us proceed to perform a numerical analysis for the model.  We first consider the scenario without gCP. This model involves 6 real free parameters besides the complex modulus $\tau$. All the six lepton masses, three lepton mixing angles and three CP violation phases rely on the following six dimensionless real parameters
\begin{equation}\label{eq:six_inputs_WO}
	\Re{\tau},  \quad \Im{\tau}, \quad |\beta v_{\xi}/(\alpha \Lambda)|,\quad \text{arg}{\left(\beta v_{\xi}/(\alpha \Lambda)\right)}\,, \quad
	  |g_2v_{\chi}/(g_1v_{\varphi})|,\quad \text{arg}{(g_2v_{\chi}/(g_1v_{\varphi}))}\,,
\end{equation}
and two overall scales $\alpha v_{d}v_{\phi}/\Lambda$ and $g_1v^2_uv_{\varphi}/\Lambda^2$ which can be taken to real. In order to quantitatively estimate how well the model can describe the experimental data on the following seven dimensionless observable quantities
\begin{equation}\label{eq:obs_qua}
	m_{e}/m_{\mu}, \quad m_{\mu}/m_{\tau},\quad	\sin^2\theta_{12},\quad \sin^2\theta_{13},\quad \sin^2\theta_{23}, \quad \delta_{CP},\quad \Delta m^2_{21}/\Delta m^2_{31}\,,
\end{equation}
which only depend on six dimensionless real parameters in Eq.~\eqref{eq:six_inputs_WO}, we perform a conventional $\chi^2$ analysis to determine the best fit values of the free parameters as well as the corresponding predictions for the flavor observables. We adopt the following the experimental data and errors to build the $\chi^2$ function:
\begin{eqnarray}
\nonumber &&	\frac{m_e}{m_{\mu}}\Big|_{\text{bf}+1\sigma}=0.0048^{+0.0002}_{-0.0002}\,, \quad \frac{m_{\mu}}{m_{\tau}}\Big|_{\text{bf}+1\sigma}0.059^{+0.002}_{-0.002}\,,\quad \sin^2\theta_{12}|_{\text{bf}+1\sigma}=0.303^{+0.012}_{-0.012}\,,\\
\nonumber &&\sin^2\theta_{13}|_{\text{bf}+1\sigma}=0.02225^{+0.00056}_{-0.00059}\,, \qquad  \sin^2\theta_{23}|_{\text{bf}+1\sigma}=0.451^{+0.019}_{-0.016}\,,\\
&& \frac{\delta_{CP}}{\pi}\Big|_{\text{bf}+1\sigma} =1.289^{+0.20}_{-0.14}\,,\qquad  \frac{\Delta m^2_{21}}{\Delta m^2_{31}}\Big|_{\text{bf}+1\sigma}=0.0294^{+0.00088}_{-0.00088}\,.
\end{eqnarray}
where the data of the lepton mixing angles and CP phase $\delta_{CP}$ is taken from NuFIT5.2~\cite{Esteban:2020cvm}, the charged lepton mass ratios are taken from~\cite{Antusch:2013jca} with $M_{\text{SUSY}}=500\,\text{GeV}$ and $\tan\beta=10$. We notice that this model yields a successful fit to current experimental data for certain values of the input parameters. We find the minimum value of the $\chi^2$ function is $\chi^2_{\text{min}}=10.68$, and the best fit values of the free parameters are
\begin{eqnarray}
	\nonumber &&	\Re\langle\tau\rangle=0.00177, ~~~~ \Im\langle\tau\rangle=1.120, ~~~~|\beta v_{\xi}/(\alpha \Lambda)|=0.0480,~~~~\text{arg}{(\beta v_{\xi}/(\alpha \Lambda))}=1.04\pi, \\
	\nonumber &&|g_2v_\chi/(g_1v_{\varphi})|=0.9787, \qquad \text{arg}{(g_2v_\chi/(g_1v_{\varphi}))}=1.005\pi,  \\
	&&	\alpha v_{\xi}v_{\phi}v_d/\Lambda^2=262.5\,\text{MeV}\,, \qquad
	g_1v_{\varphi}v^2_u/\Lambda^2=5.401\,\text{meV}\,.
\end{eqnarray}
The overall scales $\alpha v_{d}v_{\phi}/\Lambda$ and $g_1v^2_uv_{\varphi}/\Lambda^2$ are fixed by the measured electron mass and the solar neutrino mass square difference $\Delta m^2_{21}$. We see that the real part of modulus $\Re\langle\tau\rangle$ is quite small, and both phases $\text{arg}{(\beta v_{\xi}/(\alpha \Lambda))}$ and $\text{arg}{(g_2v_\chi/(g_1v_{\varphi}))}$ are very close to $\pi$. At the best fit point, the lepton masses and mixing parameters are determined to be
\begin{eqnarray}
	\nonumber &&\sin^2\theta_{13}=0.02251, \quad \sin^2\theta_{12}=0.3284, \quad \sin^2\theta_{23}=0.4954, \quad  \delta_{CP}=1.434\pi\,, \\
	\nonumber&&\alpha_{21}=0.961\pi,\quad \alpha_{31}=0.926\pi, \quad  m_1=15.13\,\text{meV} ,\quad m_2=17.40\,\text{meV}\,, \\
	\nonumber 	&&  m_3=52.31\,\text{meV}, \quad \sum^3_{i=1}m_{i}=84.84\,\text{meV}, \quad 	m_{\beta\beta}=5.619\,\text{meV}\,,\\
	&& m_{e}=0.511\,\text{MeV} , \quad  m_{\mu}=106.5\,\text{MeV} , \quad  m_{\tau}=1.803\,\text{GeV} \,.
\end{eqnarray}
where $m_{i}\,(i=1,2,3)$, $m_{j}\,(j=e,\mu,\tau)$ and $m_{\beta\beta}$ refers to the absolute masses of the three light neutrinos, the masses of the three charged leptons and the effective mass in neutrinoless double beta decay, respectively. The effective Majorana mass is determined to be $m_{\beta\beta}=5.611\,\text{meV}$ which is much below the current most stringent limit $m_{\beta\beta}<36-156$ meV from KamLAND-Zen~\cite{KamLAND-Zen:2022tow}. The next generation neutrinoless double decay experiments of ton-scale can improve the sensitivity to $m_{\beta\beta}$ considerably. For example, the SNO+ Phase II could probe the effective mass $m_{\beta\beta}$ in the range of $19-46$ meV~\cite{SNO:2015wyx}. The LEGEND experiment is expected to reach the sensitivity of (15-50)~meV by operating 1000 kg of detectors for 10 years~\cite{LEGEND:2017cdu}. The experiment nEXO will be able to probe $m_{\beta\beta}$ down to $5.7-17.7$~meV after 10 years of data taking~\cite{nEXO:2017nam}. Thus our prediction for $m_{\beta\beta}$
is a bit below but close to the future sensitivity. Moreover, the three neutrino mass sum is predicted to be $84.84\,\text{meV}$ which is compatible with the upper bound  $\sum^3_{i=1}m_{i}<120\,\text{meV}$ on neutrino mass sum from Planck ~\cite{Planck:2018vyg}.

If the gCP symmetry is imposed on the model, all the couplings would be enforced to be real, as discussed in section~\ref{sec:mod_general}. Moreover, certain residual CP symmetry require the flavons vacuum parameters $v_{\xi}$, $v_{\phi}$, $v_{\varphi}$ and $v_{\chi}$ in Eq.~\eqref{eq:M_WO_VEVs} be real, as shown in  table~\ref{tab:VEVs_subs}. Thus the two phase parameters $\text{arg}{\left(\beta v_{\xi}/(\alpha \Lambda)\right)}$ and $\text{arg}{(g_2v_{\chi}/(g_1v_{\varphi}))}$ in Eq.~\eqref{eq:six_inputs_WO} are $0$ or $\pi$ because of the gCP symmetry. Furthermore, if the modulus $\tau$ lies on the imaginary axis with $\Re\tau=0$, the VEV of $\tau$ preserves the gCP symmetry $K_{*}$, and $q=e^{2 \pi i\tau}$ as well as the modular forms $Y_{1}(\tau)$ and $Y_{2}(\tau)$ in Eq.~\eqref{eq:q_expressions} would be real. Hence all the integer weight modular forms of level 3 are real for pure imaginary $\tau$, as can be seen from Appendix~\ref{sec:Tp_modu_forms}. As a consequence, the neutrino mass matrix $m^\prime_{\nu}$ in the charged lepton diagonal basis in Eq.~\eqref{eq:nu_mass_WO2} satisfies
\begin{equation}
	(m^\prime_{\nu})_{ee}=(m^\prime_{\nu})^*_{ee}, \quad (m^\prime_{\nu})_{\mu\tau}=(m^\prime_{\nu})^*_{\mu\tau}, \quad (m^\prime_{\nu})_{e\tau}=(m^\prime_{\nu})^*_{e\mu}, \quad (m^\prime_{\nu})_{\mu\mu}=(m^\prime_{\nu})^*_{\tau\tau}\,,
\end{equation}
which is exactly the requirement of a theory invariant under $\mu-\tau$ reflection symmetry~\cite{Harrison:2002kp,Grimus:2003yn,Farzan:2006vj}. In other words, the neutrino mass matrix $m^\prime_{\nu}$ in Eq.~\eqref{eq:nu_mass_WO2} really fulfills the $\mu-\tau$ reflection symmetry
\begin{equation}
P^{T}_{\nu\tau}\,m^\prime_{\nu }\,P_{\nu\tau}=(m^\prime_{\nu})^{*}\,,
\end{equation}
with
\begin{equation}
	P_{\nu\tau}=\begin{pmatrix}
		1 & 0 & 0 \\
		0& 0 &1 \\
		0& 1 & 0\end{pmatrix}\,.
\end{equation}
It is known that the $\mu-\tau$ reflection symmetry leads to maximal atmospheric angle, maximal Dirac CP phase and trivial Majorana CP phases: $\theta_{23}=\pi/4, \; \delta_{CP}=\pm \pi/2,\; \alpha_{21},\alpha_{31}=0, \pi$~\cite{Harrison:2002kp,Grimus:2003yn,Farzan:2006vj}
After imposing the gCP symmetry and the assumption of $\Re \tau=0$, we find that all the seven dimensionless flavor observables in Eq.~\eqref{eq:obs_qua} would only depend on three dimensionless real parameters in the model,
\begin{equation}\label{eq:three_inputs_WO}
\Im{\tau}, \qquad \beta v_{\xi}/(\alpha \Lambda),\qquad
	  g_2v_{\chi}/(g_1v_{\varphi})\,.
\end{equation}
Analogously performing a numerical analysis, we find that the minimum value of $\chi^2$ function is $\chi^2_{\text{min}}=11.76$  when the free parameters take the following values
\begin{eqnarray}
	\nonumber &&  \langle\tau\rangle=1.120i, \qquad \beta v_{\xi}/(\alpha \Lambda)= -0.0484, \qquad g_2v_\chi/(g_1v_{\varphi})=-0.981, \\
	&& \qquad
	\alpha v_dv_{\xi}v_{\phi}/\Lambda^2=263.0\,\text{MeV}\,, \qquad
	g_1v^2_uv_{\varphi}/\Lambda^2=5.409\,\text{meV}\,.
\end{eqnarray}
Accordingly the lepton masses and mixing parameters are determined to be
\begin{eqnarray}
	\nonumber &&\sin^2\theta_{13}=0.02238, \quad \sin^2\theta_{12}=0.3266, \quad \sin^2\theta_{23}=0.5, \quad  \delta_{CP}=1.5\pi\,, \\
	\nonumber&&\alpha_{21}=\pi,\quad \alpha_{31}=\pi, \quad  m_1=15.18\,\text{meV} ,\quad m_2=17.44\,\text{meV}\,, \\
	\nonumber 	&&  m_3=52.43\,\text{meV}, \quad \sum_{i}m_{i}=85.05\,\text{meV}, \quad 	m_{\beta\beta}=5.595\,\text{meV}\,,\\
	&& m_{e}=0.511\text{MeV} , \quad  m_{\mu}=106.5\text{MeV} , \quad  m_{\tau}=1.807\text{GeV} \,.
\end{eqnarray}
\begin{figure}[t!]
\centering
\begin{tabular}{c}
\includegraphics[width=0.95\linewidth]{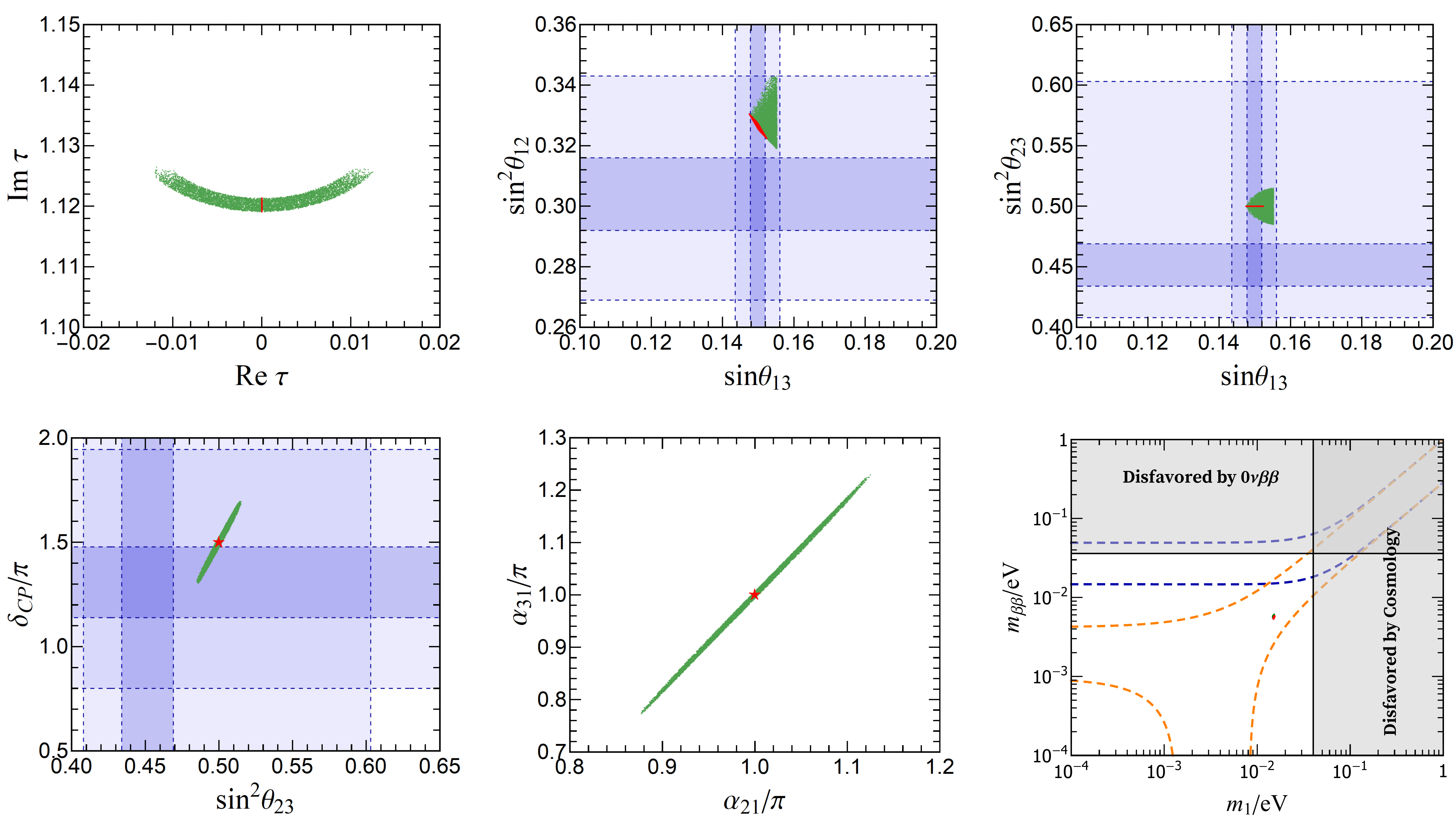}
\end{tabular}
\caption{\label{fig:M2} The predicted correlations among the input free parameters, neutrino mixing angles and CP violation phases in the example model 1. The green and red  regions denote points compatible with experimental data for models without gCP and with gCP plus $\Re \tau=0$, respectively. The light blue bounds represent the $1\sigma$ and $3\sigma$ ranges respectively with SK atmospheric data~\cite{Esteban:2020cvm}. The  orange (blue) dashed lines in the last panel represent the most general  allowed regions for NO (IO) neutrino masses respectively, where the neutrino oscillation parameter are varied within their $3\sigma$ ranges. The present upper limit $m_{\beta\beta}<(36-156)$ meV from KamLAND-Zen~\cite{KamLAND-Zen:2022tow} is shown by horizontal grey band. The vertical grey exclusion band denotes the current bound from the cosmological data of $\sum_im_i<0.120$eV at 95\% confidence level given by the Planck collaboration~\cite{Planck:2018vyg}. Notice that we have sharp prediction for the effective mass $m_{\beta\beta}$, the green region (without gCP) almost coincides with the red region (with gCP) in the last panel. }
\end{figure}

In order to show the viability and predictions of this model with/without gCP, we scan over the parameter space of the model, the six (three) dimensionless real parameters in Eq.~\eqref{eq:six_inputs_WO} (in Eq.~\eqref{eq:three_inputs_WO}) are taken as random numbers. Then we can evaluate the predictions for lepton masses and mixing parameters and only keep the points for which the flavor observables are within their experimentally allowed $3\sigma$ regions. Some interesting correlations among the observables and free parameters are obtained. The results are shown in figure~\ref{fig:M2}. Here the results without gCP and with gCP plus $\Re \tau=0$ are displayed in green and red, respectively. In the case that gCP is considered with $\Re \tau=0$, the $\mu-\tau$ reflection symmetry is reproduced, the predictions for $\theta_{23}$, $\delta_{CP}$, $\alpha_{21}$ and $\alpha_{31}$ are labelled with a red pentagram. From figure~\ref{fig:M2}, we see that all the mixing parameters and $m_{\beta\beta}$ are predicted to lie in narrow ranges:
\begin{eqnarray}
\nonumber && \sin^2\theta_{13}\in[0.02173,0.02398], \qquad   \sin^2\theta_{12}\in[0.319,0.341], \qquad  \sin^2\theta_{23}\in[0.485,0.515]\,, \\
\nonumber &&\delta_{CP}\in[1.3\pi,1.7\pi], \qquad  \alpha_{21}\in[0.877\pi,1.125\pi], \qquad  \alpha_{31}\in[0.773,1.232]\,, \\
&&  m_{\beta\beta}\in[5.347\,\text{meV},6.168\,\text{meV}]\,.
\end{eqnarray}
for the case without gCP, and
\begin{eqnarray}
\nonumber &\sin^2\theta_{23}=0.5,\quad \delta_{CP}=1.5\pi,\quad \alpha_{21}=\alpha_{31}=\pi\,, \\
& \sin^2\theta_{13}\in[0.02173,0.02322],\quad   \sin^2\theta_{12}\in[0.322,0.330], \quad  m_{\beta\beta}\in[5.347\,\text{meV},5.834\,\text{meV}]\,.
\end{eqnarray}
for the case with gCP plus $\Re \tau=0$. Furthermore, this model predicts a nearly maximal atmospheric mixing angle $\theta_{23}$ and a nearly maximal Dirac CP phase $\delta_{CP}$ when gCP is not imposed.
Notice that the models predicts a nearly maximal atmospheric mixing angle $\theta_{23}$ even if gCP isn't imposed. It is remarkable that the variation region of the effective Majorana neutrino mass $m_{\beta\beta}$ is very small because of the linear correlation between $\alpha_{21}$ and $\alpha_{31}$ shown in figure~\ref{fig:M2}.

The NLO correction to the minimal K\"ahler potential is of the form
\begin{eqnarray}
\nonumber\mathcal{K}_{\text{NLO}}&=&\sum_{n, \bm{r}}\frac{1}{\Lambda}(-i \tau+ i \bar \tau)^{n}\,\left[Y^{(n)\dagger}_{\bm{r}}(\tau) Y^{(n-1)}_{\bm{r}}(\tau) L^{\dagger} L\xi\right]_{(\bm{1_{0,0}},\bm{1})}\\
&&\quad +\frac{1}{\Lambda}(-i \tau+ i \bar \tau)^{n}\,\left[Y^{(n)\dagger}_{\bm{r}}(\tau)Y^{(n-1)}_{\bm{r}}(\tau)E^{c\dagger}E^c\xi\right]_{(\bm{1_{0,0}},\bm{1})}\,.
\end{eqnarray}
The corresponding K\"ahler metric for both LH lepton doublet $L$ and RH charged leptons $E^c$ is a unit matrix so that its contribution can be absorbed into the minimal K\"ahler potential. Hence the non-diagonal contributions of K\"ahler metric come from NNLO corrections and they are suppressed by $\langle\Phi\rangle^2/\Lambda^2$. Therefore the contributions of K\"ahler potential to the lepton mixing parameters and masses are suppressed by $\langle\Phi\rangle^2/\Lambda^2$ and  can be neglected.

\subsection{\label{sec:_seesaw_model}Example model 2 }

\begin{table}[t!]
	\renewcommand{\tabcolsep}{1.05mm}
	\centering
	\begin{tabular}{|c|c|c|c|c|c||c|c|c|c|}
		\hline \hline
		Fields &   $L$    &  $E^{c}$  & $N^c$   &  $H_{u}$ & $H_{d}$ &  $\phi$  &  $\varphi$  &  $\xi$   & $Y^{(k_Y)}_{\bm{r}}$\\ \hline
		
		$\text{SU}(2)_L\times \text{U}(1)_{Y}$ & $(\bm{2},-\frac{1}{2})$ & $(\bm{1},1)$ & $(\bm{1},0)$ & $(\bm{2},\frac{1}{2})$ & $(\bm{2},-\frac{1}{2})$ &  $(\bm{1},0)$ &  $(\bm{1},0)$ &  $(\bm{1},0)$ &    $(\bm{1},0)$ \\ \hline
		
		$\Delta(27)$  &   $\bm{\bar{3}}$  &  $\bm{\bar{3}}$  &  $\bm{\bar{3}}$  &  $\bm{1_{0,0}}$   & $\bm{1_{0,0}}$   &  $\bm{\bar{3}}$   & $\bm{\bar{3}}$  & $\bm{1_{0,0}}$  &   $\bm{1_{0,0}}$   \\ \hline

		$\Gamma^\prime_3\cong T^\prime$ & $\bm{\bar{3}_0}$ & $\bm{\bar{3}_0}$ & $\bm{\bar{3}_0}$ & $\bm{1}$  & $\bm{1}$ & $\bm{\bar{3}_0}$ & $\bm{\bar{3}_1}$ & $\bm{1^{\prime\prime}}$   & $\bm{r}$ \\ \hline
		
		modular weight & $0$ & $0$ &$1$ &  $0$ & $0$ & $3$ & $-1$ & $7$  & $k_Y$ \\  \hline
		$Z_2$ & $1$ & $1$ & $-1$ & $1$ & $1$ & $1$ & $-1$    & $1$ & $1$  \\ \hline
		$Z_3$ & $\omega$ & $\omega$ & $\omega$ & $1$ & $1$ & $\omega$ &  $\omega$  & $1$  & $1$  \\ \hline \hline
		
	\end{tabular}
	\caption{\label{tab:seesaw_model_fields}The matter fields, flavon fields of example model 2 and their transformation properties under the eclectic flavor symmetry $\Delta(27)\rtimes T^\prime$ and the auxiliary symmetry $Z_{2}\times Z_{3}$.	}
\end{table}

\begin{figure}[t!]
\centering
\begin{tabular}{c}
\includegraphics[width=0.95\linewidth]{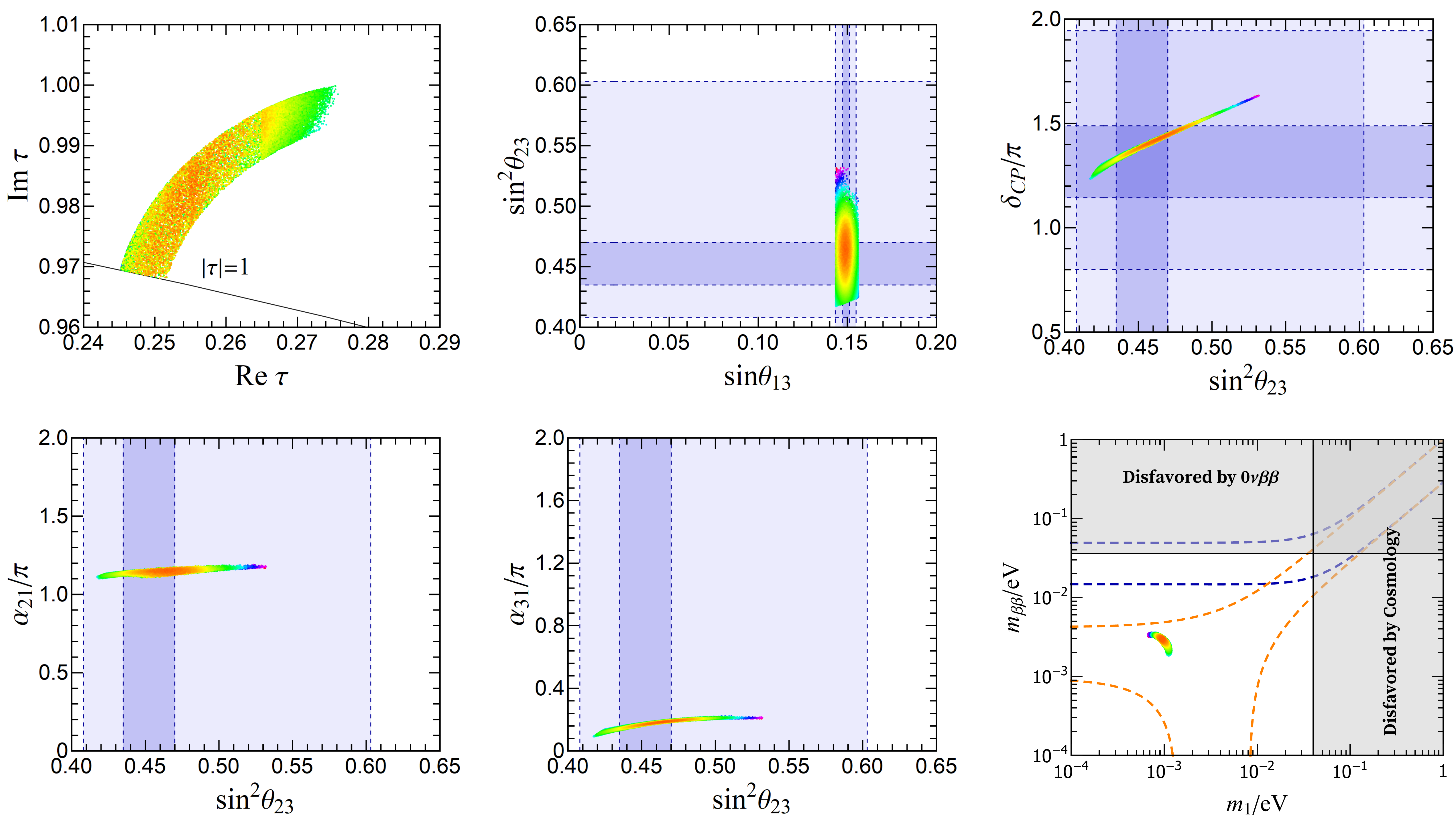}\\
\includegraphics[width=0.38\linewidth]{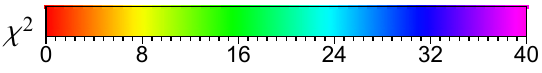}\\
\end{tabular}
\caption{\label{fig:M_seesaw} The predicted correlations among the input free parameters, neutrino mixing angles, CP violation phases and the effective mass in neutrinoless double beta decay  in the seesaw model.}
\end{figure}

In the second model, the neutrino masses are described by the type I seesaw mechanism with three RH neutrinos.  This model is also based on the eclectic flavor group $\Omega(1)$ with the extra symmetry $Z_{2}\times Z_{3}$ to eliminate undesired operators. The three generations of LH doublet leptons $L$, the RH charged leptons $E^{c}$ and the RH neutrinos $N^{c}$ are all embedded into the triplet $\bm{\bar{3}}$ of $\Delta(27)$. We introduce three flavons $\phi$, $\varphi$ and $\xi$ which transform as $\bm{\bar{3}}$, $\bm{\bar{3}}$ and $\bm{1_{0,0}}$ under $\Delta(27)$ respectively.The fields content and their classification under the eclectic flavor group $\Omega(1)$ and the auxiliary symmetry $Z_{2}\times Z_{3}$ are listed in table~\ref{tab:seesaw_model_fields}.
The superpotential for the leptons can be written as
\begin{eqnarray}
	\nonumber \mathcal{W}&=&\frac{\alpha}{\Lambda}  \left(E^c L\phi Y^{(3)}_{\bm{2^{\prime}}}\right)_{(\bm{1_{0,0}},\bm{1})} H_d+\frac{\beta}{\Lambda^2}  \left(E^c L\phi\xi Y^{(10)}_{\bm{1}}\right)_{(\bm{1_{0,0}},\bm{1})} H_d+ \frac{g_1}{\Lambda}  \left(N^c L\varphi \right)_{(\bm{1_{0,0}},\bm{1})} H_u \\
	\label{eq:seesaw_SP}	&&+ \frac{g_2}{\Lambda^2}  \left(N^c L\varphi \xi Y^{(7)}_{\bm{2}}\right)_{(\bm{1_{0,0}},\bm{1})} H_u+\frac{h}{2}\left(N^cN^c\phi Y^{(5)}_{\bm{2^{\prime}}}\right)_{(\bm{1_{0,0}},\bm{1})}\,.
\end{eqnarray}
In this model we assume that the VEV of flavon $\phi$ breaks the traditional flavor symmetry $\Delta(27)$ down to $Z^{A^2B}_{3}$ while the subgroup $Z^{A}_{3}$ is preserved  by vacuum of flavon $\varphi$. Hence the VEVs of flavons $\xi$, $\phi$ and $\varphi$ take the following form:
\begin{equation}\label{eq:seesaw_VEVs}
\langle\xi\rangle=v_{\xi}, \qquad	\langle\phi\rangle=(1,1,\omega^2)^Tv_{\phi}, \qquad \langle\varphi\rangle=(1,1,1)^Tv_{\varphi}\,.
\end{equation}
 After symmetry breaking, we can straightforwardly read out the charged lepton mass matrix, the Dirac neutrino mass matrix and the Majorana neutrino  mass matrix,
\begin{eqnarray}
\nonumber &&m_{l}=\frac{ \alpha v_dv_{\phi}}{\Lambda}\left(
\begin{array}{ccc}
\sqrt{2} Y^{(3)}_{\bm{2^{\prime}},2} & -\omega Y^{(3)}_{\bm{2^{\prime}},1} & - \omega ^2  Y^{(3)}_{\bm{2^{\prime}},1} \\
-\omega Y^{(3)}_{\bm{2^{\prime}},1} & \sqrt{2}  Y^{(3)}_{\bm{2^{\prime}},2} & - \omega ^2  Y^{(3)}_{\bm{2^{\prime}},1}  \\
- \omega ^2  Y^{(3)}_{\bm{2^{\prime}},1}  & - \omega ^2  Y^{(3)}_{\bm{2^{\prime}},1}  & \sqrt{2} \omega ^2  Y^{(3)}_{\bm{2^{\prime}},2} \\
\end{array}
\right)+\frac{i \beta Y^{(10)}_{\bm{1}}v_{\xi} v_dv_{\phi}}{\Lambda^2}\left(
\begin{array}{ccc}
0 & \omega  & -\omega ^2 \\
-\omega  & 0 & \omega ^2 \\
\omega ^2 & -\omega ^2 & 0 \\
\end{array}
\right)\,,\\
\nonumber &&m_{D}=\frac{i \omega^2 g_{1}  v_uv_{\varphi}}{\Lambda^2}\left(
\begin{array}{ccc}
0 & 1 & -1 \\
-1 & 0 & 1 \\
1 & -1 & 0 \\
\end{array}
\right)+\frac{g_2v_uv_{\xi} v_{\varphi}}{\Lambda}\left(
\begin{array}{ccc}
\sqrt{2}  Y^{(7)}_{\bm{2},2} &- \omega ^2  Y^{(7)}_{\bm{2},1} & - \omega ^2 Y^{(7)}_{\bm{2},1} \\
-\omega ^2  Y^{(7)}_{\bm{2},1}& \sqrt{2}  Y^{(7)}_{\bm{2},2} & - \omega ^2 Y^{(7)}_{\bm{2},1}\\
			-\omega ^2  Y^{(7)}_{\bm{2},1}& -\omega ^2  Y^{(7)}_{\bm{2},1}& \sqrt{2}  Y^{(7)}_{\bm{2},2} \\
		\end{array}
		\right)\,, \\
		\label{eq:seesaw_mass} && m_{N}=hv_{\phi}\left(
		\begin{array}{ccc}
			\sqrt{2} Y^{(5)}_{\bm{2^{\prime}},2} & -\omega  Y^{(5)}_{\bm{2^{\prime}},1} &- \omega ^2 Y^{(5)}_{\bm{2^{\prime}},1} \\
			-\omega  Y^{(5)}_{\bm{2^{\prime}},1}   & \sqrt{2} Y^{(5)}_{\bm{2^{\prime}},2} & - \omega ^2 Y^{(5)}_{\bm{2^{\prime}},1}  \\
			- \omega ^2 Y^{(5)}_{\bm{2^{\prime}},1} &- \omega ^2 Y^{(5)}_{\bm{2^{\prime}},1}  & \sqrt{2} \omega ^2 Y^{(5)}_{\bm{2^{\prime}},2} \\
\end{array}
\right)\,.
\end{eqnarray}
The light neutrino mass matrix is then given by the seesaw relation:
\begin{equation}
	m_{\nu}=-m^T_{D}m^{-1}_{N}m_{D}\,.
\end{equation}
Since both flavons $\phi$ and $\varphi$ enter into the neutrino mass terms and their VEVs preserve different $Z_{3}$ subgroups of $\Delta(27)$, the  traditional flavor group $\Delta(27)$ is completely broken in the neutrino sector. One can easily check that the hermitian combination $m^\dagger_{l}m_{l}$ is invariant under the action of $\rho_{\bm{\bar{3}}}(A^2 B)$. It implies that the residual flavor symmetry is $Z^{A^2B}_{3}$ in the charged lepton sector.  We find that the experimental data can be accommodated rather well for certain values of the input parameters. The best fit point is found to be given by
\begin{eqnarray}
	\nonumber &&	\Re\langle\tau\rangle=0.257, ~~~~ \Im\langle\tau\rangle=0.988, ~~~~|\beta  v_{\xi}/(\alpha\Lambda)|=0.127,~~~~\text{arg}{(\beta  v_{\xi}/(\alpha\Lambda))}=0.458\pi, \\
	\nonumber &&|g_2v_\xi/(g_1\Lambda)|=0.0418, \qquad \text{arg}{(g_2v_\xi/(g_1\Lambda))}=0.551\pi,  \\
	&&	\alpha v_dv_{\xi}v_{\phi}/\Lambda^2=222.1\,\text{MeV}\,, \qquad
	g_1v^2_uv_{\varphi}/\Lambda^2=29.26\,\text{meV} \,,
\end{eqnarray}
with $\chi^2_{\text{min}}=2.027$. Accordingly the lepton masses and mixing parameters are
\begin{eqnarray}
	\nonumber &&\sin^2\theta_{13}=0.02221, \qquad \sin^2\theta_{12}=0.291, \qquad \sin^2\theta_{23}=0.464, \qquad  \delta_{CP}=1.429\pi\,, \\
	\nonumber&&\alpha_{21}=1.160\pi,\qquad \alpha_{31}=0.185\pi, \qquad  m_1=0.954\,\text{meV} ,\qquad m_2=8.649\,\text{meV}\,, \\
	\nonumber 	&&  m_3=50.05\,\text{meV}, \qquad \sum_{i}m_{i}=59.65\,\text{meV}, \qquad 	m_{\beta\beta}=2.977\,\text{meV}\,,\\
	&& m_{e}=0.511\,\text{MeV} , \qquad  m_{\mu}=106.5\,\text{MeV} , \qquad  m_{\tau}=1.801\,\text{GeV} \,.
\end{eqnarray}
We perform a numerical analysis for this model, and the results are plotted in figure~\ref{fig:M_seesaw}. It is remarkable that the phenomenologically viable parameter space is actually small. We find that the mixing angles $\theta_{13}$ and $\theta_{12}$ can take any value in their $3\sigma$ ranges, the atmospheric mixing angle $\sin^2\theta_{23}$ is predicted to be in the range $[0.418,0.532]$, the Dirac CP violation phase $\delta_{CP}$ is limited in the range $[1.234\pi,1.634\pi]$, and the two Majorana CP violation phases lie in narrow regions $\alpha_{21}\in[1.102\pi,1.191\pi]$ and $\alpha_{31}\in[0.092\pi,0.223\pi]$. The forthcoming long-baseline
neutrino oscillation experiments DUNE~\cite{DUNE:2015lol} and T2HK~\cite{Hyper-KamiokandeProto-:2015xww} can significantly improve the sensitivity to $\theta_{23}$ and $\delta_{CP}$. It is expected that a $5\sigma$ discovery of CP violation can be reached after ten years of data taking. We expect that the predictions for $\theta_{23}$ and $\delta_{CP}$ can be tested at DUNE and T2HK.

\section{\label{sec:conclusion} Conclusion and outlook}

In the present work, we elaborate how to construct models based on the so-called eclectic flavor group~\cite{Nilles:2020nnc} which is in general a semi-direct product of the traditional flavor group with the finite modular group in the bottom-up approach. The concept of eclectic flavor group originated in top-down model building motivated by string theory. In the scheme of eclectic flavor groups, which allows us to combine the advantages of modular symmetries and traditional flavor symmetries and to avoid the limitations of them, the superpotential and the  K\"ahler potential are both severely restricted at the same time. However, there are no remaining exact flavor symmetries in nature, flavon fields should be introduced in this kind of models, and their VEVs breaks the flavor symmetries. It turns out that the choices of such eclectic symmetries are quite restrictive, with only three choices of
flavor group $Z_3\times Z_3$, $\Delta(27)$ and $\Delta(54)$ catalogued so far allowing for a non-trivial eclectic extension with a CP-like
transformation~\cite{Nilles:2020nnc,Nilles:2020kgo}.

We perform a comprehensive analysis of the superpotential and K\"ahler potential of models based on the eclectic flavor group $\Omega(1)$ which is semi-direct product of the traditional flavor group $\Delta(27)$ with the finite modular group $T^\prime$, and find three generations of quarks and leptons can only be assigned to triplet, antitriplet or trivial singlet of $\Delta(27)$.  From the discussion in section~\ref{sec:mod_general}, the general form of fermion mass matrices can be obtained for any modular weight and any number flavons. If a $\Delta(27)$ octet flavon which is invariant under auxiliary cyclic symmetries is imposed in a model, we find that the contributions to the fermion masses and fermion mixings from higher order corrections of K\"ahler potential are suppressed by powers of $\langle\Phi\rangle/\Lambda$, or they will be suppressed by $\langle\Phi\rangle^2/\Lambda^2$.

In order to extract general lessons from the eclectic flavor symmetry scheme, we construct two concrete lepton models invariant under the eclectic flavor group $\Omega(1)\cong \Delta(27)\rtimes T^\prime$ in the framework of $\mathcal{N}=1$ global supersymmetry. In the two models, all lepton fields are assigned to $\Delta(27)$ triplet and  the three light neutrino masses are generated by the Weinberg operator and the type I seesaw mechanism, respectively. To break the traditional flavor group $\Delta(27)$, three (two) triplet flavons and a trivial singlet flavon are introduced  in the  first (second) model.  Then the higher order corrections from K\"ahler potential to the lepton mixing parameters and masses are suppressed by $\langle\Phi\rangle^2/\Lambda^2$ and  can be neglected. For the first model, we assume that the traditional flavor group $\Delta(27)$ is broken down to the subgroup $Z_3^{A^2 BA^2}$ in the charged lepton sector and is broken completely in the neutrino sector. For the second model, the subgroup $Z^{A^2B}_{3}$ is preserved by the VEVs of the flavon in the charged lepton mass terms  and neutrino Majorana mass term, while  the subgroup $Z^{A}_{3}$ is preserved  by vacuum of flavon in the neutrino Dirac mass terms. Furthermore, we perform a comprehensive analysis of the two models for lepton masses and mixing parameters. In both of the two models, the six lepton masses and six lepton mixing parameters depend on two complex coupling constants, two real coupling constants and the complex modulus $\tau$ in the case of without gCP. From the numerical scan of the parameter space, we find that all six mixing parameters vary in very small regions, and a nearly maximal atmospheric mixing angle $\theta_{23}$ and  Dirac CP phase $\delta_{CP}$ are obtained in the first model. When the conditions gCP and $\Re \tau=0$ are imposed in the first model,  the experimental data of lepton masses and mixing parameters can be successfully described in terms of five real parameters, and the $\mu-\tau$ reflection symmetry is preserved in the charged lepton diagonal basis. Then atmospheric mixing angle and  Dirac CP phase are predicted to be maximal, and two Majorana CP phases are predicted to be $\pi$. Furthermore, the predictions for the effective mass in neutrinoless double beta decay are obtained in both of the two models. All the predictions of the two models are compatible with the experimental data and could be tested at forthcoming long baseline neutrino experiments and neutrinoless double beta decay experiments.

In general, the main advantage of eclectic flavor symmetry is that is leads to predictive lepton mass and mixing models with the K\"ahler potential under control. Also, the representation of the matter fields under the flavor symmetry is constrained by the eclectic flavor symmetry. For example, in the considered models, this means that non-trivial singlet representations are not permitted since they would have to form a reducible octet, which is not phenomenologically possible. Also, we have seen that the modular symmetry transformation is determined by the flavor symmetry transformation through the consistency condition.

Looking ahead, it would be interesting to explore how to implement top-down ingredients, such as the constrained choice of modular weights, in the future bottom-up eclectic models. Also, the problem of modulus stabilisation in theory based on modular symmetry still persists, and the presence of flavons only exacerbates the problem. Such problems are left for future work. In the present paper, we have developed the formalism of eclectic flavor symmetry from the bottom-up point of view, focusing on an attractive class of models based on $\Omega(1)$, where the imposition of gCP symmetry leads to a predictive model with $\mu-\tau$ reflection symmetry.

\section*{Acknowledgements}

GJD and XGL are grateful to professor Ferruccio Feruglio for sharing the his insightful notes on eclectic flavor group. CCL is supported by the National Natural Science Foundation of China under Grant Nos. 12005167, 12247103. XGL is supported in part by U.S. National Science Foundation under Grant No.~PHY-2210283. JNL is supported by the Grants No. NSFC-12147110 and the China Post-doctoral Science Foundation under Grant No. 2021M70. GJD is supported by the National Natural Science Foundation of China under Grant Nos.~11975224, 11835013. SFK acknowledges the STFC Consolidated Grant ST/L000296/1 and the European Union's Horizon 2020 Research and Innovation programme under Marie Sk\l{}odowska-Curie grant agreement HIDDeN European ITN project (H2020-MSCA-ITN-2019//860881-HIDDeN).


\section*{Appendix}

\setcounter{equation}{0}
\renewcommand{\theequation}{\thesection.\arabic{equation}}

\begin{appendix}

\section{\label{sec:Delta27_group}Traditional flavor group $\Delta(27)$}

The $\Delta(27)$ group as flavor symmetry has been studied in the literature, see~\cite{deMedeirosVarzielas:2006fc,Branco:2015gna,deMedeirosVarzielas:2015amz,Chen:2015jta} and references therein. It is a non-Abelian group of order 27 with \texttt{GAP} ID $[27,3]$~\cite{Ishimori:2010au}, and the group structure of $\Delta(27)$ is $\Delta(27)\cong(Z_3\times Z_3)\rtimes Z_3$ . In detail, $\Delta(27)$ can be generated by two generators
$A$ and $B$ which satisfy the multiplication rules
\begin{equation}
A^3=B^3=(AB)^3=(AB^2)^3=1\,.
\end{equation}
The multiplication rules of $\Delta(27)$ in Refs.~\cite{Ishimori:2010au,Chen:2015jta} are given by:
\begin{equation}
a^3=a'^3=b^3=1, \qquad aa'=a'a\,, \qquad
bab^{-1}=a^{-1}a'^{-1}, \qquad ba'b^{-1}=a\,.
\end{equation}
Note that the chosen generators $A$ and $B$ in the present work are related to the generators $a$, $a^\prime$ and $b$ of Refs.~\cite{Ishimori:2010au,Chen:2015jta} via
\begin{equation}
\label{eq:gens_relation}
a=BAB^2, \qquad  a^\prime=A, \qquad b=B\,.
\end{equation}
In the same fashion, the following relations can also be obtained
\begin{equation}
A=a^\prime, \qquad  B=b\,.
\end{equation}
The traditional flavor group $\Delta(27)$ has eleven conjugacy classes and they are
\begin{eqnarray}
	\nonumber &1C_1=\{1\}\,, \qquad &3C^{(1)}_3=\left\{A^2B^2,B^2A^2,AB^2A\right\}\,,\\
	\nonumber 	&3C^{(2)}_3=\left\{A^2B,ABA,BA^2\right\}\,, \qquad & 3C^{(3)}_3=\left\{A,BAB^2,B^2AB\right\}\,, \\
	\nonumber & 3C^{(4)}_3=\left\{AB^2,BAB,B^2A\right\}\,, \qquad & 3C^{(5)}_3=\left\{AB,BA,A^2BA^2\right\}\,, \\
	\nonumber & 3C^{(6)}_3=\left\{A^2,B^2A^2B,BA^2B^2\right\}\,, \qquad &3C^{(7)}_3=\left\{B^2,AB^2A^2,A^2B^2A\right\}\,, \\
	\label{eq:D27CC}&3C^{(8)}_3=\left\{B,A^2BA,ABA^2\right\}\,,\qquad & 1C^{(1)}_3=\left\{BAB^2A^2\right\}\,, \quad 1C^{(2)}_3=\left\{ABA^2B^2\right\}\,,
\end{eqnarray}
where $kC_{n}$ denotes a conjugacy class which contains $k$ elements with order $n$.   Group $\Delta(27)$ contains seventeen nontrivial abelian subgroups in total: thirteen $Z_3$ subgroups and four $Z_3\times Z_3$ subgroups.  In terms of the generators $A$ and $B$, the concrete forms of these abelian subgroups are as follows
\begin{itemize}[leftmargin=1.5em]
\item{$Z_{3}$ subgroups}
\begin{eqnarray}
\nonumber && Z^{A}_3=\{1, A, A^2\}, \qquad  Z^{BAB^2}_{3}=\{1, BAB^2,BA^2B^2\}, \qquad  Z^{B^2AB}_3=\{1,B^2AB, B^2A^2B\}, \\
\nonumber &&  Z^{B}_{3}=\{1, B,B^2\}, \qquad Z^{A^2BA}_3=\{1, A^2BA, A^2B^2A\},\qquad Z^{ABA^2}_{3}=\{1,ABA^2, AB^2A^2\}, \\
\nonumber &&  Z^{AB}_3=\{1, AB, B^2A^2\},\qquad  Z^{BA}_{3}=\{1, BA, A^2B^2\}, \qquad  Z^{AB^2A}_3=\{1,AB^2A, A^2BA^2\}\, \\
\nonumber&& Z^{A^2B}_{3}=\{1, A^2B,B^2A\}, \qquad  Z^{AB^2}_3=\{1,AB^2,BA^2\}, \qquad  Z^{ABA}_3=\{1,ABA,BAB\}\, \\
\nonumber&&Z^{BAB^{2}A^2}_{3}=\{1, BAB^{2}A^2,  ABA^{2}B^2\}\,.
\end{eqnarray}
The thirteen $Z_{3}$ subgroups in each line above are related with each other by group conjugation.
\item{$Z_{3}\times Z_{3}$ subgroups}
\begin{eqnarray}
\nonumber &&Z^{A}_{3}\times Z^{BAB^2}_{3}=\{1, A,BAB^2A^2,A^2,BAB^2,ABA^2B^2,B^2A^2B,B^2AB,BA^2B^2\}, \\
\nonumber &&Z^{B}_{3}\times Z^{ABA^2}_{3}=\{1, B,BAB^2A^2,B^2,A^2BA,ABA^2B^2,AB^2A^2,ABA^2,A^2B^2A\},\\
\nonumber &&Z^{BA}_{3}\times Z^{B^2A^2}_{3}=\{1, BAB^2A^2,AB,ABA^2B^2,BA,A^2B^2,A^2BA^2,B^2A^2,AB^2A\}, \\
\nonumber &&Z^{BA^2}_{3}\times Z^{B^2A}_{3}=\{1, BAB^2A^2,ABA^2B^2,A^2B,AB^2,ABA,BAB,BA^2,B^2A\}\,.
\end{eqnarray}
All the four $Z_{3}\times Z_{3}$ subgroups are  related with each other by group conjugation.
\end{itemize}
As the number of irreducible representation is equal to the number of conjugacy class, the traditional flavor group $\Delta(27)$ has eleven inequivalent irreducible representations with nine singlets labelled as $\bm{1_{(r,s)}}$ ($r,s=0,1,2$) and two triplets labelled as $\bm{3}$ and $\bm{\bar{3}}$. The character table of $\Delta(27)$ is shown in table~\ref{tab:character_Delta27}. The first line of table~\ref{tab:character_Delta27} indicates representatives of the eleven conjugacy classes in the second line.
\begin{table}
\centering
\begin{tabular}{c|ccccccccccc}	\\
&\tabnode{$E$}&\tabnode{$A^2B^2$}& \tabnode{$A^2B$} &\tabnode{$A$}&\tabnode{$AB^2$}&\tabnode{$AB$}&\tabnode{$A^2$}&\tabnode{$B^2$}&\tabnode{$B$}&\tabnode{$BAB^2A^2$}&\tabnode{$ABA^2B^2$} \\ \hline
& $1C_{1}$ & $3C_{3}^{(1)}$ & $3C_{3}^{(2)}$ & $3C_{3}^{(3)}$ & $3C_{3}^{(4)}$ & $3C_{3}^{(5)}$ & $3C_{3}^{(6)}$ & $3C_{3}^{(7)}$ & $3C_{3}^{(8)}$ & $1C_{3}^{(1)}$ & $1C_{3}^{(2)}$ \\ \hline
~~~$\bm{1_{0,0}}$~~~&     1 & 1&   1 & 1&  1&  1&  1 & 1&  1 & 1 & 1\\
~~~\tabnode{$\bm{1_{0,1}}$}~~~&      1 &$\omega^2$ & $\omega$ &  1 &$\omega^2$& $\omega$ & 1 &$\omega^2$& $\omega$ & 1 & 1\\
		~~~\tabnode{$\bm{1_{0,2}}$}~~~&     1&  $\omega$ & $\omega^2$ & 1 & $\omega$ & $\omega^2$& 1 & $\omega$ & $\omega^2$ & 1&  1\\
		~~~\tabnode{$\bm{1_{1,0}}$}~~~&      1& $\omega^2$ &$\omega^2$ & $\omega$ & $\omega$ &  $\omega$ &$\omega^2$ & 1&  1 & 1&  1\\
		~~~\tabnode{$\bm{1_{1,1}}$}~~~&      1 & $\omega$ & 1 & $\omega$ & 1 &$\omega^2$ &$\omega^2$ &$\omega^2$& $\omega$ & 1 & 1\\
		~~~\tabnode{$\bm{1_{1,2}}$}~~~&     1 & 1 & $\omega$ & $\omega$ &$\omega^2$& 1 &$\omega^2$& $\omega$ & $\omega^2$ & 1 & 1\\
		~~~\tabnode{$\bm{1_{2,0}}$}~~~&     1  &$\omega$ &  $\omega$ & $\omega^2$ &$\omega^2$ &$\omega^2$ & $\omega$ & 1&  1&  1&  1\\
		~~~\tabnode{$\bm{1_{2,1}}$}~~~&     1  &1 &$\omega^2$ &$\omega^2$ & $\omega$ & 1&  $\omega$ & $\omega^2$ & $\omega$ & 1&  1\\
		~~~\tabnode{$\bm{1_{2,2}}$}~~~&     1 &$\omega^2$ & 1& $\omega^2$ & 1&  $\omega$ & $\omega$ & $\omega$ &$\omega^2$& 1 & 1\\
		~~~\tabnode{$\bm{3}$}~~~&     3&  0&  0& 0&  0& 0&  0&  0& 0&  $3 \omega$ &$3\omega^2$\\
		~~~\tabnode{$\bm{\bar{3}}$}~~~&    3  &0&  0& 0&  0&  0 & 0&  0 &0& $3 \omega^2$&  $3 \omega$\\
\end{tabular}
\caption{The character table of $\Delta(27)$, where $\omega$ is the cube root of unity $\omega=e^{2\pi i/3}$. The arrowed lines show the transformation of the irreducible representations and conjugacy classes of $\Delta(27)$ under the actions of the outer automorphisms $u_S$(blue), $u_T$(red) and $u_{K_*}$(green).\label{tab:character_Delta27}}
\begin{tikzpicture}[overlay]
\node [above=.3cm,minimum width=0pt] at (1) (c1){};
\node [above=.3cm,minimum width=0pt] at (2) (c2){};
\node [above=.3cm,minimum width=0pt] at (3) (c3){};
\node [above=.3cm,minimum width=0pt] at (4) (c4){};
\node [above=.3cm,minimum width=0pt] at (5) (c5){};
\node [above=.3cm,minimum width=0pt] at (6) (c6){};
\node [above=.3cm,minimum width=0pt] at (7) (c7){};
\node [above=.3cm,minimum width=0pt] at (8) (c8){};
\node [above=.3cm,minimum width=0pt] at (9) (c9){};
\node [above=.3cm,minimum width=0pt] at (10) (c10){};
\node [above=.3cm,minimum width=0pt] at (11) (c11){};
\node [left=.3cm,minimum width=0pt] at (12) (r12){};
\node [left=.3cm,minimum width=0pt] at (13) (r13){};
\node [left=.3cm,minimum width=0pt] at (14) (r14){};
\node [left=.3cm,minimum width=0pt] at (15) (r15){};
\node [left=.3cm,minimum width=0pt] at (16) (r16){};
\node [left=.3cm,minimum width=0pt] at (17) (r17){};
\node [left=.3cm,minimum width=0pt] at (18) (r18){};
\node [left=.3cm,minimum width=0pt] at (19) (r19){};
\node [left=.1cm,minimum width=0pt] at (20) (r31){};
\node [left=.1cm,minimum width=0pt] at (21) (r32){};
\node [right=.2cm,minimum width=0pt] at (12) (r12ri){};
\node [right=.2cm,minimum width=0pt] at (13) (r13ri){};
\node [right=.2cm,minimum width=0pt] at (14) (r14ri){};
\node [right=.2cm,minimum width=0pt] at (15) (r15ri){};
\node [right=.2cm,minimum width=0pt] at (16) (r16ri){};
\node [right=.2cm,minimum width=0pt] at (17) (r17ri){};
\node [right=.2cm,minimum width=0pt] at (18) (r18ri){};
\node [right=.2cm,minimum width=0pt] at (19) (r19ri){};
\node [right=.1cm,minimum width=0pt] at (20) (r31ri){};
\node [right=.1cm,minimum width=0pt] at (21) (r32ri){};
\draw [->,out=45,in=135,blue!50, very thick,below=1cm] (c2) to (c8);
\draw [->,out=35,in=105,blue!50, very thick,below=1cm] (c8) to (c6);
\draw [->,out=45,in=135,blue!50, very thick,below=1cm] (c6) to (c9);
\draw [->,out=40,in=140,blue!50, very thick,below=1cm] (c9) to (c2);
\draw [->,out=45,in=135,blue!50, very thick,below=1cm] (c3) to (c4);
\draw [->,out=45,in=135,blue!50, very thick,below=1cm] (c4) to (c5);
\draw [->,out=45,in=135,blue!50, very thick,below=1cm] (c5) to (c7);
\draw [->,out=45,in=135,blue!50, very thick,below=1cm] (c7) to (c3);
\draw [->,out=45,in=135,red!50, very thick,below=1cm] (c2) to (c3);
\draw [->,out=45,in=135,red!50, very thick,below=1cm] (c3) to (c7);
\draw [->,out=45,in=135,red!50, very thick,below=1cm] (c7) to (c2);
\draw [->,out=40,in=140,red!50, very thick,below=1cm] (c4) to (c6);
\draw [->,out=30,in=150,red!50, very thick,below=1cm] (c6) to (c5);
\draw [->,out=30,in=150,red!50, very thick,below=1cm] (c5) to (c4);
\draw [<-,out=-45,in=45,blue!50, very thick,below=1cm] (r12ri) to (r15ri);
\draw [<-,out=215,in=145,blue!50, very thick,below=1cm] (r15) to (r13);
\draw [<-,out=-70,in=70,blue!50, very thick,below=1cm] (r13ri) to (r19ri);
\draw [<-,out=195,in=165,blue!50, very thick,below=1cm] (r19) to (r12);
\draw [<-,out=225,in=135,blue!50, very thick,below=1cm] (r14) to (r18);
\draw [<-,out=-35,in=-35,blue!50, very thick,below=1cm] (r18ri) to (r17ri);
\draw [<-,out=180,in=180,blue!50, very thick,below=1cm] (r17) to (r16);
\draw [<-,out=-55,in=-55,blue!50, very thick,below=1cm] (r16ri) to (r14ri);
\draw [<-,out=215,in=145,red!50, very thick,below=1cm] (r12) to (r18);
\draw [<-,out=-30,in=30,red!50, very thick,below=1cm] (r18ri) to (r15ri);
\draw [<-,out=215,in=145,red!50, very thick,below=1cm] (r15) to (r12);
\draw [<-,out=-30,in=30,red!50, very thick,below=1cm] (r13ri) to (r16ri);
\draw [<-,out=215,in=145,red!50, very thick,below=1cm] (r16) to (r19);
\draw [<-,out=-50,in=50,red!50, very thick,below=1cm] (r19ri) to (r13ri);
\draw [<->,out=45,in=135,green!50, very thick,below=1cm] (c10) to (c11);
\draw [<->,out=180,in=180,green!50, very thick,below=1cm] (r31) to (r32);
		\draw [<->,out=215,in=145,green!50, very thick,below=1cm] (r12) to (r15);
		\draw [<->,out=215,in=145,green!50, very thick,below=1cm] (r13) to (r19);
		\draw [<->,out=215,in=145,green!50, very thick,below=1cm] (r14) to (r17);
		\draw [<->,out=45,in=135,green!50, very thick,below=1cm] (c2) to (c6);
		\draw [<->,out=45,in=135,green!50, very thick,below=1cm] (c3) to (c4);
		\draw [<->,out=40,in=140,green!50, very thick,below=1cm] (c5) to (c7);
	\end{tikzpicture}
\end{table}
In our working basis, the eleven irreducible representations of $\Delta(27)$ are taken to be
\begin{eqnarray}\label{eq:Delta27_irre}
	\nonumber \bm{1_{r,s}}&:&~~\rho_{\bm{1_{r,s}}}(A)=\omega^{r}, \qquad \rho_{\bm{1_{r,s}}}(B)=\omega^{s} \,, \quad \text{with} \quad r,s=0,1,2\,, \\
	\nonumber  \bm{3}&:&~~\rho_{{\bf 3}}(A)=\left(\begin{array}{ccc}
		0&1&0\\
		0&0&1\\
		1&0&0
	\end{array}\right),  \qquad
	\rho_{{\bf 3}}(B)=\left(\begin{array}{ccc}
		1 & 0 & 0 \\
		0 & \omega & 0 \\
		0 & 0 & \omega^{2}
	\end{array}\right) \,, \\
	 \bm{\bar{3}}&:&~~\rho_{\bm{\bar{3}}}(A)=\left(\begin{array}{ccc}
		0&1&0\\
		0&0&1\\
		1&0&0
	\end{array}\right),  \qquad
	\rho_{\bm{\bar{3}}}(B)=\left(\begin{array}{ccc}
		1 & 0 & 0 \\
		0 & \omega^{2} & 0 \\
		0 & 0 & \omega
	\end{array}\right) \,.
\end{eqnarray}
From the character table, it is easy to calculate the Kronecker products of two different irreducible representations of $\Delta(27)$ as
\begin{eqnarray}
	\nonumber &&\bm{1_{r,s}}\otimes \bm{1_{r^\prime,s^\prime}}=\bm{1_{[r+r^\prime],\,\,[s+s^\prime]}} , \qquad  \bm{1_{r,s}}\otimes \bm{3}=\bm{3}, \qquad \bm{1_{r,s}}\otimes \bm{\bar{3}}=\bm{\bar{3}}\,,  \\
\label{eq:KP_Delta27}	&& \bm{3}\otimes \bm{3}=\bm{\bar{3}_{S,1}}\otimes\bm{\bar{3}_{S,2}}\otimes\bm{\bar{3}_{A}} \,,   \qquad  \bm{\bar{3}}\otimes\bm{\bar{3}}=\bm{3_{S,1}}\otimes\bm{3_{S,2}}\otimes\bm{3_{A}} \,,  \qquad \bm{3}\otimes \bm{\bar{3}}=\sum^{2}_{r,s=0} \bm{1_{r,s}}\,,
\end{eqnarray}
where $r,s,r^\prime,s^\prime=0,1,2$, the integer $[n]\equiv n~(\text{mod}~3)$,  and the subscript $S$ ($A$) denotes symmetric (antisymmetric) combination. In the following, all CG coefficients are presented in the form
$\alpha\otimes\beta$, where $\alpha_{i}$ denotes for the elements of the
first representation and $\beta_{j}$ stands for the elements of the second representation of the tensor product. In the
following, we shall adopt the convention $\beta_{[3]}=\beta_{0}\equiv\beta_{3}$.
\begin{eqnarray*}
	\begin{array}{|c|c|c|c|c|c|c|}\hline\hline
		~\bm{1_{r,s}}\otimes\bm{1_{r^\prime,s^\prime}}=  \bm{1_{[r+r^{\prime}], [s+s^{\prime}]}}  ~ &  ~\bm{1_{r,s}} \otimes \bm{3}  = \bm{3}~ & ~\bm{1_{r,s} }\otimes\bm{\overline{3}}   = \bm{\overline{3}}   \\ \hline
		& &   \\[-0.22in]
		\bm{1_{[r+r^{\prime}], [s+s^{\prime}]}} :
		\alpha_1  \beta_1
		~ &~\bm{3}: \alpha _1\left(
		\begin{array}{c}
			\beta_{[1-s]}  \\
			\omega^r \beta_{[2-s]}  \\
			\omega^{2r}  \beta_{[3-s]}
		\end{array}
		\right) ~
		& ~ \bm{\bar{3}}: \alpha _1\left(
		\begin{array}{c}
			\beta_{[1+s]}
			\\
			\omega^r \beta_{[2+s]}
			\\
			\omega^{2r} \alpha_1\beta_{[3+s]}
		\end{array}
		\right) \\
		&  &  \\[-0.22in] \hline \hline
		~\bm{3}\otimes\bm{\overline{3}}= \sum_{k,r=0}^2 \bm{1}_{k,r} ~  & ~\bm{3}\otimes\bm{3}= \bm{\overline{3}_{S_1}} \oplus   \bm{\overline{3}_{S_2} }\oplus \bm{\overline{3}_A }~ & ~\bm{\overline{3}}\otimes\bm{\overline{3}}=\bm{3_{S_1}}      \oplus\bm{3_{S_2}}\oplus\bm{3_A }~\\ \hline
	&  &  \\[-0.28in]
	~\bm{1_{0,0}}:\alpha_1 \beta_1 + \alpha_2 \beta_2 + \alpha_3 \beta_3~  & ~ & \\
	&  &
	~  \\[-0.48in]
	~\bm{1_{0,1}}:\alpha _1 \beta _3+ \alpha _2 \beta _1+\alpha _3 \beta _2~  & ~\bm{\overline{3}_{S_1}} : \left(
	\begin{array}{c}
		\alpha_1 \beta_1
		\\
		\alpha_2 \beta_2
		\\
		\alpha_3 \beta_3
	\end{array}
	\right) ~  & ~\bm{3_{S_1}} : \left(
	\begin{array}{c}
		\alpha_1 \beta_1
		\\
		\alpha_2 \beta_2
		\\
		\alpha_3 \beta_3
	\end{array}
	\right)~\\
	& & \\[-0.54in]
	~\bm{1_{0,2}}:\alpha _1 \beta _2+\alpha _2 \beta _3+\alpha _3 \beta _1~  & ~  & ~\\

	&  &  \\[-0.2in]
	~\bm{1_{1,0}}:\alpha_1 \beta_1 + \omega^2 \alpha_2 \beta_2 + \omega \alpha_3 \beta_3~  & ~ & \\
	&  &
	~  \\[-0.48in]
	~\bm{1_{1,1}}:\alpha _1 \beta _3+\omega^2 \alpha _2 \beta _1+\omega\alpha _3 \beta _2~  & ~\bm{\overline{3}_{S_2}} : \left(
	\begin{array}{c}
		\alpha_2 \beta_3 + \alpha_3 \beta_2
		\\
		\alpha_3 \beta_1 + \alpha_1 \beta_3
		\\
		\alpha_1 \beta_2 + \alpha_2 \beta_1
	\end{array}
	\right)~  & ~\bm{3_{S_2}} : \left(
	\begin{array}{c}
		\alpha_2 \beta_3 + \alpha_3 \beta_2
		\\
		\alpha_3 \beta_1 + \alpha_1 \beta_3
		\\
		\alpha_1 \beta_2 + \alpha_2 \beta_1
	\end{array}
	\right)~\\
	& & \\[-0.54in]
	~\bm{1_{1,2}}:\alpha _1 \beta _2+\omega^2\alpha _2 \beta _3+\omega\alpha _3 \beta _1~  & ~  & ~\\

	&  &  \\[-0.2in]
	~\bm{1_{2,0}}:\alpha_1 \beta_1 +  \omega \alpha_2 \beta_2 + \omega^2 \alpha_3 \beta_3~  & ~ & \\
	&  &
	~  \\[-0.48in]
	~\bm{1_{2,1}}: \alpha _1 \beta _3+\omega\alpha _2 \beta _1+\omega^2 \alpha _3 \beta _2~  & ~\bm{\overline{3}_{A}} : \left(
	\begin{array}{c}
		\alpha_2 \beta_3 - \alpha_3 \beta_2
		\\
		\alpha_3 \beta_1 - \alpha_1 \beta_3
		\\
		\alpha_1 \beta_2 - \alpha_2 \beta_1
	\end{array}
	\right)~  & ~\bm{3_{A}} : \left(
	\begin{array}{c}
		\alpha_2 \beta_3 - \alpha_3 \beta_2
		\\
		\alpha_3 \beta_1 - \alpha_1 \beta_3
		\\
		\alpha_1 \beta_2 - \alpha_2 \beta_1
	\end{array}
	\right)~\\
	& & \\[-0.54in]
	~\bm{1_{2,2}}:\alpha _1 \beta _2+\omega\alpha _2 \beta _3+\omega^2\alpha _3 \beta _1~  & ~  & ~\\
	& & \\[-0.22in] \hline \hline
	\end{array}
\end{eqnarray*}

\section{\label{sec:Tp_group}The finite modular group $\Gamma^\prime_{3}\cong T^\prime$}

The finite modular group $\Gamma^\prime_{3}\cong T^\prime$ contains 24 elements and it can be generated by two generators $S$ and $T$ with~\cite{Liu:2019khw,Lu:2019vgm,Li:2021buv,Ding:2022aoe}
\begin{equation}\label{eq:Tp_Mult_rules}
	S^4=T^3=(ST)^3=1,\qquad S^2T=TS^2\,.
\end{equation}
The 24 elements of $T^\prime$ can be divided into the seven conjugacy classes as follows,
\begin{eqnarray}
\nonumber&1C_{1}=\{1\}\,,     \qquad   1C_{2}=\{S^2\},      &\qquad
6C_{4}=\{S,T^{2}ST,TST^{2},T^{2}S^{3}T,TS^{3}T^{2},S^{3}\}\,,   \\
\nonumber & 4C^{(1)}_{3}=\{T^{2},S^{3}T^{2},T^{2}S^{3},TS^{3}T\}\,,     &\qquad 4C^{(2)}_{3}=\{T,TS,ST,T^{2}ST^{2}\}\,,   \\
& 4C^{(1)}_{6}=\{TS^2,TS^{3},S^{3}T,T^{2}S^{3}T^{2}\}\,,   &\qquad 4C^{(2)}_{6}=\{ST^{2},T^{2}S,TST,T^{2}S^2\}\,.
\end{eqnarray}
where the conjugacy class is denoted by $nC_k$ with $n$ elements and the  subscript $k$ is the order of the elements. The finite modular group $T^\prime$ has three singlet representations $\bm{1}$, $\bm{1^\prime}$ and $\bm{1^{\prime\prime}}$, three doublet representations $\bm{2}$, $\bm{2^\prime}$ and $\bm{2^{\prime\prime}}$, and one triplet representation $\bm{3}$.  The character table of $T^\prime$ group is given in table~\ref{tab:character_Tp}. The Kronecker products between different irreducible representations can be obtained from the character table
\begin{eqnarray}
\nonumber 	&&\bm{1^a}\otimes \bm{1^b}=\bm{1^{[a+b]}} , \qquad  \bm{1^a}\otimes \bm{2^b}=\bm{2^{[a+b]}}, \qquad \bm{1^a}\otimes \bm{3}=\bm{3}, \qquad \bm{2^a}\otimes \bm{2^b}= \bm{1^{[a+b]}} \oplus \bm{3},   \\
\label{eq:Tp_KP}&& \bm{2^a}\otimes \bm{3}=\bm{2}\oplus  \bm{2^{\prime}}\oplus  \bm{2^{\prime\prime}} ,    \qquad
\bm{3}\otimes \bm{3}= \bm{1}\oplus \bm{1^{\prime}}\oplus \bm{1^{\prime\prime}}\oplus \bm{3_S}\oplus\bm{3_A} \,,
\end{eqnarray}
where $a,b=0,1,2$ and we denote $\bm{1^0}\equiv\bm{1}$, $\bm{1^1}\equiv\bm{1^\prime}$, $\bm{1^{2}}\equiv\bm{1^{\prime\prime}}$ and similarly for the doublet representations. On the right-hand-side $[a+b]$ is defined as $a+b$ modulo 3.  In the present work, the seven irreducible representations are taken to be
\begin{eqnarray}\label{eq:Tp_irre}
	\nonumber & \bm{1}^{a}:~~ \rho_{{\bm 1^a}}(S)=1, &\qquad \rho_{{\bm 1^a}}(T)=\omega^{a} \,,  \\
	\nonumber & \bm{2}^{a}:~~ \rho_{{\bm 2}^a}(S)=\frac{i}{\sqrt{3}}
	\left(\begin{array}{cc}
		1 &~ \sqrt{2} \\
		\sqrt{2} &~ -1 \\
	\end{array}\right), &\qquad \rho_{{\bm 2}^a}(T)=\omega^{a+1}\left(\begin{array}{cc}
		1 &~ 0 \\
		0 &~ \omega \\
	\end{array}\right) \,,  \\
	&
	\bm{3}:~~\rho_{{\bf 3}}(S)=\frac{1}{3}
	\left(\begin{array}{ccc}
		-1 &~ 2 &~ 2 \\
		2 &~ -1 &~ 2 \\
		2 &~ 2 &~ -1 \\
	\end{array}\right),  &\qquad
	\rho_{{\bf 3}}(T)=\left(\begin{array}{ccc}
		1 &~ 0 &~ 0 \\
		0 &~ \omega  &~ 0 \\
		0 &~ 0 &~ \omega ^2 \\
	\end{array}\right) \,,
\end{eqnarray}
with $a=0,1,2$.
\begin{table}[t!]
	\begin{center}
		\begin{tabular}{|c|c|c|c|c|c|c|c|} \hline \hline
			Classes  &  $1C_{1}$  & $1C_{2}$ &  $4C^{(1)}_{3}$   &  $4C^{(2)}_{3}$   &  $6C_{4}$   &  $4C^{(1)}_{6}$   &  $4C^{(2)}_{6}$ \\ \hline
			$\bm{1}$  & 1  &  1 &  1  &  1 &  1  &  1   &  1  \\
			$\bm{1^\prime}$  &  $1$  &  $1$ &  $\omega ^2$  &  $\omega $  &  $1$  &  $\omega$   &  $\omega ^2$  \\
			$\bm{1^{\prime\prime}}$  &  $1$  &  $1$  &  $\omega $  &  $\omega ^2$ &  $1$  &  $\omega ^2$   &  $\omega $  \\
			$\bm{2}$  &  $2$  &  $-2$ &  $-1$  &  $-1$ &  $0$  &  $1$    &  $1$   \\
			$\bm{2^\prime}$  &  $2$  &  $-2$ &  $-\omega ^2$  &  $-\omega $  &  $0$  &  $\omega $   &  $\omega ^2$   \\
			$\bm{2^{\prime\prime}}$  &  $2$  &  $-2$ &  $-\omega $  &  $-\omega ^2$ &  $0$  &  $\omega ^2$    &  $\omega $   \\
			$\bm{3}$  &  $3$  &  $3$ &  $0$  &  $0$ & $-1$  &  $0$    &  $0$  \\ \hline \hline
		\end{tabular}
		\caption{\label{tab:character_Tp}Character table of the finite modular group $T^\prime$. 
		}
	\end{center}
\end{table}
From the irreducible representation matrices of $T^\prime$ in Eq.~\eqref{eq:Tp_irre}, one can easy to check that
\begin{equation}
\begin{array}{ll}
	\rho^{*}_{\bm{2}}(T)=U_{2}^{\dagger}\rho_{\bm{2}}(T)U_{2}, & \qquad \rho^{*}_{\bm{2}}(S)=U_{2}^{\dagger}\rho_{\bm{2}}(S)U_{2}\,,\\
	\rho^{*}_{\bm{2'}}(T)=U_{2}^{\dagger}\rho_{\bm{2''}}(T)U_{2}, &\qquad \rho^{*}_{\bm{2'}}(S)=U_{2}^{\dagger}\rho_{\bm{2''}}(S)U_{2}\,,	\\
\rho^{*}_{\bm{3}}(T)=U^{\dagger}_{3}\rho_{\bm{3}}(T)U_{3}, &\qquad \rho^{*}_{\bm{3}}(S)=U^{\dagger}_{3}\rho_{\bm{3}}(S)U_{3}\,,
\end{array}
\end{equation}
with
\begin{equation}
	U_{2}=\left(\begin{array}{cc}0 & -1 \\ 1 & 0 \end{array}\right), \qquad
	U_{3}=\left(
	\begin{array}{ccc}
		1 & 0 & 0 \\
		0 & 0 & 1 \\
		0 & 1 & 0 \\
	\end{array}
	\right)\,.
\end{equation}
It implies that the irreducible representations $\bm{2^\prime}$ and $\bm{2^{\prime\prime}}$ are complex conjugated to each other by a unitary transformation $U_{2}$, and the irreducible representations $\bm{2}$ and $\bm{3}$ are  real representation. Thus we have
\begin{equation}
\begin{array}{lll}
	\alpha=(\alpha_{1},\alpha_{2})^{T}\sim  \bm{2},\qquad &\rightarrow &\qquad (-\alpha_{2}^{*},\alpha_{1}^{*})^{T}\sim \bm{2}\,,\\
	\alpha=(\alpha_{1},\alpha_{2})^{T}\sim  \bm{2'},\qquad &\rightarrow& \qquad (-\alpha_{2}^{*},\alpha_{1}^{*})^{T}\sim \bm{2''}\,,\\
	\alpha=(\alpha_{1},\alpha_{2})^{T}\sim  \bm{2''},\qquad &\rightarrow& \qquad (-\alpha_{2}^{*},\alpha_{1}^{*})^{T}\sim \bm{2'}\,, \\
		\alpha=(\alpha_{1},\alpha_{2},\alpha_{3})^{T}\sim  \bm{3},\qquad &\rightarrow& \qquad (\alpha^*_{1},\alpha_{3}^{*},\alpha_{2}^{*})^{T}\sim \bm{3}\,.
\end{array}
\end{equation}
In the end, we present the CG coefficients of $T^\prime$ in the chosen basis. As the same as the CG coefficients of $\Delta(27)$, we use the notation $\alpha_{i}$ ($\beta_{i}$) to denote the elements of the first (second) representation. Furthermore, we shall adopt the following notations facilitate the expressions of CG coefficients
\begin{equation}\label{eq:per_mat}
	P=\left(
	\begin{array}{ccc}
		0 & 0 & 1 \\
		1 & 0 & 0 \\
		0 & 1 & 0 \\
	\end{array}
	\right)\,.
\end{equation}
Then all the CG coefficients of $T^\prime$ for the decomposition of product representations can be written as
\begin{eqnarray*}
\begin{array}{|c|c|c|c|c|c|c|}\hline\hline
~\bm{1^{a}}\otimes\bm{2^{b}}=\bm{2^{[a+b]}} ~ &  ~\bm{1^{a}}\otimes\bm{3}=\bm{3}~ & ~\bm{2^{a}}\otimes\bm{2^{b}}=\bm{1^{[a+b]}}\oplus\bm{3}  \\ \hline
& & ~\bm{1^{[a+b]}}:\alpha _1 \beta _2-\alpha _2 \beta _1~ \\
	& &  \\[-0.33in]
\bm{2^{[a+b]}}:\alpha _1 \begin{pmatrix}
\beta _1\\
\beta _2
\end{pmatrix}
~ &~\bm{3}: \alpha _1P^{a}\begin{pmatrix}
\beta _1\\
\beta _2\\
\beta _3
\end{pmatrix} ~ & \\ [-0.65in]
& 	& ~ \bm{3}: P^{[a+b]}\begin{pmatrix}
\alpha _1 \beta _2+\alpha _2 \beta _1\\
\sqrt{2} \alpha _2 \beta _2\\
-\sqrt{2} \alpha _1 \beta _1
\end{pmatrix}  \\
& & \\[-0.22in] \hline \hline
\multicolumn{2}{|c|}{~\bm{3}\otimes\bm{3}=\bm{1}\oplus\bm{1^{\prime}}\oplus\bm{1^{\prime\prime}}\oplus\bm{3_{S}}\oplus\bm{3_{A}} ~}  & ~\bm{2^{a}}\otimes\bm{3}=\bm{2}\oplus\bm{2^{\prime}}\oplus\bm{2^{\prime\prime}} ~\\ \hline
\multicolumn{2}{|c|}{~\bm{1}:\alpha _1 \beta _1+\alpha _2 \beta _3+\alpha _3 \beta _2~}  & ~ \\
& &	~  \\[-0.39in]
\multicolumn{2}{|c|}{~\bm{1^{\prime}}:\alpha _1 \beta _2+\alpha _2 \beta _1+\alpha _3 \beta _3~}  & ~\bm{2^{a}}: \begin{pmatrix}
	\sqrt{2} \alpha _2 \beta _3+\alpha _1 \beta _1 \\
	\sqrt{2} \alpha _1 \beta _2-\alpha _2 \beta _1
\end{pmatrix} ~  \\
& & \\[-0.46in]
\multicolumn{2}{|c|}{~\bm{1^{\prime\prime}}:\alpha _1 \beta _3+\alpha _2 \beta _2+\alpha _3 \beta _1~}  & ~ \\
\multicolumn{2}{|c|}{ }&  \\[-0.18in]
\multicolumn{2}{|c|}{~\bm{3_{S}}:\begin{pmatrix}
	2 \alpha _1 \beta _1-\alpha _2 \beta _3-\alpha _3 \beta _2\\
	2 \alpha _3 \beta _3-\alpha _1 \beta _2-\alpha _2 \beta _1\\
	2 \alpha _2 \beta _2-\alpha _1 \beta _3-\alpha _3 \beta _1
\end{pmatrix}~}  & ~\bm{2^{[1+a]}}: \begin{pmatrix}
	\sqrt{2} \alpha _2 \beta _1+\alpha _1 \beta _2\\
	\sqrt{2} \alpha _1 \beta _3-\alpha _2 \beta _2
\end{pmatrix} ~\\
\multicolumn{2}{|c|}{ } & \\[-0.18in]
\multicolumn{2}{|c|}{~\bm{3_{A}}:\begin{pmatrix}
	\alpha _2 \beta _3-\alpha _3 \beta _2\\
	\alpha _1 \beta _2-\alpha _2 \beta _1\\
	\alpha _3 \beta _1-\alpha _1 \beta _3
\end{pmatrix}~}  & ~\bm{2^{[2+a]}}: \begin{pmatrix}
	\sqrt{2} \alpha _2 \beta _2+\alpha _1 \beta _3\\
	\sqrt{2} \alpha _1 \beta _1-\alpha _2 \beta _3
\end{pmatrix}~\\
	\multicolumn{2}{|c|}{ } & \\[-0.22in] \hline \hline
	\end{array}
\end{eqnarray*}

\section{\label{sec:Tp_modu_forms}Modular forms of level $N=3$}

The modular forms of weight $k$ and level $3$ are holomorphic functions of the complex modulus $\tau$ and they span a linear space $\mathcal{M}_{k}(\Gamma(3))$ with dimensional
\begin{equation}
\texttt{dim}\mathcal{M}_k(\Gamma(3))=k+1 , ~~~k \geq 1 \,,
\end{equation}
where the modular weight $k$ is a generic non-negative integer. Each modular form  $f^{(k)}_i(\tau)$ of weight $k$ and level $3$ is invariant under the traditional flavor symmetry and  transforms under the action of $\Gamma(3)$ in the following way
\begin{equation}
f^{(k)}_i(\tau) \rightarrow f^{(k)}_i\left(\gamma \tau\right)=(c\tau+d)^kf^{(k)}_i(\tau), \quad\forall~\gamma = \left(
\begin{array}{cc}
a  &  b  \\
c  &  d
\end{array}
\right)\in\Gamma(3)\,.
\end{equation}
Furthermore, the modular forms $f^{(k)}_i(\tau)$  can be arranged into some modular multiplets $Y^{(k)}_{\bm{r}}(\tau)\equiv (f_1(\tau),f_2(\tau),\cdots)^T$ which transform as certain irreducible representation $\bm{r}$ of the finite modular group $T^\prime$~\cite{Feruglio:2017spp,Liu:2019khw}
\begin{equation}
Y_{\bm{r}}^{(k)}(\gamma\tau)=(c\tau+d)^k\rho_{\bm{r}}(\gamma)Y_{\bm{r}}^{(k)}(\tau), \qquad \forall~\gamma\in\Gamma\,,
\end{equation}
where $\gamma$ is the representative element of the coset $\gamma\Gamma(N)$ in $T^\prime$, and $\rho_{\bm{r}}(\gamma)$ is the representation matrix of $\gamma$ in the irreducible representation $\bm{r}$ of $T^\prime$.

The modular forms of level 3 can be generated by the products of Dedekind function $\eta(\tau)$, and there are only two linearly independent weight $1$ and level 3 modular forms which can be taken to be~\cite{Liu:2019khw}:
\begin{eqnarray}
\nonumber&& \hat{e}_1(\tau)=\frac{\eta^{3}(3\tau)}{\eta(\tau)},\qquad \hat{e}_{2}(\tau)=\frac{\eta^{3}(\tau / 3)}{\eta(\tau)}\,,
\end{eqnarray}
where the Dedekind function $\eta(\tau)$ is defined as
\begin{equation}
\eta(\tau)=q^{1/24}\prod_{n=1}^\infty \left(1-q^n \right),\qquad q\equiv e^{2 \pi i\tau}\,.
\end{equation}
Under the actions of the two generators $S$ and $T$ of $\Gamma$, $\eta(\tau)$ transforms as
\begin{equation}
\eta(\tau)\stackrel{S}{\longrightarrow} \eta(-1/\tau)=\sqrt{-i \tau}~\eta(\tau)\,, \qquad \eta(\tau)\stackrel{T}{\longrightarrow}\eta(\tau+1)=e^{i \pi/12}\eta(\tau) \,,
\end{equation}
which lead to the following transformation properties~\cite{Liu:2019khw}
\begin{equation}
\begin{array}{ll}
\hat{e}_1(\tau) \stackrel{S}{\longrightarrow} 3^{-3/2}(-i\tau)\hat{e}_2(\tau),& \qquad \hat{e}_{2}(\tau)\stackrel{S}{\longrightarrow} 3^{3/2}(-i\tau)\hat{e}_1(\tau) \,. \\
\hat{e}_1(\tau)\stackrel{T}{\longrightarrow} \omega \hat{e}_1(\tau),& \qquad \hat{e}_{2}(\tau)\stackrel{T}{\longrightarrow} 3(1-\omega)\hat{e}_1 + \hat{e}_2 \,.
 \end{array}
\end{equation}
The two independent weight 1 modular forms of level 3 can be arranged into a doublet $\bm{2^{\prime\prime}}$ of $T^\prime$ in our basis given in Appendix~\ref{sec:Tp_group} and the corresponding modular multiplet $Y^{(1)}_{\bm{2^{\prime\prime}}}(\tau)$ is defined as
\begin{equation}\label{eq:modular_space}
Y^{(1)}_{\bm{2^{\prime\prime}}}(\tau)\equiv\begin{pmatrix}
Y_1(\tau) \\
Y_2(\tau)
\end{pmatrix}
=\begin{pmatrix}
3\hat{e}_1(\tau)+\hat{e}_2(\tau) \\
3\sqrt{2}\hat{e}_1(\tau)
\end{pmatrix}\,.
\end{equation}
 The expressions of the q-expansions of $Y_{i}(\tau)$ are given by
\begin{eqnarray}\label{eq:q_expressions}
\nonumber   Y_1(\tau) &=& 1+6 q+6 q^3+6 q^4+12 q^7+6 q^9\,, \\
 Y_2(\tau)  &=& 3 \sqrt{2} q^{1/3} (1+q+2 q^2+2 q^4+q^5+2 q^6+q^8+2 q^9)\,.
\end{eqnarray}
The higher weight modular multiplets can be constructed by the tensor products of lower weight modular multiplets. One can find the modular multiplets of any integral weight by using the weight 1 modular multiplet $Y^{(1)}_{\bm{2^{\prime\prime}}}(\tau)$ and the CG coefficients of $T^\prime$ in Appendix~\ref{sec:Tp_group}. We summarize the modular multiplets of level $N=3$ up to weight 10 in table~\ref{tab:MF_summary}. Their explicit forms of these modular multiplets are
\begin{equation}\label{eq:Yw2to10}
	\begin{array}{lll}
		Y^{(2)}_{\bm{3}}
		=\left(Y^{(1)}_{\bm{2^{\prime\prime}}}Y^{(1)}_{\bm{2^{\prime\prime}}}\right)_{\bm{3}}\,, & ~
		Y^{(3)}_{\bm{2^{\prime}}}
		=\left(Y^{(1)}_{\bm{2^{\prime\prime}}}Y^{(2)}_{\bm{3}}\right)_{\bm{2^{\prime}}} \,,& ~
		Y^{(3)}_{\bm{2^{\prime\prime}}}
		=\left(Y^{(1)}_{\bm{2^{\prime\prime}}}Y^{(2)}_{\bm{3}}\right)_{\bm{2^{\prime\prime}}}  \,,\\
		Y^{(4)}_{\bm{1}}
		=\left(Y^{(2)}_{\bm{3}}Y^{(2)}_{\bm{3}}\right)_{\bm{1}} \,, & ~
		Y^{(4)}_{\bm{1^{\prime}}}
		=\left(Y^{(2)}_{\bm{3}}Y^{(2)}_{\bm{3}}\right)_{\bm{1^{\prime}}} \,, & ~
		Y^{(4)}_{\bm{3}}
		=\left(Y^{(2)}_{\bm{3}}Y^{(2)}_{\bm{3}}\right)_{\bm{3}}\,, \\
		Y^{(5)}_{\bm{2}}
		=\left(Y^{(2)}_{\bm{3}}Y^{(3)}_{\bm{2^{\prime}}}\right)_{\bm{2}}\,, & ~
		Y^{(5)}_{\bm{2^{\prime}}}
		=\left(Y^{(2)}_{\bm{3}}Y^{(3)}_{\bm{2^{\prime}}}\right)_{\bm{2^{\prime}}} \,,& ~
		Y^{(5)}_{\bm{2^{\prime\prime}}}
		=\left(Y^{(2)}_{\bm{3}}Y^{(3)}_{\bm{2^{\prime}}}\right)_{\bm{2^{\prime\prime}}}  \,,\\
		Y^{(6)}_{\bm{1}}
		=\left(Y^{(3)}_{\bm{2^\prime}}Y^{(3)}_{\bm{2^{\prime\prime}}}\right)_{\bm{1}} \,, & ~
		Y^{(6)}_{\bm{3i}}
		=\left(Y^{(3)}_{\bm{2^\prime}}Y^{(3)}_{\bm{2^{\prime}}}\right)_{\bm{3}} \,, & ~
		Y^{(6)}_{\bm{3ii}}
		=\left(Y^{(3)}_{\bm{2^\prime}}Y^{(3)}_{\bm{2^{\prime\prime}}}\right)_{\bm{3}}\,, \\
		Y^{(7)}_{\bm{2}}
		=\left(Y^{(3)}_{\bm{2^{\prime}}}Y^{(4)}_{\bm{3}}\right)_{\bm{2}}\,, & ~
		Y^{(7)}_{\bm{2^{\prime}}}
		=\left(Y^{(3)}_{\bm{2^{\prime}}}Y^{(4)}_{\bm{1}}\right)_{\bm{2^{\prime}}} \,,& ~
		Y^{(7)}_{\bm{2^{\prime\prime}i}}
		=\left(Y^{(3)}_{\bm{2^{\prime}}}Y^{(4)}_{\bm{1^{\prime}}}\right)_{\bm{2^{\prime\prime}}}  \,,\\
		Y^{(7)}_{\bm{2^{\prime\prime}ii}}
		=\left(Y^{(3)}_{\bm{2^{\prime}}}Y^{(4)}_{\bm{3}}\right)_{\bm{2^{\prime\prime}}}  \,,&~
		Y^{(8)}_{\bm{1}}
		=\left(Y^{(4)}_{\bm{1}}Y^{(4)}_{\bm{1}}\right)_{\bm{1}} \,, & ~
		Y^{(8)}_{\bm{1^{\prime}}}
		=\left(Y^{(4)}_{\bm{1}}Y^{(4)}_{\bm{1^{\prime}}}\right)_{\bm{1^{\prime}}} \,,  \\
		Y^{(8)}_{\bm{1^{\prime\prime}}}
		=\left(Y^{(4)}_{\bm{1^\prime}}Y^{(4)}_{\bm{1^{\prime}}}\right)_{\bm{1^{\prime\prime}}} \,, & ~
		Y^{(8)}_{\bm{3i}}
		=\left(Y^{(4)}_{\bm{1}}Y^{(4)}_{\bm{3}}\right)_{\bm{3}} \,, & ~
		Y^{(8)}_{\bm{3ii}}
		=\left(Y^{(4)}_{\bm{1^\prime}}Y^{(4)}_{\bm{3}}\right)_{\bm{3}}\,, \\
		Y^{(9)}_{\bm{2}}
		=\left(Y^{(1)}_{\bm{2^{\prime\prime}}}Y^{(8)}_{\bm{1^{\prime}}}\right)_{\bm{2}}\,, & ~
		Y^{(9)}_{\bm{2^{\prime}i}}
		=\left(Y^{(1)}_{\bm{2^{\prime\prime}}}Y^{(8)}_{\bm{1^{\prime\prime}}}\right)_{\bm{2^{\prime}}} \,,& ~
		Y^{(9)}_{\bm{2^{\prime}ii}}
		=\left(Y^{(1)}_{\bm{2^{\prime\prime}}}Y^{(8)}_{\bm{3i}}\right)_{\bm{2^{\prime}}}  \,,\\
		Y^{(9)}_{\bm{2^{\prime\prime}i}}
		=\left(Y^{(1)}_{\bm{2^{\prime\prime}}}Y^{(8)}_{\bm{1}}\right)_{\bm{2^{\prime\prime}}} \,,& ~
		Y^{(9)}_{\bm{2^{\prime\prime}ii}}
		=\left(Y^{(1)}_{\bm{2^{\prime\prime}}}Y^{(8)}_{\bm{3ii}}\right)_{\bm{2^{\prime\prime}}}  \,, & ~
		Y^{(10)}_{\bm{1}}
		=\left(Y^{(2)}_{\bm{3}}Y^{(8)}_{\bm{3i}}\right)_{\bm{1}} \,,\\
		Y^{(10)}_{\bm{1^{\prime}}}
		=\left(Y^{(2)}_{\bm{3}}Y^{(8)}_{\bm{3ii}}\right)_{\bm{1^{\prime}}} \,, & ~
		Y^{(10)}_{\bm{3i}}
		=\left(Y^{(2)}_{\bm{3}}Y^{(8)}_{\bm{1}}\right)_{\bm{3}} \,, & ~
		Y^{(10)}_{\bm{3ii}}
		=\left(Y^{(2)}_{\bm{3}}Y^{(8)}_{\bm{1^{\prime}}}\right)_{\bm{3}} \,, \\
		Y^{(10)}_{\bm{3iii}}
		=\left(Y^{(2)}_{\bm{3}}Y^{(8)}_{\bm{1^{\prime\prime}}}\right)_{\bm{3}} \,.
	\end{array}
\end{equation}

\begin{table}[t!]
	\centering
	\renewcommand{\arraystretch}{1.2}
	\begin{tabular}{|c|c||c|c|}
		\hline  \hline	
		weight & Modular form $Y^{(k)}_{\bm{r}}$ & weight & Modular form $Y^{(k)}_{\bm{r}}$ \\  \hline
		
		$k=1$ &  $Y^{(1)}_{\bm{2^{\prime\prime}}}$ & $k=2$ & $Y^{(2)}_{\bm{3}}$   \\ \hline
		
		$k=3$ &   $Y^{(3)}_{\bm{2^{\prime}}}$, $Y^{(3)}_{\bm{2^{\prime\prime}}}$  & $k=4$ &  $Y^{(4)}_{\bm{1}}$, $Y^{(4)}_{\bm{1^{\prime}}}$, $Y^{(4)}_{\bm{3}}$  \\ \hline
		
		$k=5$ &  $Y^{(5)}_{\bm{2}}$, $Y^{(5)}_{\bm{2^{\prime}}}$, $Y^{(5)}_{\bm{2^{\prime\prime}}}$   & $k=6$ &  $Y^{(6)}_{\bm{1}}$, $Y^{(6)}_{\bm{3i}}$, $Y^{(6)}_{\bm{3ii}}$   \\  \hline
		
		$k=7$ &  $Y^{(7)}_{\bm{2}}$, $Y^{(7)}_{\bm{2^{\prime}}}$, $Y^{(7)}_{\bm{2^{\prime\prime}i}}$, $Y^{(7)}_{\bm{2^{\prime\prime}ii}}$   & $k=8$ &  $Y^{(8)}_{\bm{1}}$, $Y^{(8)}_{\bm{1^{\prime}}}$, $Y^{(8)}_{\bm{1^{\prime\prime}}}$, $Y^{(8)}_{\bm{3i}}$, $Y^{(8)}_{\bm{3ii}}$   \\  \hline
		
		$k=9$ &  $Y^{(9)}_{\bm{2}}$, $Y^{(9)}_{\bm{2^{\prime}i}}$, $Y^{(9)}_{\bm{2^{\prime}ii}}$, $Y^{(9)}_{\bm{2^{\prime\prime}i}}$, $Y^{(9)}_{\bm{2^{\prime\prime}ii}}$   & $k=10$ &  $Y^{(10)}_{\bm{1}}$, $Y^{(10)}_{\bm{1^{\prime}}}$,  $Y^{(10)}_{\bm{3i}}$, $Y^{(10)}_{\bm{3ii}}$, $Y^{(10)}_{\bm{3iii}}$   \\  \hline \hline

	\end{tabular}
	\caption{\label{tab:MF_summary}The higher integral weight  modular multiplets of level $3$ up to weight 10, the subscript $\bm{r}$ denotes the transformation property under  homogeneous finite group $\Gamma^\prime_3\cong T^\prime$.  }
\end{table}

\section{\label{app:Majorana-Superpotential-method2}Alternative method to derive the Majorana superpotential }

As shown in section~\ref{sec:mod_general}, it is convenient to first consider the constraints of traditional flavor symmetry and then the modular symmetry in the paradigm of eclectic flavor symmetry. Likewise we can first analyze the constraints of modular symmetry and subsequently the traditional flavor symmetry. In the following, we shall revisit the superpotential $\mathcal{W}_M$ for the Majorana fermion mass in this alternative method. The general expression of $\mathcal{W}_M$ is given in Eq.~\eqref{eq:W-Maj-expr1}, where $\psi^{c}$ and $\Phi$ are assumed to transform as $(\bm{3},\bm{3_0})$ under the eclectic symmetry group $\Delta(27)\rtimes T'$.

Let us first study the $T'$ invariance  of the superpotential $\mathcal{W}_M$. Notice that $\bm{3_0}$ can be decomposed into a singlet $\bm{1^{\prime}}$ and a doublet $\bm{2^{\prime\prime}}$ of $T'$ with $\bm{3_0}= \bm{1^{\prime}} \oplus\bm{2^{\prime\prime}}$, as shown in Eq.~\eqref{eq:delta27-triplet-decom-Tp}. With the modular form multiplets defined in Eq.~\eqref{eq:Yn_def}, the $T^\prime$ invariant
terms of $\mathcal{W}_M$ result from the following $T'$ tensor products
\begin{equation}
\bm{1} ~\subset~ \left( \bm{1}\oplus\bm{1^{\prime}}\oplus\bm{1^{\prime\prime}}\oplus\bm{2}\oplus\bm{2^{\prime}}\oplus\bm{2^{\prime\prime}}\oplus\bm{3} \right) \otimes \left( \bm{1^{\prime}} \oplus\bm{2^{\prime\prime}} \right) \otimes
	\left( \bm{1^{\prime}} \oplus\bm{2^{\prime\prime}} \right) \otimes \left( \bm{1^{\prime}} \oplus\bm{2^{\prime\prime}} \right)\;.
\end{equation}
From the Kronecker product of $T^\prime$ in Eq.~\eqref{eq:Tp_KP} we know
\begin{equation}
	\left( \bm{1^{\prime}} \oplus\bm{2^{\prime\prime}} \right) \otimes \left( \bm{1^{\prime}} \oplus\bm{2^{\prime\prime}} \right)=\bm{1^{\prime\prime}} \oplus\bm{2}\oplus\bm{2}\oplus\bm{1^{\prime}} \oplus\bm{3}\,.
\end{equation}
Since the contraction $(\psi^{c}\psi^c)_{\bm{1^{\prime}}}=0$ and the two doublet contractions $(\psi^c\psi^c)_{\bm{2}}$ should be symmetrized, the non-zero $T'$ contraction of $(\psi^{c}\psi^c)_{\bm{r}}$ are
\begin{equation}
\left( \bm{1^{\prime}} \oplus\bm{2^{\prime\prime}} \right) \otimes \left( \bm{1^{\prime}} \oplus\bm{2^{\prime\prime}} \right)=\bm{1^{\prime\prime}} \oplus\bm{2} \oplus\bm{3}\,.
\end{equation}
Consequently the trivial singlet $\bm{1}$ resulting from the tensor products of $T'$ representations are
\begin{eqnarray}
\nonumber \bm{1} ~&\subset&~ \left( \bm{1}\oplus\bm{1^{\prime}}\oplus\bm{1^{\prime\prime}}\oplus\bm{2}\oplus\bm{2^{\prime}}\oplus\bm{2^{\prime\prime}}\oplus\bm{3} \right)  \otimes
\left(\bm{1^{\prime\prime}} \oplus\bm{2} \oplus\bm{3}\right)\otimes \left( \bm{1^{\prime}} \oplus\bm{2^{\prime\prime}} \right) \\
~&\subset&~\left( \bm{1}\oplus\bm{1^{\prime}}\oplus\bm{1^{\prime\prime}}\oplus\bm{2}\oplus\bm{2^{\prime}}\oplus\bm{2^{\prime\prime}}\oplus\bm{3} \right)\otimes
\left(\bm{1}  \oplus\bm{2^{\prime}} \oplus\bm{3}\oplus\bm{2^{\prime}}\oplus\bm{1^{\prime\prime}}\oplus\bm{3}\oplus\bm{2}\oplus\bm{2^{\prime}}\oplus\bm{2^{\prime\prime}}\right)\,.
\end{eqnarray}
Hence, we can obtain nine $T'$ invariant contractions when the modular form multiplets run over all possible irreducible multiplets of $T^\prime$. The results are given by
\begin{eqnarray}
\nonumber \mathcal{W}_{M}&=&\alpha_{1}Y_{1} (\psi^c_{1}-\psi^c_{2})^2 (\phi_{1}-\phi_{2})-2 \sqrt{2}\omega\alpha_{2} Y_{2} (\psi^c_{1}-\psi^c_{2}) \left[\phi_{3} (\psi^c_{1}+\psi^c_{2})-\psi^c_{3} (\phi_{1}+\phi_{2})\right]\\
\nonumber &&+\alpha_{3}\left( \sqrt{2} \psi^c_{3} Y_{5}-\omega^2 Y_{4} (\psi^c_{1}+\psi^c_{2})\right) (\psi^c_{3} (\phi_{1}+\phi_{2})-\phi_{3} (\psi^c_{1}+\psi^c_{2}))\\
\nonumber && +\alpha_{4}\left[2 \psi^c_{3} Y_{6} \phi_{3} (\psi^c_{1}+\psi^c_{2})-\frac{\omega Y_{7} (\psi^c_{1}+\psi^c_{2})^2 (\phi_{1}+\phi_{2})}{ \sqrt{2}}+\psi^{c2}_{3} \left(Y_{6} (\phi_{1}+\phi_{2})+\sqrt{2}\omega  Y_{7} \phi_{3}\right)\right]\\
\nonumber &&-\frac{\alpha_{5}}{\sqrt{2}} (\psi^c_{1}-\psi^c_{2})^2 \left[\sqrt{2} Y_{9} \phi_{3}- \omega^2 Y_{8} (\phi_{1}+\phi_{2})\right]\\
\nonumber &&+2\alpha_{6} (\psi^c_{1}-\psi^c_{2}) (\phi_{1}-\phi_{2}) \left[Y_{8} (\psi^c_{1}+\psi^c_{2})- \sqrt{2}\omega \psi^c_{3} Y_{9}\right]\\
\nonumber &&+\frac{\alpha_7}{4} \left[\sqrt{2} Y_{8} \left((\psi^c_{1}+\psi^c_{2})^2 (\phi_{1}+\phi_{2})+4 \psi^{c2}_{3} \phi_{3}\right)+2\omega Y_{9} (\psi^c_{1}+\psi^c_{2}) (\phi_{3} (\psi^c_{1}+\psi^c_{2})+2 \psi^c_{3} (\phi_{1}+\phi_{2}))\right]\\
\nonumber &&+2\sqrt{2}\alpha_{8} (\psi^c_{1}-\psi^c_{2}) \left[ \omega Y_{11} (\phi_{3} (\psi^c_{1}+\psi^c_{2})+\psi^c_{3} (\phi_{1}+\phi_{2}))+ Y_{10} (\psi^c_{1}+\psi^c_{2}) (\phi_{1}+\phi_{2})-2\omega^2 \psi^c_{3} Y_{12} \phi_{3}\right]\\
\label{eq:Tp_Inv_WO}&&+\frac{\alpha_{9}}{2} (\phi_{1}-\phi_{2}) \left[-2\omega^2 \psi^c_{3} Y_{11} (\psi^c_{1}+\psi^c_{2})-\omega Y_{10} (\psi^c_{1}+\psi^c_{2})^2+42 \psi^{c2}_{3} Y_{12}\right]\,.
\end{eqnarray}
Furthermore, invariance under $\Delta(27)$ requires that $\mathcal{ W}_{M}$ must be a trivial singlet $\bm{1_{0,0}}$ of $\Delta(27)$. As both matter field $\psi^c$ and flavon field $\Phi$ are assigned to $\Delta(27)$ triplet $\bm{3}$, while the modular forms are invariant under the actions of  $\Delta(27)$. Consequently the trivial singlet $\bm{1_{0,0}}$ resulting from the tensor products of $\Delta(27)$ representations are
\begin{equation}
\nonumber \bm{1_{0,0}} ~\subset~ \bm{3}  \otimes  \bm{3}  \otimes
\bm{3}
=\left( \bm{\bar{3}_{S,1}}\oplus\bm{\bar{3}_{S,2}}\oplus\bm{\bar{3}_{A}}\right)\otimes  \bm{3} =\left( \bm{\bar{3}_{S,1}}\oplus\bm{\bar{3}_{S,2}}\right)\otimes\bm{3} \,.
\end{equation}
Hence there are only two $\Delta(27)$ invariant terms in $\mathcal{W}_M$,
\begin{equation}
\mathcal{W}_{M}=\sum_{i=1}^{12}\,Y_{i}\left[a_{i} \left(\left(\psi_{1}^{c}\right)^{2} \phi_{1}+\left(\psi_{2}^{c}\right)^2 \phi_{2}+\left(\psi_{3}^{c}\right)^{2} \phi_{3}\right)+a'_{i} (\psi^{c}_{1} \psi^c_{2} \phi_{3}+\psi^{c}_{1} \psi^c_{3} \phi_{2}+\psi^c_{2} \psi^c_{3} \phi_{1})\right]\,.
\end{equation}
We find that the $\Delta(27)$ flavor symmetry imposes the following restrictions on the coefficients $\alpha_i$ in
Eq.~\eqref{eq:Tp_Inv_WO}:
\begin{equation}
\alpha_{1}=\alpha_{2}=\alpha_{3}=\alpha_{4}=\alpha_{8}=\alpha_{9}=0, \qquad \alpha_{5}=\sqrt{2}\omega\alpha_{6}=\frac{\omega}{2}\alpha_{7}\equiv\omega\alpha\,.
\end{equation}
Thus the superpotential $\mathcal{W}_M$ reduces to
\begin{equation}
\label{eq:W-Maj-fin2}\mathcal{ W}_{M} =\alpha\left[\sqrt{2} Y_{8} \left(\left(\psi_{1}^{c}\right)^{2} \phi_{1}+\left(\psi_{2}^{c}\right)^2 \phi_{2}+\left(\psi_{3}^{c}\right)^{2} \phi_{3}\right)+2\omega Y_{9} (\psi^c_{1} \psi^c_{2} \phi_{3}+\psi^c_{1} \psi^c_{3} \phi_{2}+\psi^c_{2} \psi^c_{3} \phi_{1})\right]\,,
\end{equation}
where the complex coupling $\alpha$ is enforced to be real by the gCP transformation $K_*$. We obtain the same superpotential $\mathcal{ W}_{M}$ as that of Eq.~\eqref{eq:W-Maj-fin}, and only the modular form doublet $Y^{(k_Y)}_{\bm{2''}}(\tau)=\left(Y_8, Y_9\right)^{T}$ gives non-vanishing contribution.

\end{appendix}





\begin{thebibliography}{10}

\bibitem{Feruglio:2017spp}
F.~Feruglio, {\em {Are neutrino masses modular forms?}},
  \href{http://dx.doi.org/10.1142/9789813238053_0012}{pp.~227--266}.
\newblock 2019.
\newblock \href{http://arxiv.org/abs/1706.08749}{{\ttfamily arXiv:1706.08749
  [hep-ph]}}.

\bibitem{Feruglio:2019ybq}
F.~Feruglio and A.~Romanino, ``{Lepton flavor symmetries},''
  \href{http://dx.doi.org/10.1103/RevModPhys.93.015007}{{\em Rev. Mod. Phys.}
  {\bfseries 93} no.~1, (2021) 015007},
  \href{http://arxiv.org/abs/1912.06028}{{\ttfamily arXiv:1912.06028
  [hep-ph]}}.

\bibitem{Ding:2022nzn}
G.-J. Ding, X.-G. Liu, and C.-Y. Yao, ``{A minimal modular invariant neutrino
  model},'' \href{http://dx.doi.org/10.1007/JHEP01(2023)125}{{\em JHEP}
  {\bfseries 01} (2023) 125}, \href{http://arxiv.org/abs/2211.04546}{{\ttfamily
  arXiv:2211.04546 [hep-ph]}}.

\bibitem{Feruglio:2022kea}
F.~Feruglio, ``{The irresistible call of $\tau=i$},''
  \href{http://arxiv.org/abs/2211.00659}{{\ttfamily arXiv:2211.00659
  [hep-ph]}}.

\bibitem{Feruglio:2023mii}
F.~Feruglio, ``{Fermion masses, critical behavior and universality},''
  \href{http://arxiv.org/abs/2302.11580}{{\ttfamily arXiv:2302.11580
  [hep-ph]}}.

\bibitem{Chen:2019ewa}
M.-C. Chen, S.~Ramos-S\'anchez, and M.~Ratz, ``{A note on the predictions of
  models with modular flavor symmetries},''
  \href{http://dx.doi.org/10.1016/j.physletb.2019.135153}{{\em Phys. Lett. B}
  {\bfseries 801} (2020) 135153},
  \href{http://arxiv.org/abs/1909.06910}{{\ttfamily arXiv:1909.06910
  [hep-ph]}}.

\bibitem{Lu:2019vgm}
J.-N. Lu, X.-G. Liu, and G.-J. Ding, ``{Modular symmetry origin of texture
  zeros and quark lepton unification},''
  \href{http://dx.doi.org/10.1103/PhysRevD.101.115020}{{\em Phys. Rev. D}
  {\bfseries 101} no.~11, (2020) 115020},
  \href{http://arxiv.org/abs/1912.07573}{{\ttfamily arXiv:1912.07573
  [hep-ph]}}.

\bibitem{Baur:2019kwi}
A.~Baur, H.~P. Nilles, A.~Trautner, and P.~K.~S. Vaudrevange, ``{Unification of
  Flavor, CP, and Modular Symmetries},''
  \href{http://dx.doi.org/10.1016/j.physletb.2019.03.066}{{\em Phys. Lett. B}
  {\bfseries 795} (2019) 7--14},
  \href{http://arxiv.org/abs/1901.03251}{{\ttfamily arXiv:1901.03251
  [hep-th]}}.

\bibitem{Baur:2019iai}
A.~Baur, H.~P. Nilles, A.~Trautner, and P.~K.~S. Vaudrevange, ``{A String
  Theory of Flavor and $ {CP}$},''
  \href{http://dx.doi.org/10.1016/j.nuclphysb.2019.114737}{{\em Nucl. Phys. B}
  {\bfseries 947} (2019) 114737},
  \href{http://arxiv.org/abs/1908.00805}{{\ttfamily arXiv:1908.00805
  [hep-th]}}.

\bibitem{Kobayashi:2016ovu}
T.~Kobayashi, S.~Nagamoto, and S.~Uemura, ``{Modular symmetry in
  magnetized/intersecting D-brane models},''
  \href{http://dx.doi.org/10.1093/ptep/ptw184}{{\em PTEP} {\bfseries 2017}
  no.~2, (2017) 023B02}, \href{http://arxiv.org/abs/1608.06129}{{\ttfamily
  arXiv:1608.06129 [hep-th]}}.

\bibitem{Kobayashi:2018rad}
T.~Kobayashi, S.~Nagamoto, S.~Takada, S.~Tamba, and T.~H. Tatsuishi, ``{Modular
  symmetry and non-Abelian discrete flavor symmetries in string
  compactification},'' \href{http://dx.doi.org/10.1103/PhysRevD.97.116002}{{\em
  Phys. Rev. D} {\bfseries 97} no.~11, (2018) 116002},
  \href{http://arxiv.org/abs/1804.06644}{{\ttfamily arXiv:1804.06644
  [hep-th]}}.

\bibitem{Kobayashi:2018bff}
T.~Kobayashi and S.~Tamba, ``{Modular forms of finite modular subgroups from
  magnetized D-brane models},''
  \href{http://dx.doi.org/10.1103/PhysRevD.99.046001}{{\em Phys. Rev. D}
  {\bfseries 99} no.~4, (2019) 046001},
  \href{http://arxiv.org/abs/1811.11384}{{\ttfamily arXiv:1811.11384
  [hep-th]}}.

\bibitem{Kariyazono:2019ehj}
Y.~Kariyazono, T.~Kobayashi, S.~Takada, S.~Tamba, and H.~Uchida, ``{Modular
  symmetry anomaly in magnetic flux compactification},''
  \href{http://dx.doi.org/10.1103/PhysRevD.100.045014}{{\em Phys. Rev. D}
  {\bfseries 100} no.~4, (2019) 045014},
  \href{http://arxiv.org/abs/1904.07546}{{\ttfamily arXiv:1904.07546
  [hep-th]}}.

\bibitem{Ohki:2020bpo}
H.~Ohki, S.~Uemura, and R.~Watanabe, ``{Modular flavor symmetry on a magnetized
  torus},'' \href{http://dx.doi.org/10.1103/PhysRevD.102.085008}{{\em Phys.
  Rev. D} {\bfseries 102} no.~8, (2020) 085008},
  \href{http://arxiv.org/abs/2003.04174}{{\ttfamily arXiv:2003.04174
  [hep-th]}}.

\bibitem{Kikuchi:2020frp}
S.~Kikuchi, T.~Kobayashi, S.~Takada, T.~H. Tatsuishi, and H.~Uchida,
  ``{Revisiting modular symmetry in magnetized torus and orbifold
  compactifications},''
  \href{http://dx.doi.org/10.1103/PhysRevD.102.105010}{{\em Phys. Rev. D}
  {\bfseries 102} no.~10, (2020) 105010},
  \href{http://arxiv.org/abs/2005.12642}{{\ttfamily arXiv:2005.12642
  [hep-th]}}.

\bibitem{Almumin:2021fbk}
Y.~Almumin, M.-C. Chen, V.~Knapp-P\'erez, S.~Ramos-S\'anchez, M.~Ratz, and
  S.~Shukla, ``{Metaplectic Flavor Symmetries from Magnetized Tori},''
  \href{http://dx.doi.org/10.1007/JHEP05(2021)078}{{\em JHEP} {\bfseries 05}
  (2021) 078}, \href{http://arxiv.org/abs/2102.11286}{{\ttfamily
  arXiv:2102.11286 [hep-th]}}.

\bibitem{Nilles:2020nnc}
H.~P. Nilles, S.~Ramos-S\'anchez, and P.~K.~S. Vaudrevange, ``{Eclectic Flavor
  Groups},'' \href{http://dx.doi.org/10.1007/JHEP02(2020)045}{{\em JHEP}
  {\bfseries 02} (2020) 045}, \href{http://arxiv.org/abs/2001.01736}{{\ttfamily
  arXiv:2001.01736 [hep-ph]}}.

\bibitem{Nilles:2020kgo}
H.~P. Nilles, S.~Ramos-Sanchez, and P.~K.~S. Vaudrevange, ``{Lessons from
  eclectic flavor symmetries},''
  \href{http://dx.doi.org/10.1016/j.nuclphysb.2020.115098}{{\em Nucl. Phys. B}
  {\bfseries 957} (2020) 115098},
  \href{http://arxiv.org/abs/2004.05200}{{\ttfamily arXiv:2004.05200
  [hep-ph]}}.

\bibitem{Nilles:2020tdp}
H.~P. Nilles, S.~Ramos\textendash{}S\'anchez, and P.~K.~S. Vaudrevange,
  ``{Eclectic flavor scheme from ten-dimensional string theory \textendash{} I.
  Basic results},''
  \href{http://dx.doi.org/10.1016/j.physletb.2020.135615}{{\em Phys. Lett. B}
  {\bfseries 808} (2020) 135615},
  \href{http://arxiv.org/abs/2006.03059}{{\ttfamily arXiv:2006.03059
  [hep-th]}}.

\bibitem{Nilles:2020gvu}
H.~P. Nilles, S.~Ramos\textendash{}S\'anchez, and P.~K.~S. Vaudrevange,
  ``{Eclectic flavor scheme from ten-dimensional string theory - II detailed
  technical analysis},''
  \href{http://dx.doi.org/10.1016/j.nuclphysb.2021.115367}{{\em Nucl. Phys. B}
  {\bfseries 966} (2021) 115367},
  \href{http://arxiv.org/abs/2010.13798}{{\ttfamily arXiv:2010.13798
  [hep-th]}}.

\bibitem{King:2003xq}
S.~F. King and I.~N.~R. Peddie, ``{Canonical normalization and Yukawa
  matrices},'' \href{http://dx.doi.org/10.1016/j.physletb.2004.01.087}{{\em
  Phys. Lett. B} {\bfseries 586} (2004) 83--94},
  \href{http://arxiv.org/abs/hep-ph/0312237}{{\ttfamily arXiv:hep-ph/0312237}}.

\bibitem{King:2004tx}
S.~F. King, I.~N.~R. Peddie, G.~G. Ross, L.~Velasco-Sevilla, and O.~Vives,
  ``{Kahler corrections and softly broken family symmetries},''
  \href{http://dx.doi.org/10.1088/1126-6708/2005/07/049}{{\em JHEP} {\bfseries
  07} (2005) 049}, \href{http://arxiv.org/abs/hep-ph/0407012}{{\ttfamily
  arXiv:hep-ph/0407012}}.

\bibitem{Antusch:2007vw}
S.~Antusch, S.~F. King, and M.~Malinsky, ``{Third Family Corrections to Quark
  and Lepton Mixing in SUSY Models with non-Abelian Family Symmetry},''
  \href{http://dx.doi.org/10.1088/1126-6708/2008/05/066}{{\em JHEP} {\bfseries
  05} (2008) 066}, \href{http://arxiv.org/abs/0712.3759}{{\ttfamily
  arXiv:0712.3759 [hep-ph]}}.

\bibitem{Chen:2012ha}
M.-C. Chen, M.~Fallbacher, M.~Ratz, and C.~Staudt, ``{On predictions from
  spontaneously broken flavor symmetries},''
  \href{http://dx.doi.org/10.1016/j.physletb.2012.10.077}{{\em Phys. Lett. B}
  {\bfseries 718} (2012) 516--521},
  \href{http://arxiv.org/abs/1208.2947}{{\ttfamily arXiv:1208.2947 [hep-ph]}}.

\bibitem{Chen:2013aya}
M.-C. Chen, M.~Fallbacher, Y.~Omura, M.~Ratz, and C.~Staudt, ``{Predictivity of
  models with spontaneously broken non-Abelian discrete flavor symmetries},''
  \href{http://dx.doi.org/10.1016/j.nuclphysb.2013.04.020}{{\em Nucl. Phys. B}
  {\bfseries 873} (2013) 343--371},
  \href{http://arxiv.org/abs/1302.5576}{{\ttfamily arXiv:1302.5576 [hep-ph]}}.

\bibitem{Chen:2021prl}
M.-C. Chen, V.~Knapp-Perez, M.~Ramos-Hamud, S.~Ramos-Sanchez, M.~Ratz, and
  S.~Shukla, ``{Quasi\textendash{}eclectic modular flavor symmetries},''
  \href{http://dx.doi.org/10.1016/j.physletb.2021.136843}{{\em Phys. Lett. B}
  {\bfseries 824} (2022) 136843},
  \href{http://arxiv.org/abs/2108.02240}{{\ttfamily arXiv:2108.02240
  [hep-ph]}}.

\bibitem{Baur:2022hma}
A.~Baur, H.~P. Nilles, S.~Ramos-Sanchez, A.~Trautner, and P.~K.~S. Vaudrevange,
  ``{The first string-derived eclectic flavor model with realistic
  phenomenology},'' \href{http://dx.doi.org/10.1007/JHEP09(2022)224}{{\em JHEP}
  {\bfseries 09} (2022) 224}, \href{http://arxiv.org/abs/2207.10677}{{\ttfamily
  arXiv:2207.10677 [hep-ph]}}.

\bibitem{GAP}
The GAP~Group, {\em {GAP -- Groups, Algorithms, and Programming, Version
  4.12.2}}, 2022.
\newblock \url{{https://www.gap-system.org}}.

\bibitem{Liu:2019khw}
X.-G. Liu and G.-J. Ding, ``{Neutrino Masses and Mixing from Double Covering of
  Finite Modular Groups},''
  \href{http://dx.doi.org/10.1007/JHEP08(2019)134}{{\em JHEP} {\bfseries 08}
  (2019) 134}, \href{http://arxiv.org/abs/1907.01488}{{\ttfamily
  arXiv:1907.01488 [hep-ph]}}.

\bibitem{deAdelhartToorop:2011re}
R.~de~Adelhart~Toorop, F.~Feruglio, and C.~Hagedorn, ``{Finite Modular Groups
  and Lepton Mixing},''
  \href{http://dx.doi.org/10.1016/j.nuclphysb.2012.01.017}{{\em Nucl. Phys. B}
  {\bfseries 858} (2012) 437--467},
  \href{http://arxiv.org/abs/1112.1340}{{\ttfamily arXiv:1112.1340 [hep-ph]}}.

\bibitem{Novichkov:2019sqv}
P.~P. Novichkov, J.~T. Penedo, S.~T. Petcov, and A.~V. Titov, ``{Generalised CP
  Symmetry in Modular-Invariant Models of Flavour},''
  \href{http://dx.doi.org/10.1007/JHEP07(2019)165}{{\em JHEP} {\bfseries 07}
  (2019) 165}, \href{http://arxiv.org/abs/1905.11970}{{\ttfamily
  arXiv:1905.11970 [hep-ph]}}.

\bibitem{Ding:2021iqp}
G.-J. Ding, F.~Feruglio, and X.-G. Liu, ``{CP symmetry and symplectic modular
  invariance},'' \href{http://dx.doi.org/10.21468/SciPostPhys.10.6.133}{{\em
  SciPost Phys.} {\bfseries 10} no.~6, (2021) 133},
  \href{http://arxiv.org/abs/2102.06716}{{\ttfamily arXiv:2102.06716
  [hep-ph]}}.

\bibitem{Qu:2021jdy}
B.-Y. Qu, X.-G. Liu, P.-T. Chen, and G.-J. Ding, ``{Flavor mixing and CP
  violation from the interplay of an S4 modular group and a generalized CP
  symmetry},'' \href{http://dx.doi.org/10.1103/PhysRevD.104.076001}{{\em Phys.
  Rev. D} {\bfseries 104} no.~7, (2021) 076001},
  \href{http://arxiv.org/abs/2106.11659}{{\ttfamily arXiv:2106.11659
  [hep-ph]}}.

\bibitem{Grimus:1995zi}
W.~Grimus and M.~N. Rebelo, ``{Automorphisms in gauge theories and the
  definition of CP and P},''
  \href{http://dx.doi.org/10.1016/S0370-1573(96)00030-0}{{\em Phys. Rept.}
  {\bfseries 281} (1997) 239--308},
  \href{http://arxiv.org/abs/hep-ph/9506272}{{\ttfamily arXiv:hep-ph/9506272}}.

\bibitem{Feruglio:2012cw}
F.~Feruglio, C.~Hagedorn, and R.~Ziegler, ``{Lepton Mixing Parameters from
  Discrete and CP Symmetries},''
  \href{http://dx.doi.org/10.1007/JHEP07(2013)027}{{\em JHEP} {\bfseries 07}
  (2013) 027}, \href{http://arxiv.org/abs/1211.5560}{{\ttfamily arXiv:1211.5560
  [hep-ph]}}.

\bibitem{Ding:2013hpa}
G.-J. Ding, S.~F. King, C.~Luhn, and A.~J. Stuart, ``{Spontaneous CP violation
  from vacuum alignment in $S_4$ models of leptons},''
  \href{http://dx.doi.org/10.1007/JHEP05(2013)084}{{\em JHEP} {\bfseries 05}
  (2013) 084}, \href{http://arxiv.org/abs/1303.6180}{{\ttfamily arXiv:1303.6180
  [hep-ph]}}.

\bibitem{Holthausen:2012dk}
M.~Holthausen, M.~Lindner, and M.~A. Schmidt, ``{CP and Discrete Flavour
  Symmetries},'' \href{http://dx.doi.org/10.1007/JHEP04(2013)122}{{\em JHEP}
  {\bfseries 04} (2013) 122}, \href{http://arxiv.org/abs/1211.6953}{{\ttfamily
  arXiv:1211.6953 [hep-ph]}}.

\bibitem{Chen:2014tpa}
M.-C. Chen, M.~Fallbacher, K.~T. Mahanthappa, M.~Ratz, and A.~Trautner, ``{CP
  Violation from Finite Groups},''
  \href{http://dx.doi.org/10.1016/j.nuclphysb.2014.03.023}{{\em Nucl. Phys. B}
  {\bfseries 883} (2014) 267--305},
  \href{http://arxiv.org/abs/1402.0507}{{\ttfamily arXiv:1402.0507 [hep-ph]}}.

\bibitem{Ishiguro:2020tmo}
K.~Ishiguro, T.~Kobayashi, and H.~Otsuka, ``{Landscape of Modular Symmetric
  Flavor Models},'' \href{http://dx.doi.org/10.1007/JHEP03(2021)161}{{\em JHEP}
  {\bfseries 03} (2021) 161}, \href{http://arxiv.org/abs/2011.09154}{{\ttfamily
  arXiv:2011.09154 [hep-ph]}}.

\bibitem{Novichkov:2022wvg}
P.~P. Novichkov, J.~T. Penedo, and S.~T. Petcov, ``{Modular flavour symmetries
  and modulus stabilisation},''
  \href{http://dx.doi.org/10.1007/JHEP03(2022)149}{{\em JHEP} {\bfseries 03}
  (2022) 149}, \href{http://arxiv.org/abs/2201.02020}{{\ttfamily
  arXiv:2201.02020 [hep-ph]}}.

\bibitem{Chen:2014wxa}
P.~Chen, C.-C. Li, and G.-J. Ding, ``{Lepton Flavor Mixing and CP Symmetry},''
  \href{http://dx.doi.org/10.1103/PhysRevD.91.033003}{{\em Phys. Rev. D}
  {\bfseries 91} (2015) 033003},
  \href{http://arxiv.org/abs/1412.8352}{{\ttfamily arXiv:1412.8352 [hep-ph]}}.

\bibitem{Chen:2015nha}
P.~Chen, C.-Y. Yao, and G.-J. Ding, ``{Neutrino Mixing from CP Symmetry},''
  \href{http://dx.doi.org/10.1103/PhysRevD.92.073002}{{\em Phys. Rev. D}
  {\bfseries 92} no.~7, (2015) 073002},
  \href{http://arxiv.org/abs/1507.03419}{{\ttfamily arXiv:1507.03419
  [hep-ph]}}.

\bibitem{Esteban:2020cvm}
I.~Esteban, M.~C. Gonzalez-Garcia, M.~Maltoni, T.~Schwetz, and A.~Zhou, ``{The
  fate of hints: updated global analysis of three-flavor neutrino
  oscillations},'' \href{http://dx.doi.org/10.1007/JHEP09(2020)178}{{\em JHEP}
  {\bfseries 09} (2020) 178}, \href{http://arxiv.org/abs/2007.14792}{{\ttfamily
  arXiv:2007.14792 [hep-ph]}}.

\bibitem{Antusch:2013jca}
S.~Antusch and V.~Maurer, ``{Running quark and lepton parameters at various
  scales},'' \href{http://dx.doi.org/10.1007/JHEP11(2013)115}{{\em JHEP}
  {\bfseries 11} (2013) 115}, \href{http://arxiv.org/abs/1306.6879}{{\ttfamily
  arXiv:1306.6879 [hep-ph]}}.

\bibitem{KamLAND-Zen:2022tow}
{\bfseries KamLAND-Zen} Collaboration, S.~Abe {\em et~al.}, ``{Search for the
  Majorana Nature of Neutrinos in the Inverted Mass Ordering Region with
  KamLAND-Zen},'' \href{http://dx.doi.org/10.1103/PhysRevLett.130.051801}{{\em
  Phys. Rev. Lett.} {\bfseries 130} no.~5, (2023) 051801},
  \href{http://arxiv.org/abs/2203.02139}{{\ttfamily arXiv:2203.02139
  [hep-ex]}}.

\bibitem{SNO:2015wyx}
{\bfseries SNO+} Collaboration, S.~Andringa {\em et~al.}, ``{Current Status and
  Future Prospects of the SNO+ Experiment},''
  \href{http://dx.doi.org/10.1155/2016/6194250}{{\em Adv.High Energy Phys.}
  {\bfseries 2016} (2016) 6194250},
  \href{http://arxiv.org/abs/1508.05759}{{\ttfamily arXiv:1508.05759
  [physics.ins-det]}}.

\bibitem{LEGEND:2017cdu}
{\bfseries LEGEND} Collaboration, N.~Abgrall {\em et~al.},
  \href{http://dx.doi.org/10.1063/1.5007652}{``{The Large Enriched Germanium
  Experiment for Neutrinoless Double Beta Decay (LEGEND)},''} vol.~1894,
  p.~020027.
\newblock 2017.
\newblock \href{http://arxiv.org/abs/1709.01980}{{\ttfamily arXiv:1709.01980
  [physics.ins-det]}}.

\bibitem{nEXO:2017nam}
{\bfseries nEXO} Collaboration, J.~Albert {\em et~al.}, ``{Sensitivity and
  Discovery Potential of nEXO to Neutrinoless Double Beta Decay},''
  \href{http://dx.doi.org/10.1103/PhysRevC.97.065503}{{\em Phys.Rev.C}
  {\bfseries 97} no.~6, (2018) 065503},
  \href{http://arxiv.org/abs/1710.05075}{{\ttfamily arXiv:1710.05075
  [nucl-ex]}}.

\bibitem{Planck:2018vyg}
{\bfseries Planck} Collaboration, N.~Aghanim {\em et~al.}, ``{Planck 2018
  results. VI. Cosmological parameters},''
  \href{http://dx.doi.org/10.1051/0004-6361/201833910}{{\em Astron. Astrophys.}
  {\bfseries 641} (2020) A6}, \href{http://arxiv.org/abs/1807.06209}{{\ttfamily
  arXiv:1807.06209 [astro-ph.CO]}}. [Erratum: Astron.Astrophys. 652, C4
  (2021)].

\bibitem{Harrison:2002kp}
P.~F. Harrison and W.~G. Scott, ``{Symmetries and generalizations of tri -
  bimaximal neutrino mixing},''
  \href{http://dx.doi.org/10.1016/S0370-2693(02)01753-7}{{\em Phys. Lett. B}
  {\bfseries 535} (2002) 163--169},
  \href{http://arxiv.org/abs/hep-ph/0203209}{{\ttfamily arXiv:hep-ph/0203209}}.

\bibitem{Grimus:2003yn}
W.~Grimus and L.~Lavoura, ``{A Nonstandard CP transformation leading to maximal
  atmospheric neutrino mixing},''
  \href{http://dx.doi.org/10.1016/j.physletb.2003.10.075}{{\em Phys. Lett. B}
  {\bfseries 579} (2004) 113--122},
  \href{http://arxiv.org/abs/hep-ph/0305309}{{\ttfamily arXiv:hep-ph/0305309}}.

\bibitem{Farzan:2006vj}
Y.~Farzan and A.~Y. Smirnov, ``{Leptonic CP violation: Zero, maximal or between
  the two extremes},''
  \href{http://dx.doi.org/10.1088/1126-6708/2007/01/059}{{\em JHEP} {\bfseries
  01} (2007) 059}, \href{http://arxiv.org/abs/hep-ph/0610337}{{\ttfamily
  arXiv:hep-ph/0610337}}.

\bibitem{DUNE:2015lol}
{\bfseries DUNE} Collaboration, R.~Acciarri {\em et~al.}, ``{Long-Baseline
  Neutrino Facility (LBNF) and Deep Underground Neutrino Experiment (DUNE)}:
  {Conceptual Design Report, Volume 2: The Physics Program for DUNE at LBNF},''
  \href{http://arxiv.org/abs/1512.06148}{{\ttfamily arXiv:1512.06148
  [physics.ins-det]}}.

\bibitem{Hyper-KamiokandeProto-:2015xww}
{\bfseries Hyper-Kamiokande Proto-} Collaboration, K.~Abe {\em et~al.},
  ``{Physics potential of a long-baseline neutrino oscillation experiment using
  a J-PARC neutrino beam and Hyper-Kamiokande},''
  \href{http://dx.doi.org/10.1093/ptep/ptv061}{{\em PTEP} {\bfseries 2015}
  (2015) 053C02}, \href{http://arxiv.org/abs/1502.05199}{{\ttfamily
  arXiv:1502.05199 [hep-ex]}}.

\bibitem{deMedeirosVarzielas:2006fc}
I.~de~Medeiros~Varzielas, S.~F. King, and G.~G. Ross, ``{Neutrino
  tri-bi-maximal mixing from a non-Abelian discrete family symmetry},''
  \href{http://dx.doi.org/10.1016/j.physletb.2007.03.009}{{\em Phys. Lett. B}
  {\bfseries 648} (2007) 201--206},
  \href{http://arxiv.org/abs/hep-ph/0607045}{{\ttfamily arXiv:hep-ph/0607045}}.

\bibitem{Branco:2015gna}
G.~C. Branco, I.~de~Medeiros~Varzielas, and S.~F. King, ``{Invariant approach
  to $\mathcal {CP}$ in unbroken $\Delta(27)$},''
  \href{http://dx.doi.org/10.1016/j.nuclphysb.2015.07.024}{{\em Nucl. Phys. B}
  {\bfseries 899} (2015) 14--36},
  \href{http://arxiv.org/abs/1505.06165}{{\ttfamily arXiv:1505.06165
  [hep-ph]}}.

\bibitem{deMedeirosVarzielas:2015amz}
I.~de~Medeiros~Varzielas, ``{$\Delta(27)$ family symmetry and neutrino
  mixing},'' \href{http://dx.doi.org/10.1007/JHEP08(2015)157}{{\em JHEP}
  {\bfseries 08} (2015) 157}, \href{http://arxiv.org/abs/1507.00338}{{\ttfamily
  arXiv:1507.00338 [hep-ph]}}.

\bibitem{Chen:2015jta}
P.~Chen, G.-J. Ding, A.~D. Rojas, C.~A. Vaquera-Araujo, and J.~W.~F. Valle,
  ``{Warped flavor symmetry predictions for neutrino physics},''
  \href{http://dx.doi.org/10.1007/JHEP01(2016)007}{{\em JHEP} {\bfseries 01}
  (2016) 007}, \href{http://arxiv.org/abs/1509.06683}{{\ttfamily
  arXiv:1509.06683 [hep-ph]}}.

\bibitem{Ishimori:2010au}
H.~Ishimori, T.~Kobayashi, H.~Ohki, Y.~Shimizu, H.~Okada, and M.~Tanimoto,
  ``{Non-Abelian Discrete Symmetries in Particle Physics},''
  \href{http://dx.doi.org/10.1143/PTPS.183.1}{{\em Prog. Theor. Phys. Suppl.}
  {\bfseries 183} (2010) 1--163},
  \href{http://arxiv.org/abs/1003.3552}{{\ttfamily arXiv:1003.3552 [hep-th]}}.

\bibitem{Li:2021buv}
C.-C. Li, X.-G. Liu, and G.-J. Ding, ``{Modular symmetry at level 6 and a new
  route towards finite modular groups},''
  \href{http://dx.doi.org/10.1007/JHEP10(2021)238}{{\em JHEP} {\bfseries 10}
  (2021) 238}, \href{http://arxiv.org/abs/2108.02181}{{\ttfamily
  arXiv:2108.02181 [hep-ph]}}.

\bibitem{Ding:2022aoe}
G.-J. Ding, F.~R. Joaquim, and J.-N. Lu, ``{Texture-zero patterns of lepton
  mass matrices from modular symmetry},''
  \href{http://arxiv.org/abs/2211.08136}{{\ttfamily arXiv:2211.08136
  [hep-ph]}}.

\end{thebibliography}

\providecommand{\href}[2]{#2}\begingroup\raggedright\endgroup

\end{document}